\documentclass{svproc}
\usepackage{url}

\usepackage{graphicx}
\usepackage{multicol}
\usepackage{footmisc}
\usepackage{xcolor}
\usepackage[breaklinks=true,colorlinks=true,linkcolor=blue,urlcolor=blue,citecolor=blue]{hyperref}
\usepackage{amsmath}
\usepackage{amssymb}

\newcommand{\e}{{\mathrm{e}}}
\newcommand{\sech}{{\mathrm{sech}}}
\newcommand{\cn}{{\mathrm{cn}}}

\begin{document}
\mainmatter              
\title{
Symbolic computation of solitary wave solutions and solitons through homogenization of degree
}
\titlerunning{Simplified Hirota Method}  
\author{
Willy Hereman\inst{1} \and \"{U}nal G\"{o}kta\c{s}\inst{2}
}
\authorrunning{Willy Hereman, \"{U}nal G\"{o}kta\c{s}} 
%
\tocauthor{Willy Hereman, and \"{U}nal G\"{o}kta\c{s}}
\institute{
Department of Applied Mathematics and Statistics,
Colorado School of Mines,
Golden CO 80401-1887, USA, \\
\email{whereman@mines.edu},\\ WWW home page:
\texttt{http://inside.mines.edu/\homedir whereman}
\and
Department of Computer Science and Engineering,
Texas A\&M University,
College Station, TX 77843-3112, USA, \\
\email{ugoktas@tamu.edu}
}

\maketitle              

\begin{abstract}

A simplified Hirota method for the computation of solitary waves 
and solitons of nonlinear partial differential equations (PDEs) is presented.
A change of dependent variable transforms the PDE into an equation 
that is homogeneous of degree.
Solitons are then computed using a perturbation-like scheme
involving linear and nonlinear operators in a finite number of steps.
The method is applied to fifth-order Korteweg-de Vries (KdV) equations
due to Lax, Sawada-Kotera, and Kaup-Kupershmidt.

The method works for non-quadratic homogeneous equations for
which the bilinear form might be unknown.
Furthermore, homogenization of degree allows one to compute solitary wave
solutions of nonlinear PDEs that do not have solitons.
Examples include the Fisher and FitzHugh-Nagumo equations,
and a combined KdV-Burgers equation.
When applied to a wave equation with a cubic source term,
one gets a ``bi-soliton" solution describing the coalescence of
two wavefronts.

The method is largely algorithmic and implemented in {\em Mathematica}.


\end{abstract}
\vfill
\newpage
\phantom{.}
\vskip 2.0cm
\noindent
\begin{center}
\includegraphics[width=4.2in,height=5.34in] 
{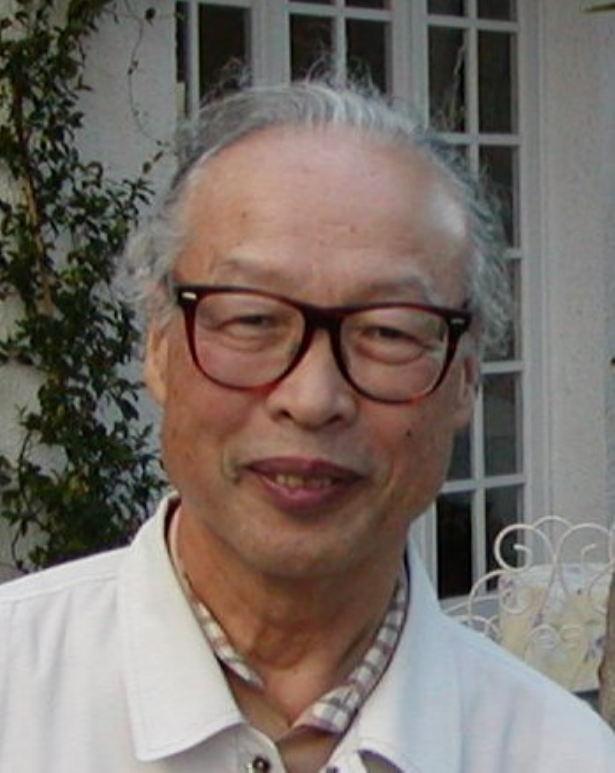}
\end{center}
\vspace*{1mm}
\begin{center}
{In memory of Prof.\ R.\ Hirota (1932-2015)} \\
{Photograph courtesy of J.\ Hietarinta.}
\end{center}
\vfill
\newpage
\section{Introduction}
\label{introduction}

In the 1970s, Hirota \cite{hirota-physrevlett-1972,hirota-jpsjpn-1972}
started working on an algebraic method to compute solitons of completely 
integrable nonlinear partial differential equations (PDEs).
His method has three major steps.
Given a nonlinear PDE,
(i) change the dependent variable (a.k.a.\ apply Hirota's transformation)
so that the transformed PDE is homogeneous of degree in a new dependent
variable (or variables),
(ii) express that homogeneous equation into one or more bilinear
equations using the Hirota operators,
(iii) solve the bilinear equation(s) using a perturbation-like
scheme that terminates after a finite number of steps.

Finding the Hirota transformation is quite challenging and
often requires insight and ingenuity.
Based on experience, Hietarinta \cite{hietarinta-kluwer-1990}
provides some useful tips for finding a suitable candidate
thereby reducing the guesswork.

Next, finding the appropriate bilinear form for the
homogeneous equation can also be a difficult task.
In particular in cases where the homogeneous equation is cubic
or quartic in the new dependent variable
and would have to be decoupled into a pair of bilinear equations,
either involving an extra independent variable or an
additional function \cite{hietarinta-lnp767-2009}.
To circumvent this difficulty, we will not use the bilinear form
of the homogeneous equation but include it for completeness.

To compute solitons, the type of solutions one seeks for the
homogeneous equation is quite specific.
They are a {\em finite} sums of polynomials in exponential
functions with different traveling wave arguments.
The terms in that sum are computed order-by-order, using a
``tracking" or ``bookkeeping" parameter $(\epsilon)$ which is set equal
to one\footnote{Unlike the {\em small} parameter $\epsilon$ used in
perturbation methods where one seeks {\em approximate} solutions up to
some order in $\epsilon$.}
after the exact solutions are computed.

Hirota's method \cite{hirota-book-miura-1976,hirota-solitons-1980,hirota-book-jimbo-miwa-1983,hirota-book-takeno-1985,hirota-book-2004}
can be found in many books on solitons and complete integrability \cite{ablowitz-clarkson-book-1991,ablowitz-segur-book-1981,drazin-johnson-book-1989,date-miwa-etal-2000,newell-book-1985},
books on differential equations (e.g., \cite{wazwaz-handbook-2008}),
encyclopedia (e.g., \cite{wazwaz-encyclop-2015}),
and survey papers \cite{caudrey-ptrsa-2011,ma-pdeam-2022,ma-ijnsns-2021,mohanetal-hss-2022,satsuma-lnp495-1997}
most noteworthy those by
Hietarinta \cite{hietarinta-lnp495-1997,hietarinta-jmodphysa-1997,hietarinta-lnp767-2009}.

Hietarinta's papers have a wealth of information about Hirota's method:
how to use it to construct regular and oscillatory solitons (breathers),
B\"acklund transformations and Lax pairs,
and as a tool in a computer-aided search for possibly new
completely integrable systems.
His surveys have a plethora of examples
including nonlinear Schr\"odinger (NLS) equations,
the sine- and sinh-Gordon equations, shallow water wave equations,
the Sasa-Satsuma equation, and systems of coupled equations such as the
Hirota-Satsuma and Davey-Stewartson systems.

Hirota wrote a book \cite{hirota-book-2004} about his method.
As far as we know, the only other book about the bilinear method
is by Matsuno \cite{matsuno-book-1984}.
Several theses, for example, \cite{pekcan-ms-thesis-2005,willox-phd-thesis-1993,zhuang-ms-thesis-1991}
have been written about Hirota's method and it is the subject of
thousands of research papers.

Of course, there are several mathematically more rigorous methods
to compute solitons, such as the Inverse Scattering Transform (IST),
the Wronskian determinant methods,
the Riemann-Hilbert approach, the dressing method,
the Darboux and B\"acklund transformation methods, etc.
In contrast to the more advanced analytic methods that use
complex analysis, such as IST and the Riemann-Hilbert method,
Hirota's method can not solve the initial value problem for
nonlinear PDEs.
Regardless, Hirota's method is a direct, powerful, and effective method
to quickly find the explicit form of solitons.
Apart from soliton solutions, Hirota's method can be used to find
rational (lump) solutions of PDEs and the method applies to various types
of discrete equations as well.
A discussion of those is beyond the scope of this paper.

A mathematical foundation for the Hirota method by Sato and
other researchers at the Kyoto School of Mathematics can be found in,
for example,
\cite{date-bosonization-1994,date-etal-book-jimbo-miwa-1983,jimbo-miwa-publrims-1983,lambert-etal-jpsjpn-2007,ohta-etal-progtheorphyssuppl-1988,willox-satsuma-book-2004}.
There are deep connections of Hirota's method with
infinite dimensional Lie algebras, transformation groups,
Grassmanian manifolds, Wronskians, 
Gramians, Pfaffians, Bell polynomials, Pl\"ucker relations, etc.
We refer the interested reader to the literature.

This survey paper is based on one (WH) of the authors' thirty
years of experience with Hirota's method mainly from the perspective
of applications and computer implementation.
He argues that if one seeks solutions involving exponentials,
replacing a nonlinear PDE
(which usually consists of both linear and nonlinear terms)
with an equation that is homogeneous in degree in a
new dependent variable (or variables) is quite important,
perhaps more so than working with Hirota's bilinear form(s)
of the transformed equation.
Therefore, ``homogenization of degree" is at the core of what is
now called\footnote{Some authors 
\cite{lakestani-etal-compmethdiffeqs-2022,singh-saharay-intjmodphysb-2023,wei-cmmp-2009,zhang-etal-modphyslettb-2022}
call it the Hereman or Hereman-Nuseir method.}
the {\em simplified} Hirota method in which Hirota's bilinear operators
are no longer used.
Instead, we use a perturbation-like scheme involving linear and
nonlinear operators to solve the homogeneous equation without
first recasting it into bilinear form.

Although the bilinear representation of the PDE is not used
in our approach, dismissing it would be a mistake because it is 
a valuable tool in the search for completely integrable equations
\cite{hietarinta-manchester-1989,hietarinta-kluwer-1990}
and theoretical considerations
(see, e.g., \cite{willox-satsuma-book-2004} and the references therein).

The concept of homogenization of degree is illustrated for the 
Burgers equation and the ubiquitous Korteweg-de Vries (KdV) equation.
For the Burgers equation, a truncated Laurent series of its solution
yields the Cole-Hopf transformation, which allows one to transform the
Burgers equation into the heat equation.
The latter is homogeneous of degree one (linear) and can be
solved by separation of variables and other methods.
Using Hirota's method, traveling wave solutions of the heat equation
involving one or more exponentials readily lead to multiple kink
solutions of the Burgers equation.
Contrary to solitons, these do not collide elastically but coalesce
into a single wavefront.

In the case of the KdV equation,
a truncated Laurent series reveals the transformation that
Hirota used to replace the KdV by a quadratic (bilinear) equation.
The connection between Hirota's transformation and the truncated
Laurent expansion, a.k.a.\ truncated Painlev\'e expansion or singular
manifold expansion, has been long known
\cite{estevezetal-jpa-1993,musette-conte-jpa-1994,nozaki-jpsjpn-1987}.
As the examples will show, it is a crucial step in the
application of any flavor of Hirota's method.

The idea of homogenization is further illustrated 
on a class of completely integrable fifth-order KdV equations,  
including those of Lax \cite{lax-communpureapplmath-1968},
Sawada-Kotera (SK) and Caudrey-Dodd-Gibbon (CDG)
\cite{hereman-acm-1992,sawada-kotera-progtheorphys-1974},
and Kaup-Kupershmidt (KK)
\cite{fordy-gibbons-pla-1980,hirota-ramani-pla-1980,kaup-studapplmath-1980}.
Their solitons are computed with a straightforward algorithm
involving linear and nonlinear operators which are not
necessarily quadratic.
Also, the cubic operators we introduce are not the same as the
trilinear operators discussed in
\cite{grammaticosetal-pla-1994,hietarinta-lnp767-2009}
because we split off the linear operator the same way as
for quadratic equations.

The computations for the KK case are complicated, lengthy,
and nearly impossible without using a symbolic manipulation program
such as {\em Maple} or {\em Mathematica}.
One reason is that the homogeneous equation is of fourth degree.
Another reason is that the structure of the soliton solutions is
quite different from those of the KdV, Lax, and SK equations.
Although the soliton solutions of the KK equation were already
presented in \cite{hereman-nuseir-mcs-1997}, and these for the
Lax and SK equations have been computed long before that,
from time to time their computation resurfaces in the literature,
most recently in \cite{karakoc-etal-jcam-2023,kumar-mohan-physscr-2022,kumar-etal-physscr-2022,wang-modphyslettb-2017,wang-xiao-physscr-2018,wazwaz-joes-2016,wei-cmmp-2009}.

Homogenization of degree also allows one to find solitary wave solutions
of nonlinear PDEs that are either not completely integrable or
for which the bilinear form is unknown.
A couple of such examples, mainly from mathematical biology, will be shown.
We pay particular attention to
a FitzHugh-Nagumo (FHN) equation with convection term for it
has a so-called {\em bi-soliton} solution that describes the
coalescence of wavefronts.
The same happens for Burgers and wave equations with cubic source terms
which are also discussed in detail.

The simplified Hirota method
has been successfully used by many authors 
to find solitary wave and soliton solutions.
Most notably, Wazwaz
has extensively applied the method to find
bi-soliton solutions \cite{wazwaz-physscr-2010,wazwaz-jfi-2010}
and soliton solutions of a large number of PDEs 
involving one or more space variables
(see, e.g., \cite{wazwaz-mmas-2015,wazwaz-encyclop-2015,wazwaz-joes-2016}
and many of his other papers).
Additional applications to PDEs with multiple space variables 
can be found in, e.g.,  
\cite{lakestani-etal-compmethdiffeqs-2022,singh-saharay-intjmodphysb-2023,wei-cmmp-2009,zhang-etal-modphyslettb-2022}.

Before applying the (simplified) Hirota method, it is a
good idea to test if the PDE has the Painlev\'e property 
\cite{ablowitz-clarkson-book-1991,conte-book-1999} 
by running, e.g., the \verb|PainleveTest.m|
code \cite{baldwin-hereman-jnmp-2006}.
The Laurent series
used in the Painlev\'e test often provides insight in which
homogenizing transformation to use.

We developed a {\em Mathematica} package,
called \verb|PDESolitonSolutions.m| \cite{goktas-hereman-code-2023}.
It uses the homogenization method to solve several polynomial PDEs 
that are completely integrable as well as some that 
do not have soliton solutions.
In this paper we focus on $(1+1)$-dimensional PDEs although
our code already works for some PDEs involving up to three space
variables $(x,y,z)$ in addition to time $(t)$.
We cover only two examples of PDEs with multiple space variables.
One of the examples is the well-studied Kadomtsev-Petviashvili (KP)
equation.

The paper is organized as follows.
In Section~\ref{homogenization} we discuss the homogenization
of the Burgers and KdV equations using logarithmic
derivative transformations.

After a brief review of the original Hirota method, we describe the
simplified version in Section~\ref{solving-homogeneous-PDE} still
using the KdV equation as the prime example.

In Section~\ref{fifth-order-equation}, we apply the simplified
Hirota method to the Lax, SK, and KK equations.
For each we compute the one-, two- and three-soliton solutions
explicitly.

In Section~\ref{mkdv-equation} we show how the method needs to
be adjusted to find solitons for the modified KdV (mKdV) equation.

To show how the simplified method can be applied to PDEs
that are not ``solitonic" in Section~\ref{non-solitonic-equations}
we compute solitary wave solutions of the Fisher and FHN  
equations with and without convection terms.
Additional examples include a combined KdV-Burgers equation,
a Burgers and wave equation with cubic source terms,
and an equation due to Calogero.
For each of these equations we compute exact travelling wave solutions.
None has soliton solutions although some have bi-soliton solutions.

Section~\ref{two-but-not-three-solitons} covers an equation in $(1+1)$ dimensions 
which has two-soliton but not three-soliton solutions.

In Section~\ref{multiple-space-dimensions} we compute
multi-soliton solutions for the KP equation and an
equation in $(3+1)$ dimensions studied by Geng and Ma \cite{geng-ma-physlettA-2007}.

Section~\ref{software} covers software to automate Hirota's method.
In particular, we discuss the implementation and limitations of \verb|PDESolitonSolutions.m| 
and review related software packages.

Finally, some conclusions are drawn in Section~\ref{conclusions} followed by a brief 
discussion of future work.

\section{Homogenization of Nonlinear PDEs}
\label{homogenization}

\subsection{The Burgers equation}
\label{burgers-equation}
Our initial example is the Burgers (a.k.a.\ Burgers-Bateman) equation,
\begin{equation}
\label{orgburgers}
u_t + 2 u u_x - u_{xx} = 0,
\end{equation}
named after Harry Bateman (1882-1946) and Johannes Burgers (1895-1981).
The subscripts denote partial derivatives, e.g.,
$u_{xx} = \frac{\partial^2 u}{\partial x^2}$ and later on
$u_{3x} = \frac{\partial^3 u}{\partial x^3}$, etc.
Note that the coefficient of the diffusion term $(u_{xx})$
has been normalized.
Equation (\ref{orgburgers}) can be linearized with a logarithmic
derivative transformation due to Cole and Hopf.
First integrate\footnote{Alternatively, set $u = v_x$ and integrate with respect to $x$
to get $v_t + v_x^2 - v_{xx} = 0$.
Substitution of $v = c \, \ln f$ yields (\ref{homogeneousburgers}).
The same can be done for other equations in this paper.}
the Burgers equation with respect to $x$, yielding
\begin{equation}
\label{intburgers}
\partial_t \left( \int^x u \, dx \right) + u^2 - u_{x} = 0.
\end{equation}
Then substitute
\begin{equation}
\label{colehopfburgers}
u = c \, (\ln f)_x = c \,\left( \frac{f_x}{f} \right),
\end{equation}
where $c$ is a constant, to get
\begin{equation}
\label{homogeneousburgers}
f (f_{t} - f_{xx}) + (c + 1) f_x^2 = 0.
\end{equation}
Setting $c = -1$ yields the heat equation
\begin{equation}
\label{heateq}
f_{t} - f_{xx} = 0.
\end{equation}
Then,
\begin{equation}
\label{colehopftfburgers}
u(x,t) =  - (\ln f)_{x} = -\frac{f_{x}}{f}
\end{equation}
is the well-known Cole-Hopf
transformation\footnote{This transformation is consistent with
the scaling symmetry \cite{heremanetal-bookchapter-2009}
of the Burgers equation which is invariant under
$x \rightarrow \lambda^{-1} x,
\, t \rightarrow \lambda^{-2} t,
\, u \rightarrow \lambda u$
with an arbitrary constant $\lambda$.
Hence, one would expect a {\em first} derivative of $\ln f$.}.
We now show where this mysterious transformation comes from.
As in the Painlev\'e test \cite{baldwin-hereman-jnmp-2006},
substitute a Laurent series
\begin{equation}
\label{laurent}
u(x,t) = f^{\alpha}(x,t) \sum_{k=0} ^{\infty} u_k (x,t) f^k (x,t)
\end{equation}
into (\ref{orgburgers}).
Note that $f(x,t)$ is the manifold of the poles since $\alpha$ is a negative integer.
The most singular terms $f^{2\alpha-1}$ and $f^{\alpha-2}$ will balance when $\alpha = -1$
and vanish for $u_0 (x,t) = -f_x $.
Truncating (\ref{laurent}) at the constant level term in $f$ yields
an auto-B\"{a}cklund transformation,
\begin{equation}
\label{autobacklundburgers}
u(x,t) = -\frac{f_{x}}{f} + u_1 (x,t) = - (\ln f)_{x} + u_1 (x,t),
\end{equation}
provided $u_1(x,t)$ is also a solution of the Burgers equation.
For the zero solution $(u_1 = 0)$ 
(\ref{autobacklundburgers}) becomes the Cole-Hopf transformation (\ref{colehopftfburgers}).
The transformation allows us to replace the Burgers equation
which has a mismatch of linear and quadratic terms in $u$ by an equation
that is homogeneous in degree in the new field variable $f$.
The fact that the resulting equation happens to be of {\it first} degree (linear) 
is advantageous for it can be solved by separation of variables 
eventually resulting in a large class of solutions of (\ref{orgburgers}).

\begin{figure}[htb]
\begin{center}
\begin{tabular}{cc}
\includegraphics[width=2.35in, height=1.40in]{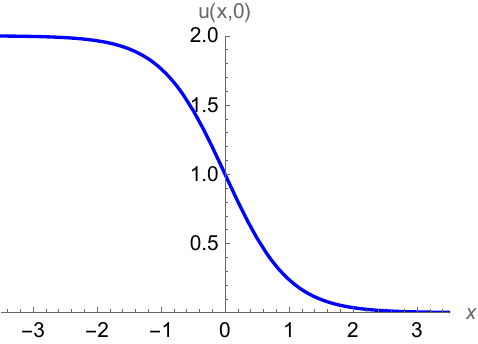}
&
\includegraphics[width=2.80in, height=1.50in]{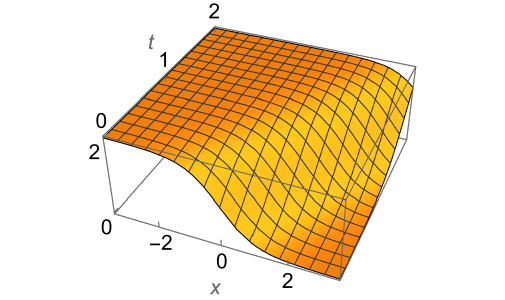}
\end{tabular}
\end{center}
\caption{2D and 3D graphs of the one-kink solution (\ref{kinkburgers3})
for $K = 1$ and $\Delta = 0$.}
\label{kinksolutionburgerseq}
\end{figure}

\begin{figure}[htb!]
\begin{center}
\begin{tabular}{ccc}
\includegraphics[width=1.67in, height=1.34in]{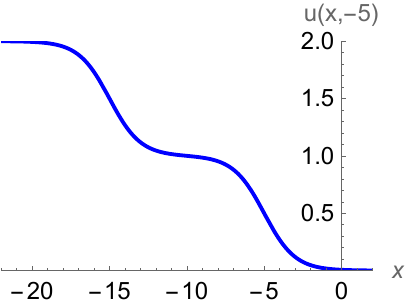}
& \phantom{x} & 
\includegraphics[width=1.875in, height=1.50in]{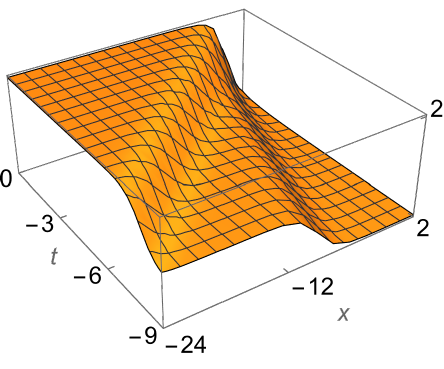}
\end{tabular}
\end{center}
\caption{2D and 3D graphs of the two-kink solution (\ref{Nkinkburgers})
for $k_1 = -1, k_2 = -2,$ and $\delta_1 = \delta_2 = 0$.}
\label{twokinksolutionburgerseq}
\end{figure}

Setting the stage for what follows, we consider a couple of
simple solutions of (\ref{heateq}).
Substituting
$f(x,t) = 1 + \e^{\theta} = 1 + \e^{k \, x - \omega \, t + \delta},$
where $k$ is the wave number, $\omega$ the angular frequency,
and $\delta$ a phase constant,
into (\ref{heateq}) yields the dispersion law $\omega = - k^2$.
Hence,
\begin{eqnarray}
\label{kinkburgers1}
u(x,t) &=& -(\ln f)_{x}
=  -\frac{f_{x}}{f}
= - k \left( \frac{\e^{\theta}}{1 + \e^{\theta}} \right)
= - k \left( \frac{\e^{\theta} {\e^{-\frac{\theta}{2}}}}
  {(1 + \e^{\theta}) {\e^{-\frac{\theta}{2}}}} \right)
\nonumber \\
&=& -k \left( \frac{\e^{\frac{\theta}{2}}}{\e^{\frac{\theta}{2}} + \e^{-\frac{\theta}{2}}} \right)
= -\tfrac{1}{2} k \left(\frac{{2} \e^{\frac{\theta}{2}}}{\e^{\frac{\theta}{2}} + \e^{-\frac{\theta}{2}}} \right)
\nonumber \\
&=&  -\tfrac{1}{2} k \left( \frac{\e^{\frac{\theta}{2}} + {\e^{-\frac{\theta}{2}}}
  + \e^{\frac{\theta}{2}} - {\e^{-\frac{\theta}{2}}}}{\e^{\frac{\theta}{2}} + \e^{-\frac{\theta}{2}}} \right)
= -\tfrac{1}{2} k \left(1 + \tanh \tfrac{\theta}{2} \right)
\end{eqnarray}
with $\theta = k x + k^2 t + \delta$, 
or, simply
\begin{equation}
\label{kinkburgers3}
u(x,t) = K \left( 1 - \tanh \Theta \right),
\end{equation}
with $\Theta = K x - 2 K^2 t + \Delta$, 
$K = -\tfrac{k}{2}$, and $\Delta = -\tfrac{\delta}{2}$.  
This kink-shaped solution (shock wave) of the Burgers equation is
pictured in Fig.~\ref{kinksolutionburgerseq}.
%
\vskip 0.001pt
\noindent
Due to its linearity,
$f(x,t) = 1 + \sum_{i=1}^N \e^{\theta_i}$ where
$\e^{\theta_i} = \e^{k_i \, x + k_i^2 \, t + \delta_i}$ with
$k_i$ and $\delta_i$ arbitrary constants,
also solves (\ref{heateq}) yielding a $N$-kink solution
\begin{equation}
\label{Nkinkburgers}
u(x,t) = -\frac{k_i \sum_{i=1}^N \e^{\theta_i}}
{1 + \sum_{i=1}^N \e^{\theta_i}}
\end{equation}
for any integer $N \ge 1$.
Fig.~\ref{twokinksolutionburgerseq} shows solution (\ref{Nkinkburgers})
for the case where two wavefronts ($N=2$) coalesce into a single
kink-shaped wavefront as time progresses.
For a more detailed analysis of solutions of type (\ref{Nkinkburgers})
we refer to \cite{wangetal-csf-2004}.
%
\subsection{The Korteweg-de Vries equation}
\label{korteweg-devries-equation}
Next we explore the homogenization of the ubiquitous KdV equation,
\begin{equation}
\label{orgkdv}
u_t + 6 u u_x + u_{3x} = 0,
\end{equation}
named after Diederik Korteweg (1848-1941) and Gustav de Vries (1866-1934).

\begin{figure}[htb]
\begin{center}
\includegraphics[width=3.65in, height=1.96in]{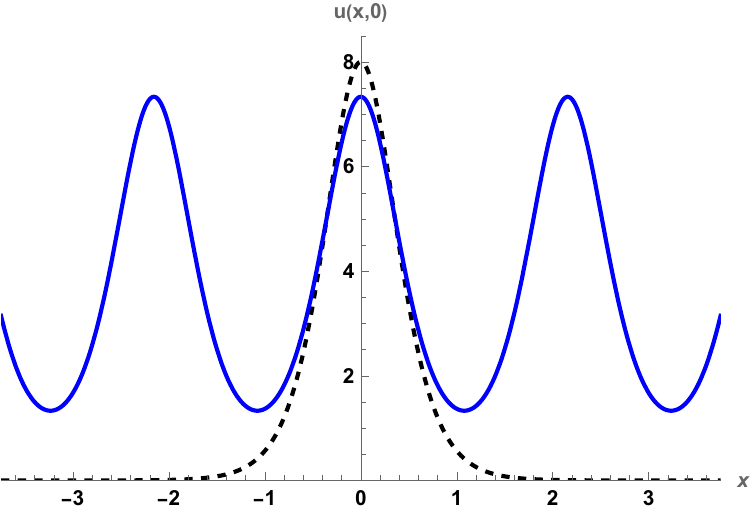}
\end{center}
\caption{Graphs of the solitary wave (dashed line) and cnoidal wave (solid line) solutions for
$k = 2,\, m =\tfrac{9}{10},$ and $\delta = 0$.}
\label{kdv-sech-cn-sols}
\end{figure}

In \cite{korteweg-devries-philmag-1895} they derived the equation and its solitary wave
and cnoidal wave solutions:
\begin{eqnarray}
\label{kdvsolitarysimpler}
u(x,t) &=& 2 k^2 \,{\sech}^2 (k x - 4 k^3 t + \delta),
\\
u(x,t) &=& \tfrac{4}{3} k^2 (1-m) + 2 k^2 m \,{\cn}^2 (k x - 4 k^3 t + \delta ; m),
\end{eqnarray}
where $m \in (0,1)$ is the modulus of the Jacobi elliptic cosine (cn) function.
Both solutions are shown in Fig.~\ref{kdv-sech-cn-sols}.
As $m$ approaches 1, the peaks of the periodic solution get a little taller, the valleys become lower and flatter before they eventually spread out horizontally to become the pulse-type hyperbolic secant solution.
\vskip 0.000001pt
\noindent
\vskip 0.000001pt
\noindent
The interaction of the more complicated soliton solutions (to be discussed later in this paper)
were first observed in numerical simulations by Norman Zabusky and Martin Kruskal \cite{zabusky-kruskal-prl-1965} in 1965.

To compute soliton solutions with Hirota's method the
original KdV equation needs to be replaced by an equation
(in a new field variable) that is homogeneous of degree.
To get a candidate transformation, again substitute a Laurent series (\ref{laurent})
into (\ref{orgkdv}).
The most singular terms $f^{2\alpha-1}$ and $f^{\alpha-3}$ will balance when $\alpha = -2$.
The terms $f^{-5}$ and $f^{-4}$ vanish when
$u_0 (x,t) = -f_x$ and $u_1 (x,t) = 2 f_{xx}$.
Hence, we obtain an auto-B\"acklund transformation for the KdV equation
\begin{equation}
\label{autobacklundkdv}
u(x,t) = -\frac{2{f_x}^2}{f^2} + \frac{2 f_{xx}}{f} + u_2 (x,t)
       = 2 (\ln f)_{xx} + u_2 (x,t),
\end{equation}
where $u_2(x,t)$ is also a solution of the KdV equation.
Taking $u_2 = 0$ yields the Hirota
transformation\footnote{Note that the KdV equation is
invariant \cite{heremanetal-bookchapter-2009} when
$x \rightarrow \lambda^{-1} x,
\, t \rightarrow \lambda^{-3} t,
\, u \rightarrow \lambda^2 u$.
Therefore, a {\em second} derivative of $\ln f$ makes sense.}
that ``bilinearizes" the KdV equation.
To see the effect of a logarithmic derivative transformation
substitute
\begin{equation}
\label{kdvtinf}
u = c \, (\ln f)_{xx}
  = c \,\left( \frac{f f_{xx} -{f_x}^2}{f^2} \right),
\end{equation}
where $c$ is an undetermined constant, into the integrated version of (\ref{orgkdv}):
\begin{equation}
\label{intkdv}
\partial_t \left( \int^x u \, dx \right) + 3 u^2 + u_{xx} = 0.
\end{equation}
This yields
\begin{equation}
\label{kdvtransformed}
f^3 ( f_{xt} + f_{4x} ) - f^2 (f_x f_t - 3 (c-1) f_{xx}^2 + 4  f_x f_{3x} )
+ 3 (c-2) f_{x}^2 (f_{x}^2 - 2 f f_{xx}) = 0.
\end{equation}
Setting $c = 2$ (confirming what we learned from the truncated Laurent series),
(\ref{kdvtransformed}) simplifies 
into\footnote{Many authors, in particular those working on the mathematical
foundation of Hirota's method, use $\tau$ instead of $f$
and investigate the rich mathematical properties of the ``tau" function.
}
\begin{equation}
\label{kdvquadratic}
f ( f_{xt} + f_{4x} ) - f_x f_t + 3 f_{xx}^2 - 4 f_x f_{3x} = 0,
\end{equation}
which is homogeneous of {\it second} degree in $f$.
Hirota \cite{hirota-book-2004} introduced the transformation
$u = 2 \, (\ln f)_{xx}$ in the early 1970s and realized that (\ref{kdvquadratic})
can be written in bilinear form
\begin{equation}
\label{hirotakdv}
\left( D_x D_t + D_x^4 \right) (f \mathbf{\cdot} f) = 0,
\end{equation}
with operators $D_x$ and $D_t$ defined
(see, e.g., \cite{hirota-solitons-1980,hirota-book-2004}) as
\begin{eqnarray}
\label{Dxoper}
D_x^m (f \mathbf{\cdot} g ) 
&=& {( \partial_{x} - \partial_{x'} )}^m  f(x,t) g(x',t) \bigg|_{x'=x},
\\
\label{Dtoper}
D_t^n (f \mathbf{\cdot} g ) 
&=& {(\partial_{t} - \partial_{t'} )}^n f(x,t) g(x,t') \bigg|_{t'=t},
\end{eqnarray}
with $m$ and $n$ positive integers. 

Working with these Hirota operators is easy because it amounts to
applying Leibniz rule for derivatives of products of functions with
every other sign flipped.
Thus,
\begin{equation}
D_x^m (f \mathbf{\cdot} g )
= \sum_{j = 0}^m \frac{(-1)^{m-j}m!} {j!(m - j)!}
  \left(\frac{\partial^j f} {\partial {x^j}} \right)
  \left(\frac{\partial^{m-j} g} {\partial {x^{m-j}}} \right),
\end{equation}
and, more general,
\begin{eqnarray}
D_x^m D_t^n (f \mathbf{\cdot} g )
&=&
{( \partial_{x} - \partial_{x'} )}^m  {( \partial_{t} - \partial_{t'} )}^n f(x,t) g(x',t') \bigg|_{x'=x, t'=t}
\\
&=& \sum_{j = 0}^m  \sum_{i=0}^n
\frac{(-1)^{n+m-i-j} m! n!} {j! (m - j)! i! (n - i)!}
\left(\frac{\!\partial^{i+j} f} {\!\partial {t^i}\partial {x^j}} \right)
\left( \frac{\!\partial^{n+m-i-j} g} {\!\partial {t^{n-i}}\partial {x^{m-j}}} \right).
\end{eqnarray}
For example,
\begin{equation}
\label{dx4}
D_x^4 (f \mathbf{\cdot} g ) 
= f_{4x} g - 4 f_{3x} g_{x} + 6 f_{xx} g_{xx} - 4 f_{x} g_{3x} + f g_{4x},
\end{equation}
and
\begin{equation}
\label{dxt}
D_x D_t (f \mathbf{\cdot} g ) = f_{xt} g - f_t g_x - f_x g_t + f g_{xt}.
\end{equation}
With the above one can readily verify that
$\left( D_x D_t + D_x^4 \right) (f \mathbf{\cdot} f) = 0$
yields (\ref{kdvquadratic}).

\section{Solving the Homogeneous PDE}
\label{solving-homogeneous-PDE}
\subsection{Hirota's method}
\label{Hirota-method}
We now show how Hirota computed soliton solutions of (\ref{hirotakdv}).
He sought a solution of the form
\begin{equation}
\label{hirotaexpansionf}
f(x,t) = 1 + \sum_{n=1}^{\infty} \epsilon^n \, f^{(n)}(x,t)
       = 1 + \epsilon f^{(1)} + \epsilon^2 f^{(2)} + \hdots,
\end{equation}
where $\epsilon$ is a formal parameter.
The building blocks of solitons are exponentials with different plane-wave arguments.
Actually, $f^{(1)}$ will be the sum of a chosen but fixed number $(N)$ of exponentials
$\e^{\theta_i} = \e^{k_i \, x - \omega_i \, t + \delta_i}
\; (i = 1, \hdots, N)$.
Then, $f^{(2)}$ will have products of just two of these exponentials such as $\e^{2 \theta_i}$ and
$\e^{\theta_i + \theta_j} \; (i,j = 1, \hdots, N)$.
In turn, $f^{(3)}$ will have products of three exponentials, for example,
$\e^{3\theta_i}, \;  \e^{2 \theta_i + \theta_j}, \; \e^{\theta_i + 2 \theta_k},$ 
and $\e^{\theta_i + \theta_j + \theta_k} \; (i,j,k = 1, \hdots, N)$.
The role of $\epsilon$ is to keep track of how many exponentials
are in the mix because terms involving products of two exponentials
can never be equated to terms with products of three exponentials, etc.
In other words, $\epsilon$ serves as a {\em bookkeeping} parameter
which can be set to one once the computations are done.
As we will see in all the examples that follow, when solitons exist 
(\ref{hirotaexpansionf})
will truncate and therefore be a {\em finite} sum of exponentials.

Substituting (\ref{hirotaexpansionf}) into (\ref{hirotakdv}) and splitting
order-by-order in $\epsilon$ gives
\begin{eqnarray}
\label{bilinearscheme}
O(\epsilon^{0}):  && B(1 {\bf \cdot } 1 ) = 0,
\nonumber \\
O(\epsilon^{1}):  && B(1 {\bf \cdot } f^{(1)} + f^{(1)} {\bf \cdot } 1) = 0,
\nonumber \\
O(\epsilon^{2}): && B(1 {\bf \cdot } f^{(2)}
+ f^{(1)} {\bf \cdot } f^{(1)} + f^{(2)} {\bf \cdot } 1 ) = 0,
\nonumber \\
O(\epsilon^{3}): && B(1 {\bf \cdot } f^{(3)}
+ f^{(1)} {\bf \cdot } f^{(2)} + f^{(2)} {\bf \cdot } f^{(1)}
+ f^{(3)} {\bf \cdot } 1) = 0,
\nonumber \\
O(\epsilon^{4}): && B(1 {\bf \cdot } f^{(4)}
+ f^{(1)} {\bf \cdot } f^{(3)} +
f^{(2)} {\bf \cdot } f^{(2)} + f^{(3)} {\bf \cdot } f^{(1)}
+ f^{(4)} {\bf \cdot } 1 ) = 0,
\nonumber \\
\vdots && \phantom{ B(1 {\bf \cdot } 1)}  \;\;\; \vdots
\nonumber \\
O(\epsilon^{n}): && B \left( \sum_{j=0}^{n}f^{(j)}{\bf \cdot} f^{(n-j)} \right) = 0,
\;\; n \ge 0,
\quad {\rm with} \,  f^{(0)} = 1,
\end{eqnarray}
where for the present example $B = D_x D_t + D_x^4$.

\begin{figure}[htb]
\begin{center}
\includegraphics[width=3.18in, height=1.68in]{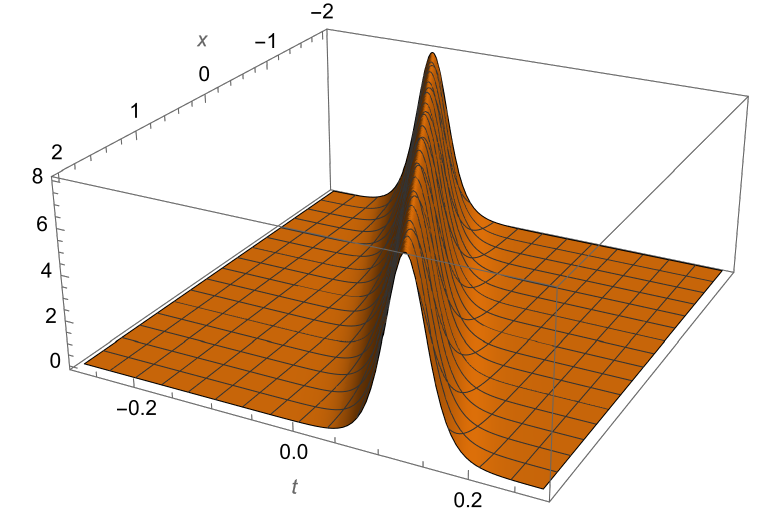}
\end{center}
\caption{3D graph of the hump-shaped solution (\ref{ukdvonesoliton}) 
for $K = 2$ and $\Delta = 0$.}
\label{humpsolutionKdVeq}
\end{figure}

\begin{figure}[htb]
\begin{center}
\begin{tabular}{ccc}
\includegraphics[width=1.52in, height=1.76in]{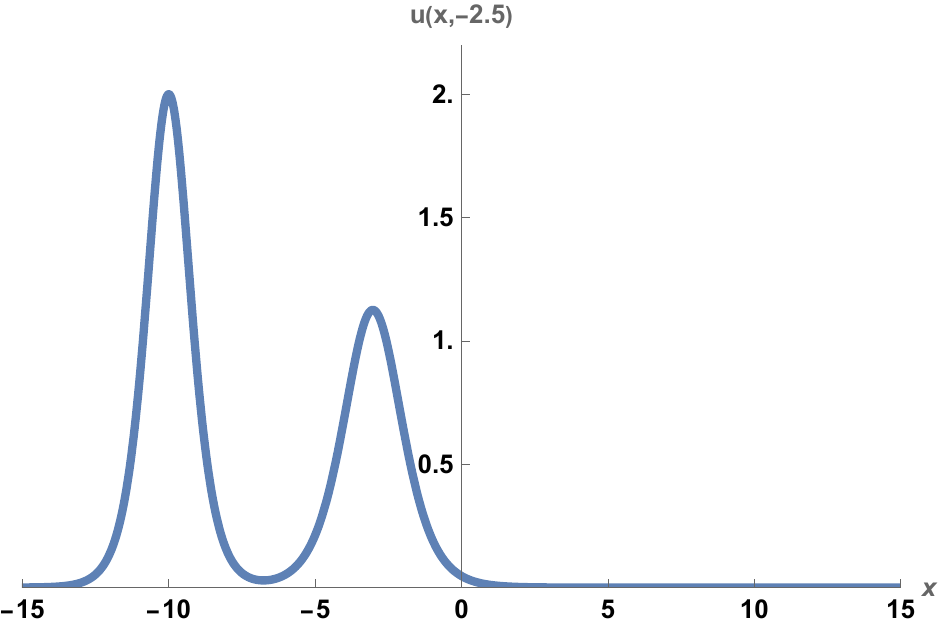} &
\includegraphics[width=1.52in, height=1.76in]{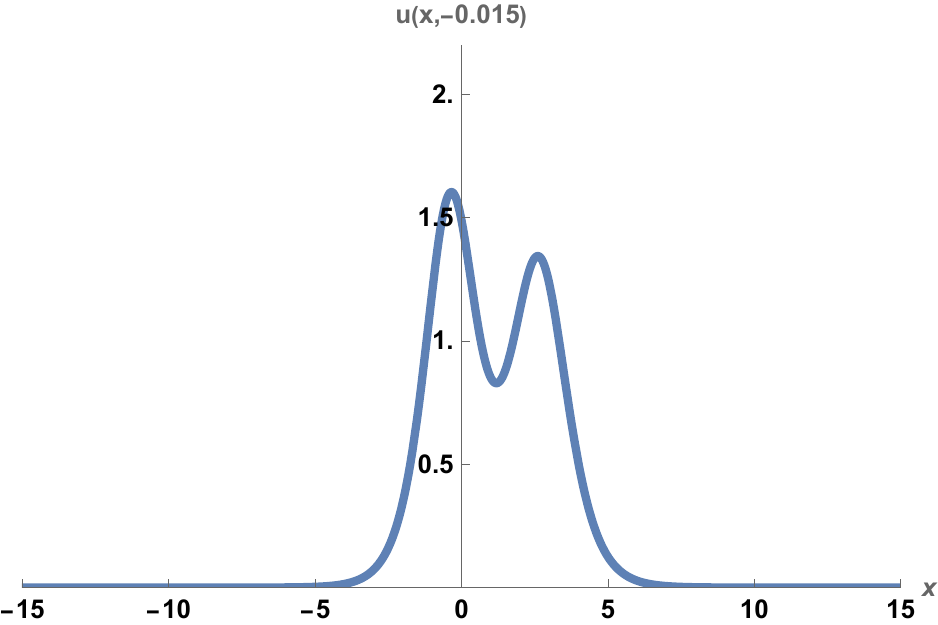} &
\includegraphics[width=1.52in, height=1.76in]{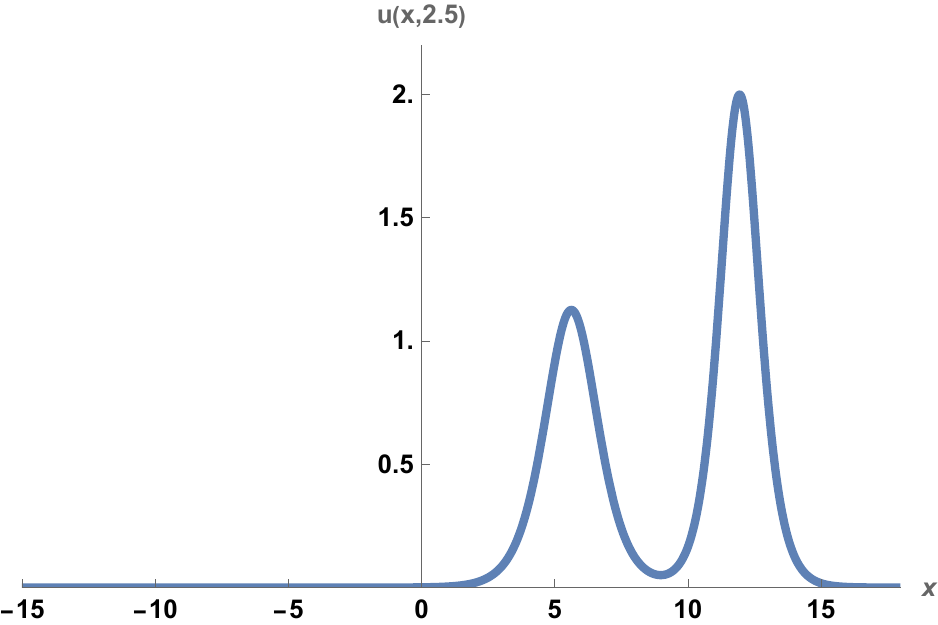}
\end{tabular}
\end{center}
\caption{Graph of the two-soliton solution (\ref{twosolitonkdvform2}) 
of the KdV equation at three different moments in time.}
\label{kdv-two-solitons-in-time}
\end{figure}

\begin{figure}[h]
\begin{center}
\includegraphics[width=3.60in, height=2.22in]{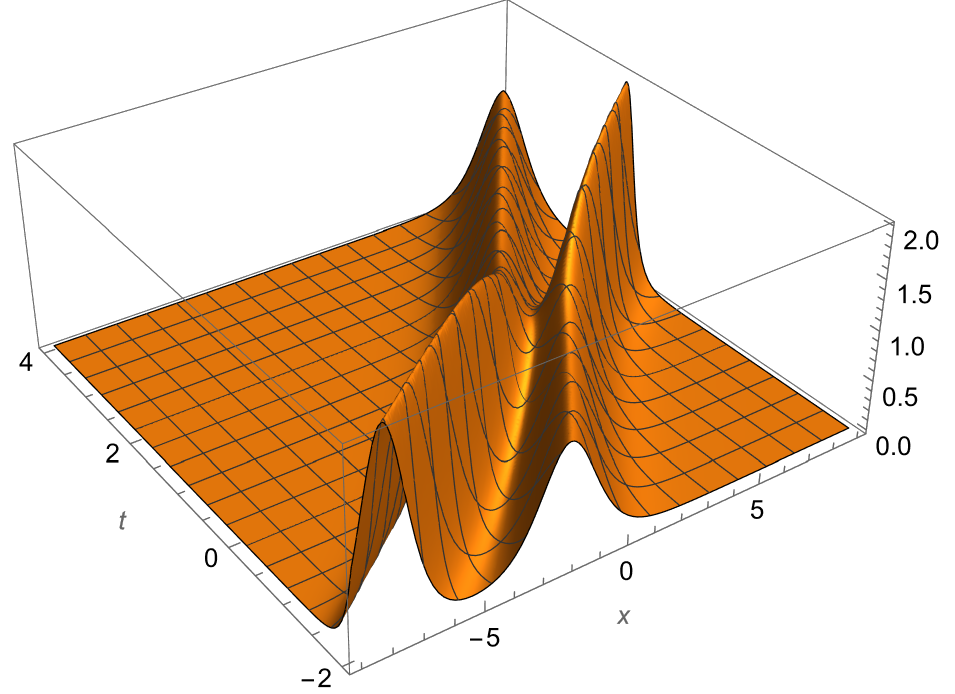}
\end{center}
\caption{Bird's eye view of a two-soliton collision for the KdV equation.
Notice the phase shift after collision: the taller (faster) soliton is
shifted forward and the shorter (slower) soliton backward relative to
where they would have been if they had not collided.}
\label{kdv-two-solitons-in-3D}
\end{figure}

To illustrate, we compute the one- and two-soliton solutions of (\ref{orgkdv}).
Note that the first equation in (\ref{bilinearscheme}) is trivially satisfied.
Using (\ref{dx4}) and (\ref{dxt}), the second equation
reduces\footnote{With $B = D_x D_t + D_x^4$, one has
$B(1.f) = B(f.1) = f_{xt} + f_{4x}$ for any function $f$.}
to $f^{(1)}_{xt} + f^{(1)}_{4x} = 0$.
\vskip 5pt
\noindent
\leftline{\bf One-soliton solution of the KdV equation}
\vskip 2pt
\noindent
If we take $f^{(1)} =  \e^{\theta} \equiv \e^{k \, x - \omega \, t + \delta}$,
that second equation yields the dispersion law $\omega = k^3$.
Next, one can readily verify that $B(f^{(1)} {\bf \cdot } f^{(1)})$ is zero.
Consequently, $f^{(2)}$ is zero and so are $f^{(3)},\, f^{(4)},$ etc.
Therefore, there are only two terms in (\ref{hirotaexpansionf}).
Explicitly,
\begin{equation}
\label{kdv1solfinalf}
f = 1 + \e^{\theta} = 1 + \e^{k \, x - k^3 \, t + \delta}
\end{equation}
after setting $\epsilon = 1$.
Hence,
\begin{eqnarray}
\label{ukdvonesoliton}
u(x,t)
  & = & 2 \left(\frac{f f_{xx} - {f_x}^2}{f^2} \right)
    = \frac{2 k^2\,\e^{\theta}}{\left( 1 + \e^{\theta} \right)^2}
    = \frac{2 k^2\,\e^{\theta} {\e^{-\theta}}}{\left[ {\e^{-\tfrac{\theta}{2}}}
    \left(1 + \e^{\theta}\right) \right]^2}
 \nonumber \\
  & = & \tfrac{1}{2} k^2 \sech^2 \left[ \tfrac{1}{2}(k x -k^3 t + \delta) \right]
    = 2 K^2 \sech^2 \left(K x - 4 K^3 t + \Delta \right),
\end{eqnarray}
where $K = \tfrac{k}{2}$ and $\Delta =\tfrac{\delta}{2}$.
Fig.~\ref{humpsolutionKdVeq} shows a 3D graph of this so-called
{\em solitary wave solution}
or {\em one-soliton solution} for $K = 2$ and $\Delta = 0$.
\vskip 0.01pt
\noindent
\vskip 0.001pt
\noindent
\leftline{\bf Two-soliton solution of the KdV equation}
\vskip 3pt
\noindent
Starting with $f^{(1)} = \e^{\theta_1} + \e^{\theta_2}$, where $\e^{\theta_i} 
= \e^{k_i \, x - \omega_i \, t + \delta_i}$,
the first nontrivial equation in (\ref{bilinearscheme}) yields $\omega_i = k_i^3$.
Then, $B(f^{(1)} {\bf \cdot } f^{(1)}) = - 6 k_1 k_2 (k_1 - k_2)^2 \e^{\theta_1 + \theta_2}$ which
determines the form of $f^{(2)}$, namely,
$f^{(2)} = a_{12} \e^{\theta_1 + \theta_2}$, with some constant coefficient 
$a_{12}$ to be computed.
Then,
$B(1 {\bf \cdot } f^{(2)} )
= B(f^{(2)} {\bf \cdot } 1)
= f^{(2)}_{xt} + f^{(2)}_{4x}
= 3 a_{12} k_1 k_2 (k_1 + k_2)^2 \e^{\theta_1 + \theta_2}$.
Substitution of the pieces into the third equation of (\ref{bilinearscheme})
then gives
\begin{equation}
\label{a12kdv}
a_{12} = \left( \frac{k_1 - k_2}{k_1 + k_2} \right)^2.
\end{equation}
One can show that from $O(\epsilon^{3})$ onward one can set $f^{(3)},\, f^{(4)}$,
etc., equal to zero.
Thus, $f$ contains only four terms.
With $\epsilon = 1$,
using 
\begin{equation}
\label{kdv2solfinalf}
f = 1 + \e^{\theta_1} + \e^{\theta_2} + a_{12} \e^{\theta_1 + \theta_2}, 
\end{equation}
and $u = 2 (\ln f)_{xx}$, this yields 
\begin{equation}
\label{twosolitonkdvform1}
u(x,t) = \frac{2 \left[ k_1^2 \e^{\theta_1} + k_2^2 \e^{\theta_2}
+ 2 (k_1 - k_2)^2 \e^{\theta_1 + \theta_2}
+ a_{12} (k_2^2 \e^{\theta_1} + k_1^2 \e^{\theta_2})\e^{\theta_1 + \theta_2} \right]}
{\left( 1 + \e^{\theta_1} + \e^{\theta_2} + a_{12}\, \e^{\theta_1+\theta_2}\right)^2 }.
\end{equation}
Setting $k_i = 2 K_i,\, 
\delta_i = 2 \Delta_i + \ln \left( \frac{K_2 + K_1}{K_2 - K_1} \right)$,
the above can be written as
\begin{eqnarray}
\label{twosolitonkdvform2}
u(x,t) & = &
       \frac{4 \left( K_2^2 - K_1^2 \right) \left[ (K_2^2 - K_1^2)
       + K_1^2 {\rm cosh}(2 \Theta_2) + K_2^2 {\rm cosh}(2 \Theta_1) \right] }
       { \left[ (K_2 - K_1) {\rm cosh}(\Theta_2 + \Theta_1)
       + (K_2 + K_1) {\rm cosh}(\Theta_2 - \Theta_1) \right]^2 }
\nonumber \\
& = & 2 \left( K_2^2 - K_1^2 \right)
    \left( \frac{K_1^2 {\rm sech}^2(\Theta_1) + K_2^2 {\rm csch}^2(\Theta_2)}
    {\left[ K_1 \tanh(\Theta_1) - K_2 \coth(\Theta_2) \right]^2} \right),
\end{eqnarray}
where $\Theta_i = K_i x - 4 K_i^3 t + \Delta_i \; (i = 1, 2).$
The elastic scattering of two solitons for the KdV equation is shown in Figs.~\ref{kdv-two-solitons-in-time}
and~\ref{kdv-two-solitons-in-3D} for $k_1 = 2, \, k_2 = \tfrac{3}{2}$, 
and $\delta_1 = \delta_2 = 0$.
\vskip 0.001pt
\noindent
\vskip 0.00001pt
\noindent
\vskip 0.000001pt
\noindent
\subsection{Simplified Hirota method}
\label{simplified-Hirota-method}
In this Section we use a simplified version of Hirota's method which does
not use the bilinear representation (\ref{hirotakdv}).
Instead, we write (\ref{kdvquadratic}) in the form
\begin{equation}
\label{kdvoperform}
f \, {\cal L} f + {\cal N} (f,f) = 0,
\end{equation}
where
\begin{equation}
\label{loperkdv}
{\cal L} f = f_{xt} + f_{4x}
\end{equation}
and
\begin{equation}
\label{nloperkdv}
{\cal N} (f, g) = - f_x g_t + 3 f_{xx} g_{xx} - 4 f_x g_{3x}
\end{equation}
define a linear differential operator ${\cal L}$ and
a quadratic differential operator ${\cal N}$.
Note that the latter is linear in each of the 
auxiliary functions $f(x,t)$ and $g(x,t)$.
So, we could also call it ``bilinear" but, of course,
${\cal N}$ differs from Hirota's bilinear operator $B$.
Substituting (\ref{hirotaexpansionf}) into (\ref{kdvoperform}),
and setting the coefficients of powers of $\epsilon$ to zero
yields\footnote{Details of the derivation are given in the Appendix.}
\begin{eqnarray}
\label{newbilinearscheme}
O(\epsilon^{1}): && {\cal L} f^{(1)} = 0,
\nonumber \\
O(\epsilon^{2}): && {\cal L} f^{(2)} = - {\cal N} (f^{(1)}, f^{(1)}),
\nonumber \\
O(\epsilon^{3}): && {\cal L} f^{(3)}
    = - \left( f^{(1)} {\cal L} f^{(2)} + {\cal N}(f^{(1)}, f^{(2)})
      + {\cal N}(f^{(2)}, f^{(1)}) \right),
\nonumber \\
\vdots && \phantom{{\cal L} f^{(2)}} \;\; \vdots
\nonumber \\
O(\epsilon^{n}): && {\cal L}f^{(n)} =
  - \sum_{j=1}^{n-1} \left( f^{(j)} {\cal L} f^{(n-j)}
   + {\cal N} (f^{(j)}, f^{(n-j)}) \right),
   \;\; n \ge 2.
\end{eqnarray}
The $N$-soliton solution of the KdV is then generated from
\begin{equation}
\label{kdvf1}
f^{(1)} = \sum_{i=1}^{N} \e^{\theta_i}
   \equiv \sum_{i=1}^{N}  \e^{k_i \, x - \omega_i \, t + \delta_i},
\end{equation}
where $N$ is a natural number, by solving the equations (\ref{newbilinearscheme})
successively to determine $f^{(2)}, f^{(3)}$, etc.
The first equation, ${\cal L} f^{(1)} = 0$, yields the dispersion relation
$\omega_i  = k_i^3$.
With (\ref{kdvf1}) one readily computes
\begin{equation}
\label{rhsquadraticscheme2}
  - {\cal N} (f^{(1)}, f^{(1)})
 = -\! \sum_{i,j=1}^{N} 3 k_i k_j^2 (k_i - k_j ) \e^{\theta_i + \theta_j}
 = \!\!\!\!
 \sum_{1 \le i< j \le N} 3 k_i k_j (k_i - k_j )^2 \e^{\theta_i + \theta_j}.
\end{equation}
Note that there are no terms $\e^{2 \theta_i}$.
Hence, $f^{(2)}$ must be of the from
\begin{equation}
\label{kdvf2}
f^{(2)} = \sum_{1 \le i< j \le N} a_{ij} \e^{\theta_i + \theta_j},
\end{equation}
with
constants\footnote{The $a_{ij}$ are often called phase factors because
they can be absorbed in the 
exponents via $a_{ij} = {\mathrm e}^{A_{ij}}$.}
$a_{ij}$ to be determined.
Next, compute
\begin{equation}
\label{lhsquadraticscheme2}
{\cal L} f^{(2)}
= \sum_{1 \le i< j \le N} 3 k_i k_j (k_i + k_j )^2 \, a_{ij}
\, \e^{\theta_i + \theta_j},
\end{equation}
and equate (\ref{rhsquadraticscheme2}) with (\ref{lhsquadraticscheme2})
to get
\begin{equation}
\label{kdvaij}
a_{ij} = \left( \frac{k_i - k_j}{k_i + k_j} \right)^2,
\quad \quad 1 \le i<j \le N.
\end{equation}
To keep matters transparent we show some details of the computation of
the three-soliton solution and the result for the four-soliton solution.
\vskip 3pt
\noindent
\leftline{\bf Three-soliton solution of the KdV equation}
\vskip 2pt
\noindent
Proceeding in a similar way with the third equation in (\ref{newbilinearscheme}) leads to
the explicit form of $f^{(3)}$.
For $N=3$, we find
\begin{equation}
\label{kdvf3}
f^{(3)} = b_{123} \e^{\theta_1 + \theta_2 + \theta_3}
\end{equation}
with
\begin{equation}
\label{kdvb123}
b_{123} = a_{12} \, a_{13} \, a_{23} =
\left[ \frac{( k_1 - k_2 ) \, ( k_1 - k_3 ) \, ( k_2 - k_3 ) }{( k_1 + k_2 )
\, ( k_1 + k_3 ) \, ( k_2 + k_3 )} \right]^2.
\end{equation}
For $N=3,$ one can verify that $f^{(n)} = 0$ for $n>3.$
Thus,
\begin{eqnarray}
\label{kdv3solfinalf}
f &=& 1 + \e^{\theta_1} + \e^{\theta_2} + \e^{\theta_3} + a_{12}\, \e^{\theta_1 + \theta_2}
+ a_{13}\, \e^{\theta_1 + \theta_3} + a_{23}\, \e^{\theta_2 + \theta_3}
\nonumber \\
&& + a_{12} \, a_{13} \, a_{23} \, \e^{\theta_1 + \theta_2 + \theta_3}
\end{eqnarray}
after setting $\epsilon = 1$.
Notice that (\ref{kdv3solfinalf}) has no terms in
$\e^{2 \theta_1}$, $\e^{2 \theta_2}$,
$\e^{2 \theta_1 + \theta_2}, \e^{\theta_1 + 2 \theta_2}$, etc.
The explicit expression of $u(x,t)$ (not shown due to length)
then follows from $u(x,t) = 2 (\ln f)_{xx}$.

\begin{figure}[htb]
\begin{center}
\begin{tabular}{ccc}
\includegraphics[width=1.52in, height=1.75in]{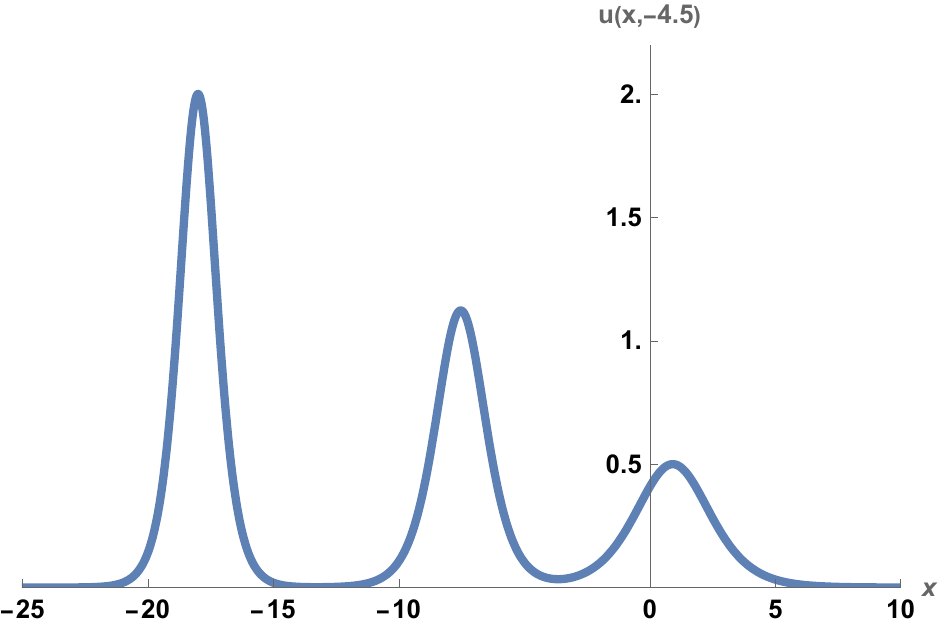} & 
\includegraphics[width=1.52in, height=1.75in]{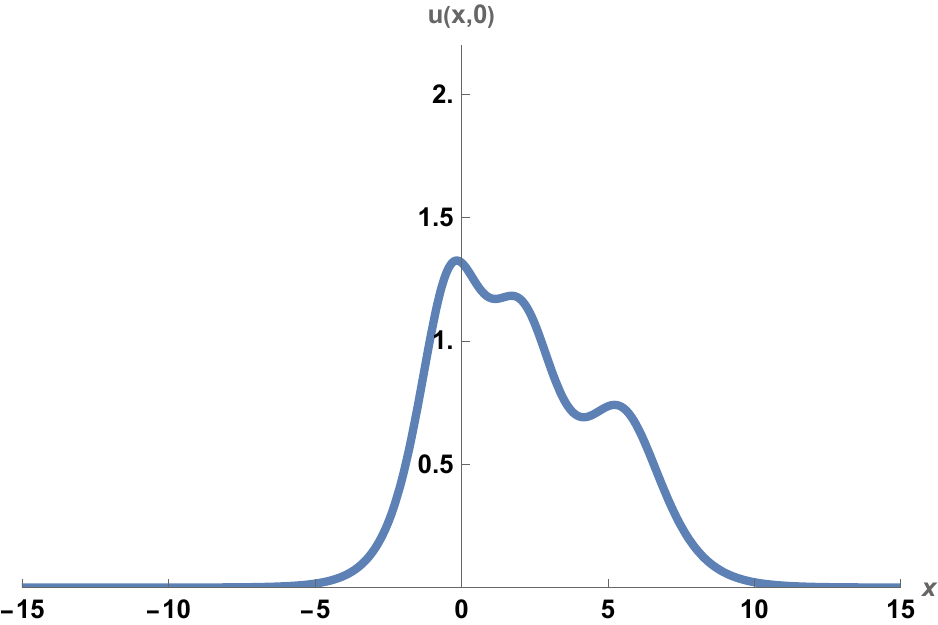} &
\includegraphics[width=1.52in, height=1.75in]{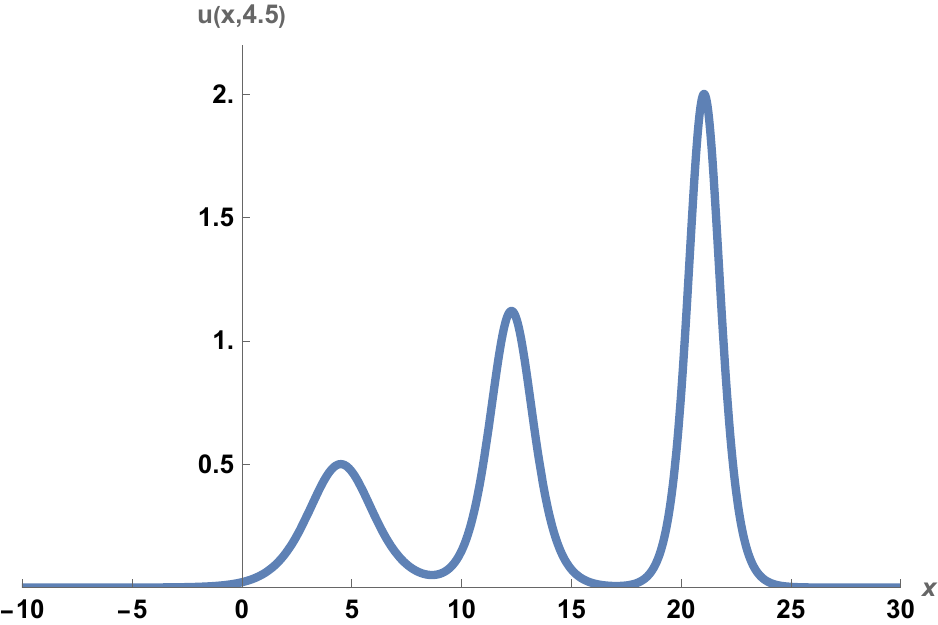}
\end{tabular}
\end{center}
\caption{Graph of the three-soliton solution of the KdV equation at three different moments in time.}
\label{kdv-three-solitons-in-time}
\end{figure}

\begin{figure}[htb!]
\begin{center}
\includegraphics[width=3.60in, height=2.22in]{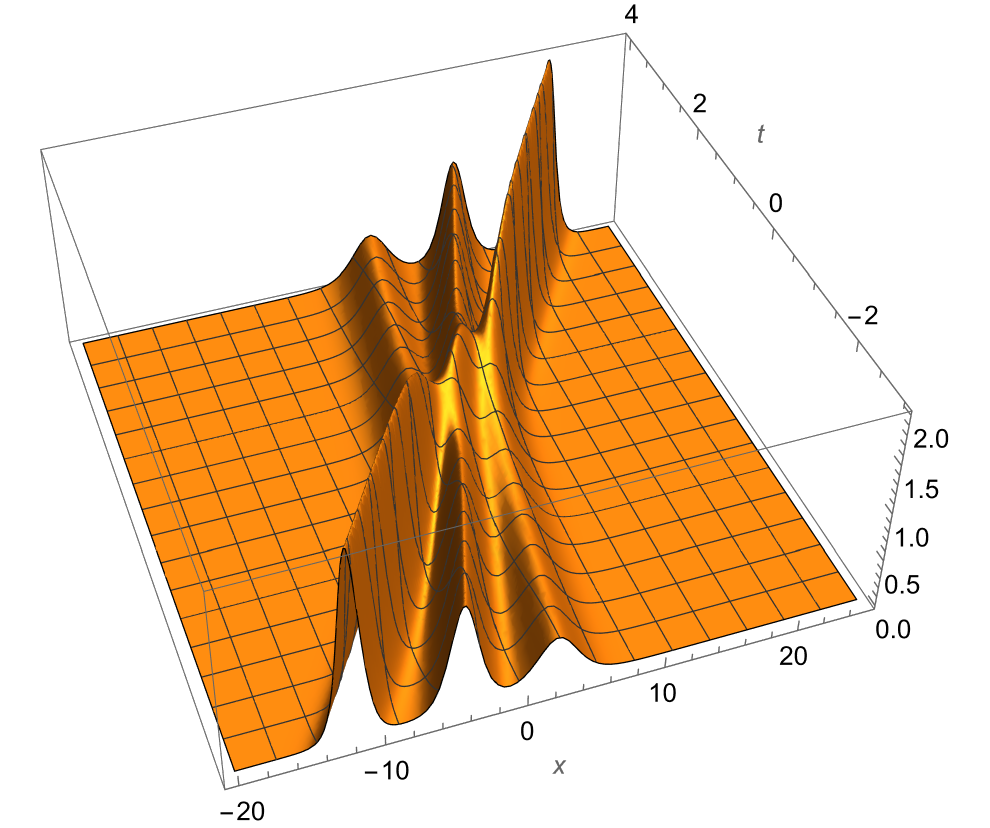}
\end{center}
\caption{Bird's eye view of three solitons colliding for the KdV equation.
Notice the phase shift after collision:
the faster soliton has advanced and the slower ones are behind.
The shortest of the three solitons is shifted the most.}
\label{kdv-three-solitons-in-3D}
\end{figure}

The elastic collision of three solitons for the KdV equation
is shown in Figs.~\ref{kdv-three-solitons-in-time}
and~\ref{kdv-three-solitons-in-3D} for
$k_1 = 2, \, k_2 = \tfrac{3}{2}, k_3 = 1,$ and $\delta_1 = \delta_2 = \delta_3 = 0$.
\vskip 0.0000001pt
\noindent
\vskip 0.0001pt
\noindent
\vskip 0.0000001pt
\noindent
\leftline{\bf Four-soliton solution of the KdV equation}
\vskip 2pt
\noindent
The computation of the four-soliton solution proceeds along the same lines.
After setting $\epsilon = 1$,
\begin{eqnarray}
\label{kdv4solfinalf}
\!\!f &=& 1 + \e^{\theta_1} + \e^{\theta_2} + \e^{\theta_3} + \e^{\theta_4}
+ a_{12}\, \e^{\theta_1 + \theta_2}
+ a_{13}\, \e^{\theta_1 + \theta_3}
+ a_{14}\, \e^{\theta_1 + \theta_4}
+ a_{23}\, \e^{\theta_2 + \theta_3}
\nonumber \\
&&
\!\!\!\!\!+ a_{24}\, \e^{\theta_2 + \theta_4}
+ a_{34}\, \e^{\theta_3 + \theta_4}
+ a_{12} a_{13} a_{23} \, \e^{\theta_1 + \theta_2 + \theta_3}
+ a_{12} a_{14} a_{24} \, \e^{\theta_1 + \theta_2 + \theta_4}
\nonumber \\
&&\!\!\!\!\!+ a_{13} a_{14} a_{34} \e^{\theta_1 \!+\! \theta_3 \!+\! \theta_4}
 +\! a_{23} a_{24} a_{34} \e^{\theta_2 \!+\! \theta_3 \!+\! \theta_4}
 +\! a_{12} a_{13} a_{14} a_{23} a_{24} a_{34}
\e^{\theta_1 \!+\! \theta_2 \!+\! \theta_3 \!+\! \theta_4},
\end{eqnarray}
with $a_{ij}$ as defined in (\ref{kdvaij}).

The four-soliton solution $u(x,t)$ of the KdV equation follows
from $u(x,t) = 2 (\ln f)_{xx}$.
Its analytic expression is not shown for it would fill pages.
\vskip 3pt
\noindent
\leftline{\bf {\it N}-soliton solution of the KdV equation}
\vskip 3pt
\noindent
Hirota introduced \cite[Eq.\ (5.38)]{hirota-solitons-1980} a concise
formula for the function $f$ leading to the $N$-soliton solution of
the KdV equation,
\begin{equation}
\label{nsolitonkdv}
f = \sum_{\mu = 0,1}
\e^{ \left[ \sum_{i < j}^{(N)} \mu_{i}\mu_{j} A_{ij}
+ \sum_{i=1}^{N} \mu_{i} \theta_{i} \right]},
\end{equation}
where $\sum_{\mu = 0,1}$ denotes the sum over the $2^N$ combinations
of $\mu_1 = 0, 1,$ $ \mu_2 = 0, 1,$ $\ldots,\;\mu_N = 0, 1$.
Furthermore, $\sum_{i < j}^{(N)}$ indicates summation over
all possible pairs $(i,j)$ with $i$ and $j$ chosen
from the $N$ elements $\{1,2,...,N\}$ but $i < j$,
and $a_{ij} = \e^{A_{ij}}.$

Inspired by the result obtained by the IST, 
the $N$-soliton solution can be written in a compact form
\cite{gardner-etal-commpureapplmath-1974,hirota-physrevlett-1972,wadati-sawada-jpsjpn-1980a,wadati-toda-jpsjpn-1972}
as
\begin{equation}
\label{kdvNsoliton}
u(x,t) = 2 \left( \ln {\mathrm{det}} (I + M) \right)_{xx}
\end{equation}
where $I$ is the $N \times N$ identity matrix and
\begin{equation}
\label{kdvmatrixm}
M_{\ell m} = \frac{\e^{\Theta_{\ell} + \Theta_m}}{K_\ell + K_m}
\;\;\; {\mathrm{with}} \; \Theta_\ell
= K_\ell x - 4 K_\ell^3 t +\Delta_\ell.
\end{equation}
Note that ${\mathrm{det}}(I + M)$ will match $f$ in (\ref{kdv1solfinalf}),
(\ref{kdv2solfinalf}), (\ref{kdv3solfinalf}), and (\ref{kdv4solfinalf})
when $k_i = 2 K_i$ and
$\delta_i = 2 \Delta_i - \ln(2 K_i)$ with $K_i > 0$.
%
\section{Application to a Class of Fifth-order Evolution Equations}
\label{fifth-order-equation}
In this section we investigate the soliton solutions of a
three-parameter family of fifth-order KdV equations,
\begin{equation}
\label{orgfifthkdveq}
u_t + \alpha u^2 u_{x} + \beta u_x u_{xx} + \gamma u u_{3x} + u_{5x} = 0,
\end{equation}
where $\alpha, \beta,$ and $\gamma$ are nonzero real parameters.
With $u = \tfrac{1}{\gamma} {\tilde{u}}$ one gets
\begin{equation}
\label{tildefifthkdveq}
{\tilde{u}}_t + \tfrac{\alpha}{\gamma^2} {\tilde{u}}^2 {\tilde{u}}_{x}
+ \tfrac{\beta}{\gamma} {\tilde{u}}_x {\tilde{u}}_{xx} + {\tilde{u}} {\tilde{u}}_{3x}
+ {\tilde{u}}_{5x} = 0,
\end{equation}
showing that the individual values of the parameters are less
important than the ratios
$\tfrac{\alpha}{\gamma^2}$ and $\tfrac{\beta}{\gamma}$.
Table~\ref{5kdvcases} shows the values of these ratios for which
(\ref{orgfifthkdveq}) is known to be completely integrable together with values
of $(\alpha,\beta,\gamma)$ used in the literature.
The names of the equations are also listed together with a couple of references.
Using scales on $u, x,$ and $t$, the named equations cannot be
transformed into one another; they are fundamentally
different\footnote{After a trivial scaling the CDG equation becomes
the SK equation. They are the {\em same} equations which often
goes unnoticed in the literature (see, e.g., \cite{kumar-etal-physscr-2022,saleem-hussain-jcam-2023}).}.
\vskip 0.01pt
\noindent
\begin{table}[htb]
\begin{center}
\begin{tabular}{rrrrrrrrr}
\hline \rule{0pt}{12pt}
$\tfrac{\alpha}{\gamma^2}$ & \phantom{xxxx} & $\tfrac{\beta}{\gamma}$ & \phantom{xxxx}
 & $(\alpha,\beta,\gamma)$ & Name & & & References
\\ [5pt]
\hline \rule{0pt}{12pt}
$\frac{3}{10}$  & & $2$ & & $(30, 20, 10)$ & & & & \cite{lax-communpureapplmath-1968}
\\
& & & & $(120, 40, 20)$ & Lax & & &
\cite{satsuma-kaup-jpsjpn-1977}
\\
& & & & & & & &
\\
& & & & $(5, 5, 5)$ & & & &
\\
$\frac{1}{5}$ & & $1$ & & $(45, 15, 15)$ & Sawada-Kotera & & &
    \cite{sawada-kotera-progtheorphys-1974}
\\
& & &  & $(180, 30, 30)$ & \;\;\;\, Caudrey-Dodd-Gibbon & & & \cite{caudrey-dodd-gibbon-prsa-1976,dodd-gibbon-prsa-1977}
\\
& & & & & & & &
\\
$\frac{1}{5}$ & & $\frac{5}{2}$ & & $(20, 25, 10)$ & Kaup-Kupershmidt & & &
\cite{fordy-gibbons-pla-1980,hirota-ramani-pla-1980,kaup-studapplmath-1980}
\\ [4pt]
\hline
\end{tabular}
\end{center}
\caption{Completely integrable fifth-order evolutions equations
  of type (\ref{orgfifthkdveq}).}
\label{5kdvcases}
\end{table}
\vskip 0.01pt
\noindent
Integrate (\ref{orgfifthkdveq}),
\begin{equation}
\label{intfifthkdv}
\partial_t \left( \int^x u \, dx \right) + \tfrac{1}{3} \alpha u^3 +
\tfrac{1}{2} (\beta - \gamma) u_x^2 + \gamma u u_{xx} + u_{4x} = 0,
\end{equation}
and substitute (\ref{kdvtinf})
where $c$ is a constant, to get
\begin{eqnarray}
\label{fkdvfifthdegree}
\!\!\!\!&&\!\!\!\!\!6 f^5 ( f_{xt} + f_{6x} ) - 3 f^4 (2 f_x f_t + \ldots + 12 f_x f_{5x})
   + 2 f^3 \left( (...) f_{xx}^3 + \ldots + (...) f_{x}^2 f_{4x} \right)
\nonumber \\
\!\!\!\!&&\!\!\!\!\!\! +3 f^2 f_{x}^2 \left( (...)f_{xx}^2 \!+\! (...)f_x f_{3x} \right)
   +\!2 f_{x}^4 (360 \!-\!6 \beta c\!+\!\alpha c^2\!-\!12 \gamma c)(3 f f_{xx}\!-\! f_{x}^2)
   \!=\!0,
\end{eqnarray}
which is of sixth degree.
In the next subsections we investigate the integrable cases listed in
Table~\ref{5kdvcases}.
For each case the constant $c$ can be obtained from substituting a Laurent series into (\ref{orgfifthkdveq}).

\subsection{The Lax equation}
\label{lax-equation}
Using $\alpha = \tfrac{3}{10} \gamma^2,\; \beta = 2 \gamma$, and $c = \tfrac{20}{\gamma}$,
(\ref{fkdvfifthdegree}) reduces to a homogeneous {\it trilinear} equation
\begin{equation}
\label{cubiclax}
f^2 ( f_{xt} + f_{6x} ) - f ( f_x f_t - 5 f_{xx} f_{4x} + 6  f_x f_{5x} )
+ 10 ( f_{xx}^3 - 2 f_x f_{xx} f_{3x} + f_x^2 f_{4x} ) \!=\! 0,
\end{equation}
which can be written in bilinear form consisting of two coupled equations
\cite[p.\ 56]{hirota-book-2004},
\cite{hirota-solitons-1980,satsuma-kaup-jpsjpn-1977}:
\begin{eqnarray}
\label{bilinearlaxeq}
&& \left(D_x D_{s} + D_x^4 \right) (f \mathbf{\cdot} f) = 0,
\nonumber \\
&& \left( D_x D_{t} + D_x^6 \right) (f {\bf \cdot} f)
- \tfrac{5}{3}\left( D_{s}^2 + D_{s} D_x^3 \right) (f \mathbf{\cdot} f)
= 0,
\end{eqnarray}
for only one function $f$ but with an extra independent variable $s$
which corresponds to the time variable in the KdV equation.
This comes as no surprise because the Lax equation belongs
to the family of KdV flows \cite[p. 114]{newell-book-1985}
each with its own time variable.
Upon elimination of $s$ via suitable cross differentiations
one obtains (\ref{cubiclax}).

Note that (\ref{cubiclax}) can also be recast in terms of
Hirota trilinear operators
\cite[Eq.\ (8.113)]{hietarinta-lnp767-2009}.
Completely integrable trilinear equations have been studied
\cite{grammaticosetal-pla-1994,hietarinta-lnp767-2009,hietarintaetal-needs-1995,ma-fmc-2013}
but are less common than their bilinear counterparts.
Specific examples can be found in, for example,  \cite{matsukidairaetal-pla-1990,satsuma-etal-jpsjpn-1992,schiff-bookchapter-1992}.

We will not use (\ref{bilinearlaxeq}) in the subsequent
computation of solitons.
Instead, we write the cubic equation (\ref{cubiclax}) as
\begin{equation}
\label{laxoperform}
f^2 {\cal L} f + f {\cal N}_1 (f,f) + {\cal N}_2 (f,f,f) = 0,
\end{equation}
with operators
\begin{eqnarray}
\label{laxoperators}
{\cal L} f &=&   f_{xt} + f_{6x},
\nonumber \\
{\cal N}_1 (f,g) &=& -( f_t g_x - 5 f_{xx} g_{4x} + 6 f_x g_{5x} ),
\\
{\cal N}_2 (f,g,h) 
&=& 10 (f_{xx} g_{xx} h_{xx} - 2 f_x g_{xx} h_{3x} + f_x g_x h_{4x}),
\nonumber
\end{eqnarray}
where $f, g$, and $h$ are auxiliary functions. 

Upon substitution of (\ref{hirotaexpansionf}) into (\ref{laxoperform})
the first four equations of the perturbation scheme
become\footnote{The derivation is given in the Appendix.}
\begin{eqnarray}
\label{trilinearscheme}
\nonumber
O(\epsilon^{1}):  {\cal L} f^{(1)} &=& 0,
\nonumber \\
O(\epsilon^{2}): {\cal L} f^{(2)}
   &=&  - {\cal N}_1 (f^{(1)},f^{(1)}),
\nonumber \\
O(\epsilon^{3}): {\cal L} f^{(3)} &=&
  - \left( 2 f^{(1)} {\cal L} f^{(2)}
  + {\cal N}_1 (f^{(1)},f^{(2)} ) + {\cal N}_1 (f^{(2)},f^{(1)}) \right.
\nonumber \\
&& \left. + f^{(1)} {\cal N}_1 (f^{(1)}, f^{(1)})
    + {\cal N}_2 (f^{(1)},f^{(1)}, f^{(1)}) \right),
\nonumber \\
O(\epsilon^{4} ): {\cal L} f^{(4)}
 &=& - \left(
     2 f^{(1)} {\cal L} f^{(3)}
     + \left( 2 f^{(2)} + {f^{(1)}}^2 \right) {\cal L} f^{(2)}
     + {\cal N}_1 (f^{(1)},f^{(3)})
     \right.
 \nonumber \\
 && \left.
    + \, {\cal N}_1 (f^{(3)},f^{(1)})
    + {\cal N}_1 (f^{(2)},f^{(2)})
    + f^{(1)} \left( {\cal N}_1 (f^{(1)},f^{(2)}) \right.
    \right.
 \nonumber \\
 && \left. \left.
    + \, {\cal N}_1 (f^{(2)},f^{(1)}) \right)
    + f^{(2)} {\cal N}_1 (f^{(1)},f^{(1)})
    + {\cal N}_2 (f^{(1)},f^{(1)},f^{(2)})
    \right.
  \nonumber \\
 && \left.
    + \, {\cal N}_2 (f^{(1)},f^{(2)},f^{(1)})
    + {\cal N}_2 (f^{(2)},f^{(1)},f^{(1)})
    \right),
\end{eqnarray}
where we used the first equation to simplify the other ones.
Starting from (\ref{kdvf1}),
one can proceed as in KdV case to construct soliton solutions of any order $N$.
The only difference is that for the Lax equation $\omega_i = k_i^5$ instead of
$\omega_i = k_i^3$.
For example, the one-soliton solution
\begin{equation}
\label{Lax1solu}
u(x,t) = \tfrac{5}{\gamma} k^2 \sech^2 \left[ \tfrac{1}{2}(k x -k^5 t +\delta) \right]
       = \tfrac{20}{\gamma} K^2 \sech^2 \left( K x- 16 K^5 t + \Delta \right),
\end{equation}
where
$K = \frac{k}{2}$ and $\Delta = \frac{\delta}{2}$, solves
\begin{equation}
\label{Laxeq}
u_t + \tfrac{3}{10} \gamma^2 u^2 u_{x} + 2 \gamma u_x u_{xx}
+ \gamma u u_{3x} + u_{5x} = 0.
\end{equation}
\subsection{The Sawada-Kotera equation}
\label{sawada-kotera-equation}
Using
$\alpha = \tfrac{1}{5} \gamma^2,\; \beta = \gamma,$ and
$c = \tfrac{30}{\gamma}$
one gets a quadratic equation,
\begin{equation}
\label{skinf}
f (f_{xt} + f_{6x}) - f_x f_t - 10 f_{3x}^2 + 15 f_{xx} f_{4x}
- 6 f_x f_{5x} = 0,
\end{equation}
which can be written in bilinear form \cite{hirota-solitons-1980} as
\begin{equation}
\label{bilinearformsk}
\left( D_x D_t + D_x^6 \right) (f \mathbf{\cdot} f) = 0.
\end{equation}
Ignoring the bilinear representation, we write
(\ref{skinf}) in the form (\ref{kdvoperform}) with
\begin{eqnarray}
\label{sklinop}
{\cal L} f &=& f_{xt} + f_{6x},
\\
\label{sknlop}
{\cal N} (f, g) &=& - f_x g_t - 10 f_{3x} g_{3x} + 15 f_{xx} g_{4x}
- 6 f_x g_{5x},
\end{eqnarray}
and proceed as in the KdV case, leading to the following soliton solutions.
\vskip 3pt
\noindent
\leftline{\bf One-soliton solution of the SK equation}
\vskip 2pt
\noindent
The solitary wave solution
\begin{eqnarray}
u(x,t) & = & \tfrac{15}{2 \gamma} k^2 \sech^2 \left[\tfrac{1}{2}(k x - k^5 t + \delta) \right]
\nonumber \\
       & = & \tfrac{30}{\gamma} K^2 \sech^2 \left( K x -16 K^5 t + \Delta \right),
\end{eqnarray}
where $K = \tfrac{k}{2}$ and $\Delta = \tfrac{\delta}{2}$, solves
\begin{equation}
\label{skorg}
u_t + \tfrac{1}{5} \gamma^2 u^2 u_{x} + \gamma u_x u_{xx}
+ \gamma u u_{3x} + u_{5x} = 0.
\end{equation}
\vskip 3pt
\noindent
\leftline{\bf Higher-order soliton solutions of the SK equation}
\vskip 2pt
\noindent
The computation of higher-order soliton solutions is analogous to the KdV equation;
see (\ref{kdv2solfinalf}), (\ref{kdv3solfinalf}), and (\ref{kdv4solfinalf}).
Except that the dispersion relation is now quintic, $\omega_i = k_i^5,$
and the $a_{ij}$ must be replaced by
\begin{equation}
a_{ij} =
\frac{{(k_i - k_j)}^2 \, (k_i^2- k_i k_j + k_j^2 )}
    {{(k_i + k_j )}^2  \,(k_i^2 + k_i k_j + k_j^2 )}
  = \frac{{(k_i - k_j)}^3 \, (k_i^3 + k_j^3 )}
    {{(k_i + k_j)}^3 \, (k_i^3 - k_j^3 )}.
\end{equation}
The actual two- and three-soliton solutions $u(x,t)$ of the
SK equation are very long expressions (not shown).
%
\subsection{The Kaup-Kupershmidt equation}
\label{kaup-kupershmidt-equation}
Using $\alpha = \tfrac{1}{5} \gamma^2,\; \beta = \tfrac{5}{2} \gamma,$
and $c = \tfrac{15}{\gamma}$, (\ref{fkdvfifthdegree}) becomes a
{\em quartic} equation,
\begin{eqnarray}
\label{kkeqinf}
&& 4 f^3 ( f_{xt} + f_{6x} )
- f^2 ( 4 f_t f_x - 5 {f_{3x}}^2 + 24 f_x f_{5x} )
\nonumber \\
&& - 30 f f_x ( f_{xx} f_{3x} -  2 f_x f_{4x} )
+ 15 {f_x}^2 ( 3 {f_{xx}}^2 - 4 f_x f_{3x} ) = 0,
\end{eqnarray}
which can be written as a coupled system of bilinear equations
\cite[p.\ 36]{hirota-book-2004}, \cite{parker-physicad-2000,wang-modphyslettb-2017},
\begin{eqnarray}
\label{bilineareq1kk}
&& \left( D_x D_{t} + \tfrac{1}{16} D_x^6 \right) (f \mathbf{\cdot} f)
+ \tfrac{15}{4} D_x^2 (f \mathbf{\cdot} g) = 0,
\\
\label{bilineareq2kk}
&& D_x^4 (f \mathbf{\cdot} f) - 4 f g = 0,
\end{eqnarray}
for two unknown functions $f$ and $g$.
One can verify that upon elimination of $g$ in (\ref{bilineareq1kk})
and (\ref{bilineareq2kk}) indeed yields (\ref{kkeqinf}).

In what follow, we will ignore the bilinear system and write (\ref{kkeqinf})
in operator form as
\begin{equation}
\label{kkoperform}
f^3 {\cal L} f + f^2 {\cal N}_1 (f,f) + f {\cal N}_2 (f,f,f) + {\cal N}_3 (f,f,f,f) = 0.
\end{equation}
This homogeneous equation involves one linear operator and
three nonlinear operators defined as
\begin{eqnarray}
\label{linearoperkk}
{\cal L} f & = & 4 ( f_{xt} + f_{6x}),
\\
\label{nonlinearoper1kk}
{\cal N}_1 (f,g) & = & - ( 4 f_t g_x  - 5 f_{3x} g_{3x} + 24 f_x g_{5x} ),
\\
\label{nonlinearoper2kk}
{\cal N}_2 (f,g,h) & = & - 30 f_x ( g_{xx} h_{3x} - 2 g_x h_{4x} ),
\\
\label{nonlinearoper3kk}
{\cal N}_3 (f,g,h,j) & =& 15 f_x g_x ( 3 h_{xx} j_{xx} - 4 h_x j_{3x} ),
\end{eqnarray}
for auxiliary functions $f(x,t), g(x,t), h(x,t),$ and $j(x,t)$. 
The nonlinear operators are bilinear, trilinear, and quadrilinear, respectively.

\begin{figure}[htb]
\begin{center}
\begin{tabular}{cc}
\includegraphics[width=2.29in, height=1.552in]{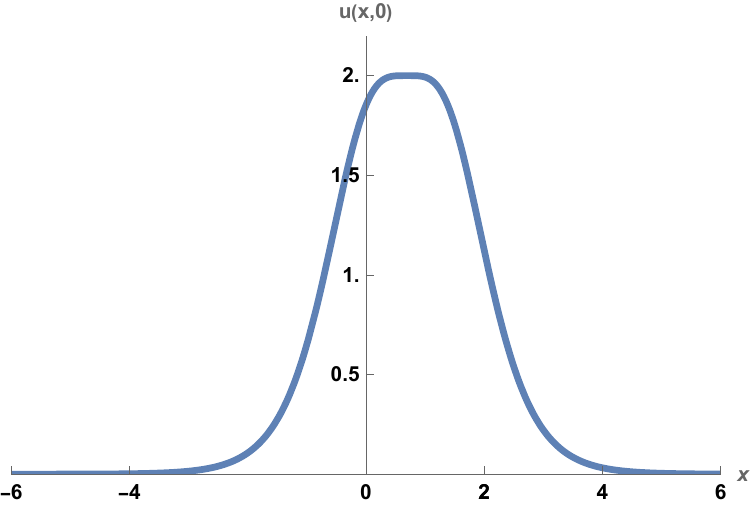}
&
\includegraphics[width=2.29in, height=1.552in]{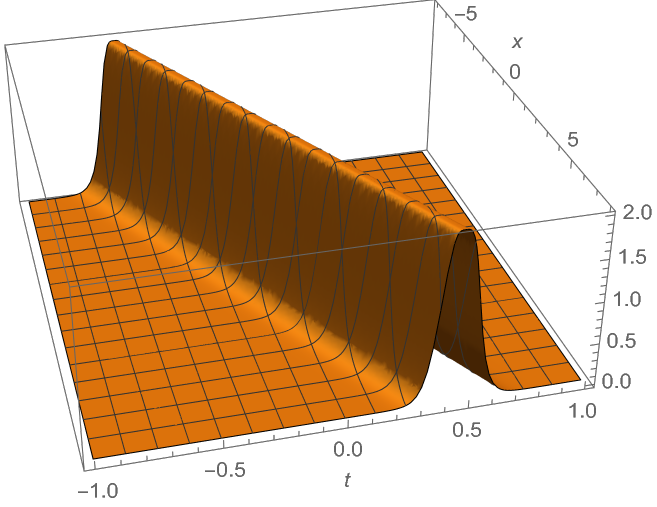}
\end{tabular}
\end{center}
\caption{2D and 3D graphs of solution (\ref{kku1solform2}) with
$\gamma = 10, k = 2,$ and $\delta = 0$.}
\label{graphs1solKKeq}
\end{figure}

\begin{figure}[htb!]
\begin{center}
\begin{tabular}{ccc}
\includegraphics[width=1.57in, height=1.8in]{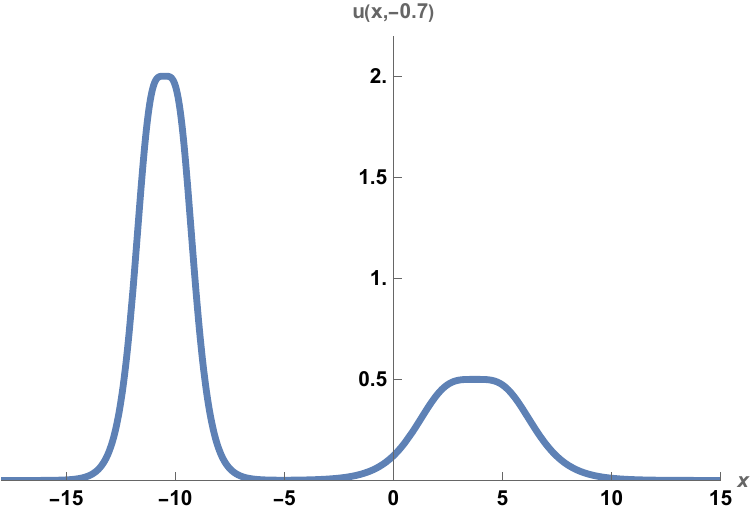} &
\includegraphics[width=1.57in, height=1.8in]{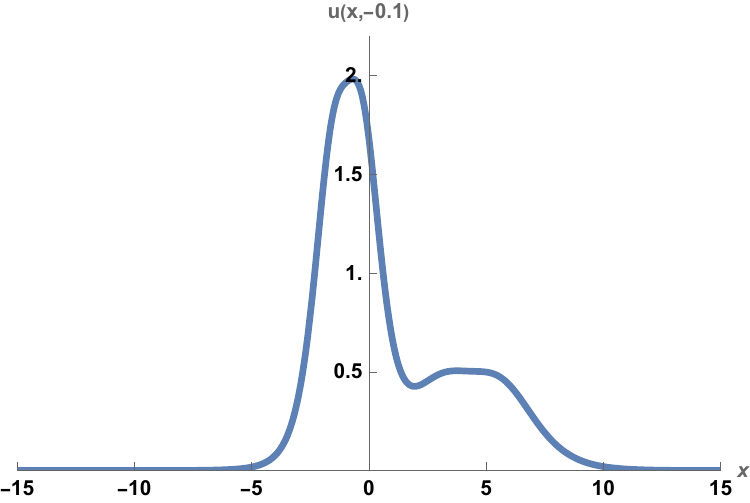} &
\includegraphics[width=1.57in, height=1.8in]{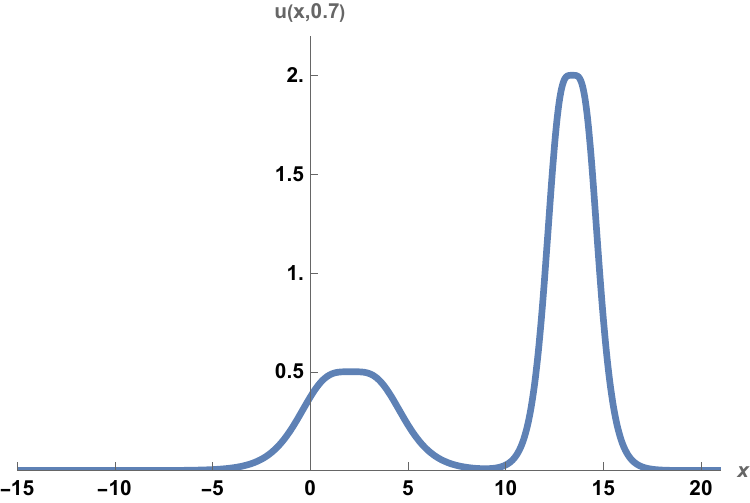}
\end{tabular}
\end{center}
\caption{Graph of the two-soliton solution of the KK equation at three
different moments in time.}
\label{kktwosolitonsintime}
\end{figure}

\begin{figure}[t!]
\begin{center}
\includegraphics[width=3.60in, height=2.22in]{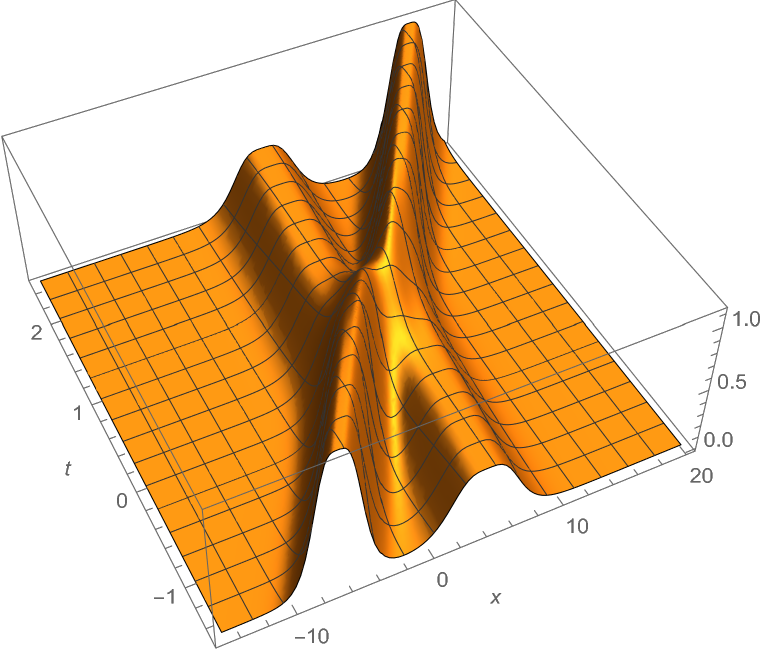}
\end{center}
\caption{Bird's eye view of the collision of two solitons for the KK equation.
Notice the phase shift after the collision.}
\label{3Dkktwosolitons}
\end{figure}

Substituting (\ref{hirotaexpansionf}) into (\ref{kkoperform})
and equating the coefficients of powers of $\epsilon$ to zero
yields\footnote{Details of the derivation are given in the Appendix.}
the perturbation scheme of which the first four equations read
\begin{eqnarray}
\label{quadrilinearscheme}
O(\epsilon^{1}): {\cal L} f^{(1)} &=& 0,
\nonumber \\
O(\epsilon^{2}): {\cal L} f^{(2)}
    &=& - {\cal N}_1 (f^{(1)}, f^{(1)}),
\nonumber \\
O(\epsilon^{3}): {\cal L} f^{(3)}
 &=& - \left(
     3 f^{(1)} {\cal L} f^{(2)} + 2 {f^{(1)}} {\cal N}_1 (f^{(1)}, f^{(1)})
     + {\cal N}_1 (f^{(2)}, f^{(1)})
     \right.
 \nonumber \\
 &&
    \left.
    + {\cal N}_1 (f^{(1)}, f^{(2)}) + {\cal N}_2 (f^{(1)}, f^{(1)},f^{(1)})
    \right),
\nonumber \\
O(\epsilon^{4} ): {\cal L} f^{(4)}
 &=& - \left(
     3 f^{(1)} {\cal L} f^{(3)}
     + 3 \left( f^{(2)} + {f^{(1)}}^2 \right) {\cal L} f^{(2)}
     + {\cal N}_1 (f^{(1)}, f^{(3)})
     \right.
 \nonumber \\
 && \left.
    + \, {\cal N}_1 (f^{(3)}, f^{(1)})
    + {\cal N}_1 (f^{(2)}, f^{(2)})
    + 2 {f^{(1)}} \left( {\cal N}_1 (f^{(1)},f^{(2)}) \right.
    \right.
 \nonumber \\
 && \left. \left.
    + \, {\cal N}_1 (f^{(2)},f^{(1)}) \right)
    + \left( 2 {f^{(2)}} + {f^{(1)}}^2 \right) \,{\cal N}_1 (f^{(1)},f^{(1)})
    \right.
  \nonumber \\
 && \left.
    + \, {\cal N}_2 (f^{(1)},f^{(1)},f^{(2)})
    + {\cal N}_2 (f^{(1)},f^{(2)},f^{(1)})
    + {\cal N}_2 (f^{(2)},f^{(1)},f^{(1)})
    \right.
    \nonumber \\
 && \left.
    + \, f^{(1)} {\cal N}_2 (f^{(1)}, f^{(1)},f^{(1)})
    + {\cal N}_3 (f^{(1)}, f^{(1)},f^{(1)},f^{(1)})
    \right),
\end{eqnarray}
where we used the first equation to simplify the subsequent ones.
Clearly, the number of terms grows at each order in $\epsilon$
and the computational complexity increases accordingly.
Full details of the step-by-step solution of the perturbation
scheme for the KK equation with coefficients
$\alpha = 20, \beta = 25,$ and $\gamma = 10,$
can be found \cite{hereman-nuseir-mcs-1997,nuseir-phd-thesis-1995}
where we used {\em Macsyma} to perform the lengthy computations.
Here we summarize the results for general $\alpha, \beta,$ and $\gamma$ 
subject to the conditions 
$\alpha = \tfrac{1}{5} \gamma^2,$ and $\beta = \tfrac{5}{2} \gamma$.
\vskip 3pt
\noindent
\leftline{\bf One-soliton solution of the KK equation}
\vskip 2pt
\noindent
Taking $f^{(1)} = \e^{\theta} = \e^{k x -\omega t + \delta}, \;$
${\cal L} f^{(1)} = 0$ yields $ \omega = k^5.$
In contrast to the KdV case, the right hand side of the second equation,
\begin{equation}
\label{kkn1f1f1sol1}
-{\cal N}(f^{(1)},f^{(1)}) = 15 k^6 \, \e^{2 \theta},
\end{equation}
does not vanish but has a term in $\e^{2 \theta}$.
Thus, $f^{(2)}$ must be of the form
\begin{equation}
\label{kkf2sol1}
f^{(2)} = a \, \e^{2 \theta},
\end{equation}
with undetermined constant coefficient $a$.
Then,
\begin{equation}
\label{kklf2sol1}
{\cal L} f^{(2)} = 240 a k^6 \, \e^{2 \theta}
\end{equation}
and equating the right hand sides of (\ref{kkn1f1f1sol1}) and (\ref{kklf2sol1})
yields $a=\tfrac{1}{16}.$
Next, we check that we can set $f^{(n)} = 0$ for $n \ge 3$ by
verifying that the right hand sides of the subsequent equations
in (\ref{quadrilinearscheme}) are all zero.
This is indeed the case and the perturbation scheme terminates after two steps.
Setting $\epsilon = 1$,
\begin{equation}
\label{kkfsol1}
f = 1 + \e^{\theta} + \tfrac{1}{16} \, \e^{2\theta},
\end{equation}
and $u = \tfrac{15}{\gamma} (\ln f)_{xx} $ yields
\begin{equation}
\label{kku1solform1}
u =  \tfrac{240}{\gamma} k^2 \,
     \left[ \frac{\e^{\theta} (16 + 4 \e^{\theta} + \e^{2\theta})}
     {(16 + 16 \e^{\theta} + \e^{2\theta})^2} \right]
\end{equation}
which solves
\begin{equation}
\label{orgfifthkk}
u_t + \tfrac{1}{5} \gamma^2 u^2 u_{x} + \tfrac{5}{2} \gamma u_x u_{xx} +
    \gamma u u_{3x} + u_{5x} = 0.
\end{equation}
The one-soliton solution can also be written as
\begin{eqnarray}
\label{kku1solform2}
u &=& \tfrac{240}{\gamma} k^2 \,
\left( \frac{ \left[ 1 - \tanh^2 (\frac{\theta}{2})\right]
 \left[ 21 - 30 \tanh \frac{\theta}{2} + 13\tanh^2 (\frac{\theta}{2}) \right] }
{\left[ 33 - 30 \tanh \frac{\theta}{2} + \tanh^2 (\frac{\theta}{2}) \right]^2} \right)
\\
\label{kku1solform2bis}
&=& \tfrac{240}{\gamma} k^2 \,
\left( \frac{ 4 + 17 \cosh \theta - 15 \sinh \theta }
{\left[ 16 + 17 \cosh \theta - 15 \sinh \theta \right]^2} \right),
\end{eqnarray}
where $\theta = k x - k^5 t + \delta$.
Fig.~\ref{graphs1solKKeq} shows the 2D and 3D graphs of the one-soliton
solution for $\gamma = 10, k = 2,$ and $\delta = 0$.
In comparison with the solitary wave solution of the KdV equation shown in
Figs.~\ref{kdv-sech-cn-sols} and~\ref{humpsolutionKdVeq},
the solution of the KK equation is wider and flatter at the top.
\vskip 0.00001pt
\noindent
\vskip 0.0001pt
\noindent
\leftline{\bf Two-soliton solution of the KK equation}
\vskip 2pt
\noindent
Starting from
\begin{equation}
\label{kkf1sol2}
f^{(1)} = \e^{\theta_1} + \e^{\theta_2},
\end{equation}
where $\theta_i= k_i x - k_i^5 t + \delta_i \; (i = 1,2)$,
we compute
\begin{equation}
\label{kkn1f1f1sol2}
- {\cal N}_1(f^{(1)},f^{(1)})
 = 15 k_1^6 \, \e^{2{\theta}_1} + 15 k_2^6 \, \e^{2{\theta}_2}
+ 10 k_1 k_2 (2 k_1^4 - k_1^2 k_2^2 + 2 k_2^4) \, \e^{\theta_1+\theta_2}.
\end{equation}
Thus $f^{(2)}$ must be of the form
\begin{equation}
\label{kkf2sol2}
f^{(2)} = a_1 \, \e^{2 {\theta}_1} + a_2 \, \e^{2 {\theta}_2}
          + a_{12} \, \e^{{\theta}_1 + {\theta}_2},
\end{equation}
with the (constant) coefficients $a_1, a_2,$ and $a_{12}$ to be determined.
Then,
\begin{eqnarray}
\label{kklf2sol2}
{\cal L} f^{(2)}
 &=& 240 a_1 k_1^6 \, \e^{2 \theta_1} + 240 a_2 k_2^6 \, \e^{2 \theta_2}
\nonumber \\
 & & +\, 20 a_{12} k_1 k_2 (k_1 + k_2)^2 ( k_1^2 + k_1 k_2 + k_2^2) \, \e^{\theta_1 +\theta_2}.
\end{eqnarray}
Equating (\ref{kkn1f1f1sol2}) with (\ref{kklf2sol2}) determines
$ a_1 = a_2 = \tfrac{1}{16}$, 
as expected, and 
\begin{equation}
\label{kka12sol2}
a_{12} = \frac{2 k_1^4 - k_1^2 k_2^2 + 2 k_2^4 }{2{(k_1+k_2)}^2 (k_1^2+k_1 k_2+k_2^2)}.
\end{equation}
Therefore,
\begin{equation}
\label{kkf2sol2fin}
f^{(2)}= \tfrac{1}{16} \, \e^{2 \theta_1} + \tfrac{1}{16} \, \e^{2 \theta_2}
         + \frac{(2 k_1^4 -k_1^2 k_2^2 + 2 k_2^4) }{2{(k_1+k_2)}^2 (k_1^2+k_1 k_2+k_2^2)}
         \, \e^{\theta_1 +\theta_2}.
\end{equation}
The main difference with the the KdV, Lax, and SK equations is
that the terms $\e^{2 \theta_1}$ and $\e^{2 \theta_2}$
in $f^{(2)}$ no longer drop out.
At $O(\epsilon^3)$ one gets
\begin{equation}
\label{kkf3sol2}
f^{(3)} = b_{12} \, \left( \e^{\theta_1 + 2\theta_2} + \e^{2\theta_1 +\theta_2} \right),
\end{equation}
with
\begin{equation}
\label{kkb12sol2}
 b_{12} = \frac{(k_1-k_2)^2 (k_1^2 -k_1 k_2 +k_2^2)}{16 (k_1+k_2)^2 (k_1^2+k_1 k_2+k_2^2)}.
\end{equation}
At the next order
\begin{equation}
\label{kkf4sol2}
f^{(4)}
= b_{12}^2 \, \e^{2(\theta_1 +\theta_2)}
= \frac{(k_1-k_2)^4 (k_1^2-k_1 k_2 + k_2^2)^2}
  {256 (k_1+k_2)^4 (k_1^2 +k_1 k_2 +k_2^2)^2}
  \, \e^{2 (\theta_1 + \theta_2)}.
\end{equation}
After verification that all $f^{(n)}$ are zero for $n \ge 5$
and setting $\epsilon = 1$,
\begin{eqnarray}
\label{kkfsol2fin}
f & = &
  1 + \e^{\theta_1} + \e^{\theta_2} + \tfrac{1}{16} \, \e^{2 \theta_1}
  + \tfrac{1}{16} \, \e^{2 \theta_2} + a_{12} \, \e^{\theta_1 +\theta_2}
\nonumber \\
& &
+\, b_{12} \left( \e^{2 \theta_1 + \theta_2}
+ \e^{\theta_1 + 2 \theta_2} \right)
+ b_{12}^2 \, \e^{2 (\theta_1 + \theta_2)}.
\end{eqnarray}
The explicit expression of $u(x,t)$ (not shown due to length)
then follows from $u = \tfrac{15}{\gamma} (\ln f)_{xx}$.
The collision of two solitons for the KK equation is shown in
Figs.~\ref{kktwosolitonsintime} and~\ref{3Dkktwosolitons}
for $k_1 = 2, \, k_2 = 1$, and $\delta_1 = \delta_2 = 0$.
\vskip 0.0001pt
\noindent
\vskip 0.000001pt
\noindent
\vskip 0.1pt
\noindent
\leftline{\bf Three-soliton solution of the KK equation}
\vskip 3pt
\noindent
Starting with
\begin{equation}
\label{kkf1sol3}
f^{(1)} = \sum_{i=1}^3 \e^{\theta_i}
        = \e^{\theta_1} + \e^{\theta_2} + \e^{\theta_3},
\end{equation}
where $\theta_i = k_i x - k_i^5 t + \delta_i$,
the equations of the perturbation scheme are solved order-by-order yielding
expressions for 
$f^{(2)}, f^{(3)}, \ldots, f^{(6)}$ 
because, as it turns out, $f^{(n)}=0$ for $n \ge 7$.
The latter requires verification that the right hand sides at $O(\epsilon^7)$
and beyond all vanish in order for the perturbation scheme to terminate.
The computations are very lengthy, time consuming, and currently at
the limit of what {\em Mathematica} can
handle\footnote{With the code PDESolitonSolutions.m
discussed in Section~\ref{software},
the computation of the three-soliton solution takes
about 4 minutes on a Dell XPS-15 laptop with Intel Core i7 processor
at 4.7GHz and 32 GB of memory.}.
Summarizing the results:
\begin{equation}
\label{kkf2sol3}
f^{(2)} = \tfrac{1}{16} \sum_{i=1}^3 \e^{2 \theta_i}
          + \sum_{1 \le i < j \le 3} a_{ij} \, \e^{\theta_i + \theta_j},
\end{equation}
with phase factors
\begin{equation}
\label{kkaijsol3}
a_{ij} = \frac{2 k_i^4 - k_i^2 k_j^2 + 2 k_j^4}{2{(k_i+k_j)}^2
(k_i^2+k_i k_j +k_j^2)}, \quad \quad 1 \le i<j \le 3.
\end{equation}
Next,
\begin{equation}
\label{kkf3sol3}
f^{(3)} = \sum_{1 \le i < j \le 3}
    b_{ij} \, \left( \e^{2 \theta_i + \theta_j} + \e^{\theta_i + 2 \theta_j} \right)
    + c_{123} \, \e^{ {\theta}_1 + {\theta}_2+ {\theta}_3},
\end{equation}
where
\begin{equation}
\label{kkbijsol3}
b_{ij} = \frac{(k_i-k_j)^2 (k_i^2 - k_i k_j + k_j^2)}
         {16 (k_i+k_j)^2 (k_i^2 + k_i k_j + k_j^2)},
         \quad \quad 1 \le i<j \le 3,
\end{equation}
and
\begin{eqnarray}
\label{kkc123sol3}
c_{123} & = &\! \tfrac{1}{D} \left[
  (2 k_1^4 - k_1^2 k_2^2 + 2 k_2^4) (k_3^8+k_1^4 k_2^4)
  + (2 k_1^4 - k_1^2 k_3^2 + 2 k_3^4) (k_2^8+k_1^4 k_3^4) \right.
\nonumber \\
& & \left.
    \!\!+ (2 k_2^4 - k_2^2 k_3^2 + 2 k_3^4) (k_1^8+k_2^4 k_3^4) \right]
    \!-\! \tfrac{1}{2 D}\!\left[
   (k_1^2+k_2^2)(k_1^4+k_2^4) (k_3^6+ k_1^2 k_2^2 k_3^2)
   \right.
\nonumber \\
& & \left.
    \!\!+ (k_1^2+k_3^2) (k_1^4+k_3^4)(k_2^6+ k_1^2 k_2^2 k_3^2)
       + (k_2^2+k_3^2) (k_2^4+k_3^4)(k_1^6 + k_1^2 k_2^2 k_3^2) \right.
\nonumber \\
& & \left. \!\!+ 12 k_1^4 k_2^4 k_3^4 \right]
\end{eqnarray}
with
\begin{equation}
\label{kkDsol3}
D = 4 \prod_{1 \le i< j \le 3} (k_i + k_j)^2 \, (k_i^2 + k_i k_j + k_j^2).
\end{equation}
Carrying on,
\begin{eqnarray}
\label{kkf4sol3}
f^{(4)} &=&
\sum_{1 \le i < j \le 3} b_{ij}^2 \, \e^{2 (\theta_i + \theta_j)}
+ 16 \left( a_{23} b_{12} b_{13} \, \e^{2 \theta_1 + \theta_2 + \theta_3} \right.
\nonumber \\
& &  \left. + \, a_{13} b_{12} b_{23} \e^{\theta_1 + 2 \theta_2 + \theta_3}
+  a_{12} b_{13} b_{23} \, \e^{\theta_1 +  \theta_2 + 2 \theta_3} \right),
\\
f^{(5)} &=&
   256\, b_{12} b_{13} b_{23} \left( b_{12} \,\e^{2 \theta_1 + 2 \theta_2 + \theta_3}
   + b_{13} \e^{2 \theta_1 + \theta_2 + 2 \theta_3} \right.
\nonumber \\
& & \left. + \, b_{23} \, \e^{\theta_1 + 2 \theta_2 + 2 \theta_3} \right),
\\
f^{(6)} & = &
  16 \, (16 b_{12} b_{13} b_{23})^2 \, \e^{2( \theta_1 + \theta_2 + \theta_3)}.
\end{eqnarray}
Finally, after setting $\epsilon = 1$,
\begin{equation}
\label{kkfsol3fin}
f = 1 + f^{(1)} + f^{(2)} + f^{(3)} + f^{(4)} + f^{(5)} + f^{(6)},
\end{equation}
and $u(x,t) = \tfrac{15}{\gamma} (\ln f)_{xx} \,$
(not shown due to length) then solves (\ref{orgfifthkk}).

\begin{figure}[htb]
\begin{center}
\begin{tabular}{ccc}
\includegraphics[width=1.52in, height=1.76in]{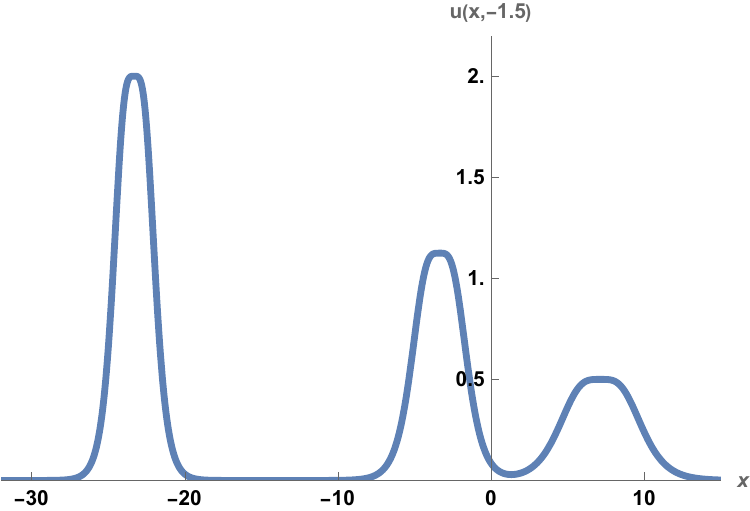} &
\includegraphics[width=1.52in, height=1.76in]{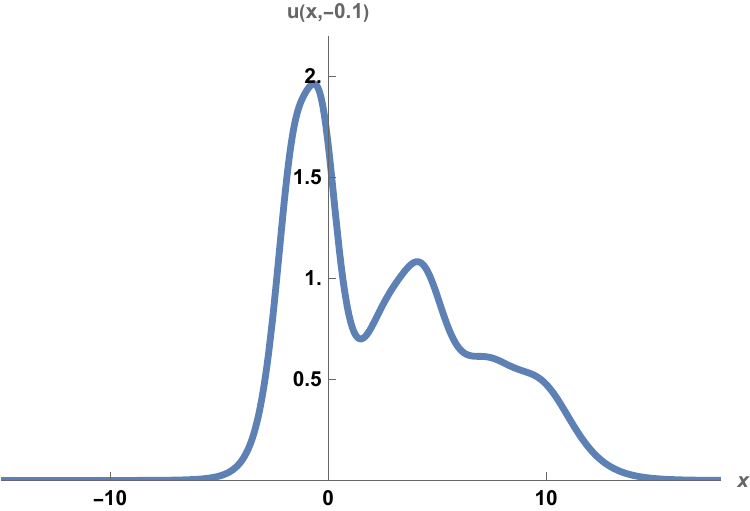} &
\includegraphics[width=1.52in, height=1.76in]{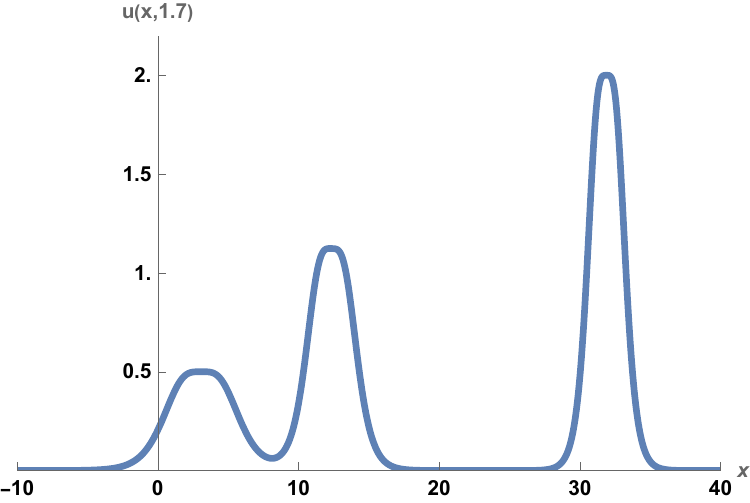}
\end{tabular}
\end{center}
\caption{Graph of the three-soliton solution of the KK equation at three different moments in time.}
\label{kk-three-solitons-in-time}
\end{figure}

\begin{figure}[htb]
\begin{center}
\includegraphics[width=3.85in, height=2.75in]{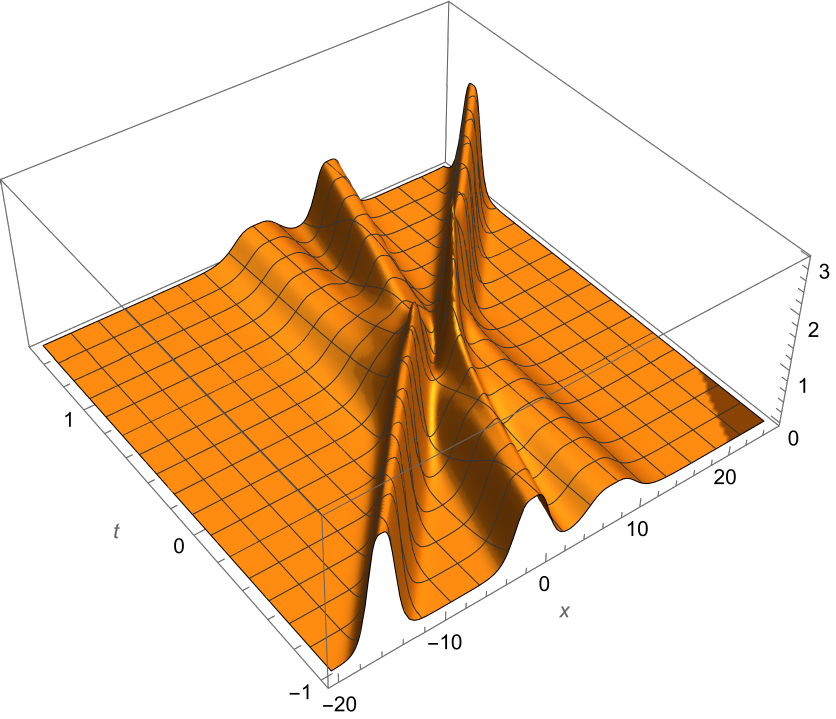}
\end{center}
\caption{Bird's eye view of the collision of three solitons for the KK equation.
Notice the phase shift after the collision.}
\label{kk-three-solitons-in-3D}
\end{figure}

The collision of three solitons for the KK equation is shown in Figs.~\ref{kk-three-solitons-in-time}
and~\ref{kk-three-solitons-in-3D} for
$k_1 = 2, k_2 = \tfrac{3}{2}, k_3 = 1,$ and $\delta_1 = \delta_2 = \delta_3 = 0$.
\vskip 0.00001pt
\noindent
\vskip 0.00001pt
\noindent
\section{The Modified KdV Equation}
\label{mkdv-equation}
Of course, not every polynomial soliton equation in $(1+1)$ dimensions
can be solved with a solution of type (\ref{hirotaexpansionf}).
Consider, for example, the mKdV equation,
\begin{equation}
\label{orgmkdv}
u_t + 24 u^2 u_x + u_{3x} = 0,
\end{equation}
which after integration becomes
\begin{equation}
\label{mkdvint}
\partial_t \left( \int^x u \, dx \right) + 8 u^3 + u_{xx} = 0.
\end{equation}
The Laurent series for (\ref{orgmkdv}) suggests the transformation
\begin{equation}
\label{laurentmkdv}
u = \pm \tfrac{1}{2} i \, (\ln {\tilde f})_x
  = \pm \tfrac{1}{2} i \, \left( \frac{{\tilde f}_x}{{\tilde f}} \right).
\end{equation}
Substitution of either of these branches into (\ref{mkdvint}) yields
\begin{equation}
\label{mkdvtransformed1}
{\tilde f} ({\tilde f}_{t} + {\tilde f}_{3x}) - 3 {\tilde f}_x {\tilde f}_{xx} = 0.
\end{equation}
Although homogeneous of second degree and deceptively simple,
it has no solution of the form
${\tilde f} = 1 + \e^{\theta}$
where $\theta = k \, x - \omega \, t + \delta $.
Indeed, the term in $\e^{\theta}$ vanishes for $\omega = k^3$ but
the term $-3 k^3 \e^{2\theta}$ is only zero when $k=0$.
It is clear from (\ref{laurentmkdv}) that to obtain a real-valued solution,
i.e., $u^{\star} = u$, ${\tilde f}$ must be a complex function.
One can readily verify that $u = \pm \tfrac{1}{2} i \, (\ln (f + i g))_x$,
for real functions $f$ and $g$ does not work either.
So, ${\tilde f}$ must be a ratio of complex functions.
Hence,
\begin{equation}
\label{mkdvtf1}
u = \pm \tfrac{1}{2} i \, \left( \ln \left(\frac{f + i\, g}{h + i\, j} \right) \right)_x,
\end{equation}
where $f, g, h,$ and $j$ are real functions.
From $u^{\star} = u$ it follows that $h = f$ and $j = -g$.
Observe that (\ref{orgmkdv}) remains invariant when $u$ is replaced by its negative.
Therefore, without loss of generality, we continue with the plus sign,
\begin{equation}
\label{mkdvtffinal}
u  =
\tfrac{1}{2} i \,\left( \ln \left( \frac{f + i\, g}{f - i\, g} \right) \right)_x
   = \left( \arctan \left( \tfrac{f}{g} \right) \right)_x
   = \frac{f_x g - f g_x}{f^2 + g^2},
\end{equation}
which is Hirota's transformation for the mKdV equation \cite{hirota-jpsjpn-1972}.
Note that the roles of $f$ and $g$ can thus be interchanged in the
computations below.

Goldstein \cite{goldstein-app-2007} gave a different
argument\footnote{
The argument is based on modified singular manifold expansion methods
\cite{estevezetal-jpa-1993,gordoa-estevez-tmp-1994,musette-conte-jpa-1994}.}
to arrive at (\ref{mkdvtffinal}).
Accounting for the $\pm$ signs in (\ref{laurentmkdv}),
he argued that the solution may have two families of
singularities and therefore
assumed\footnote{With
$u = \pm \tfrac{1}{2}\,i\,(\ln (F/G))_x$, (\ref{orgmkdv})
can be replaced by $(D_t + D_x^3) (F \mathbf{\cdot} G) = 0$ and
$D_x^2 (F \mathbf{\cdot} G) = 0$ where $G = F^{\star}$.
See, e.g., \cite{hereman-acm-1992} for explicit expressions
of $F$ and $G$ for the two- and three-soliton cases.}
\begin{equation}
\label{mkdvtfgoldstein}
u = \tfrac{1}{2} i \, \left( \frac{F_x}{F} - \frac{G_x}{G} \right)
  = \tfrac{1}{2} i \, \left( \ln \left( \frac{F}{G} \right) \right)_x.
\end{equation}
Note that the two terms (in the first equality above) indeed account for
the two branches in (\ref{laurentmkdv}).
Setting $F = f + i\, g$ and $G = f - i \, g$ then gives (\ref{mkdvtffinal}).

Applying Hirota's transformation (\ref{mkdvtffinal}) to (\ref{mkdvint}) yields
\begin{eqnarray}
\label{mkdvinfg}
&& f^3 (g_t + g_{3x}) - g^3 (f_t + f_{3x}) - f^2 (f_t g + 3 f_x g_{xx} + 3 f_{xx} g_x  + f_{3x} g )
\nonumber \\
&& + g^2 (f g_t + f g_{3x} + 3 f_x g_{xx} + 3 f_{xx} g_x ) + 6 f g_x (f_x^2 + g_x^2)
\nonumber \\
&& - 6 f_x g (f_x^2 + g_x^2) + 6 f g (f_x f_{xx} - g_x g_{xx}) = 0,
\end{eqnarray}
which is clearly not of the usual form the simplified Hirota method
applies to.
The terms in (\ref{mkdvinfg}) can be regrouped as
\begin{eqnarray}
\label{mKdVinfg2}
&& (f^2 + g^2)(f_t g - f g_t - f g_{3x} + 3 f_x g_{xx} - 3 f_{xx} g_x + f_{3x} g )
\nonumber \\
&& - 6 (f_x g - f g_x) (f f_{xx} - f_x^2 + g g_{xx} - g_x^2) = 0.
\end{eqnarray}
Taking advantage of the fact that there are two free functions
in play,
Hirota \cite{hirota-jpsjpn-1972,hirota-book-miura-1976}
then set the factors multiplying $f^2+g^2$ and $f_x g - f g_x$
separately equal to zero, to get the coupled system
\begin{eqnarray}
\label{mKdVinfgfirst}
&& f (g_t + g_{3x} ) - g (f_t + f_{3x}) - 3 (f_x g_{xx} - f_{xx} g_x) = 0,
\\
\label{mKdVinfgsecond}
&& f f_{xx} - f_x^2 + g g_{xx} - g_x^2 = 0,
\end{eqnarray}
which can be written in bilinear form as
\begin{eqnarray}
\label{mkdvbilinearfirst}
&& (D_t + D_x^3) (f \mathbf{\cdot} g) = 0,
\\
\label{mkdvbilinearsecond}
&& D_x^2 (f \mathbf{\cdot} f + g \mathbf{\cdot} g) = 0.
\end{eqnarray}
Ignoring the bilinear form, one could write (\ref{mKdVinfgfirst})
and (\ref{mKdVinfgsecond}) as
\begin{eqnarray}
\label{eq1opermkdv}
&& f \, {\cal L} g - g \, {\cal L} f + {\cal N}_1 (f,g) = 0,
\\
\label{eq2opermkdv}
&& {\cal N}_2 (f,f) + {\cal N}_2 (g,g) = 0,
\end{eqnarray}
with
\begin{equation}
\label{lopermkdv}
{\cal L} f = f_{t} + f_{3x}
\end{equation}
and
\begin{eqnarray}
\label{nloper1mkdv}
{\cal N}_1 (f, g) &=& - 3 (f_x g_{xx} - f_{xx} g_x),
\\
\label{nloper2mkdv}
{\cal N}_2 (f, g) &=& f g_{xx} - f_x g_x.
\end{eqnarray}
With a suitable adaptation of the method in Section~\ref{simplified-Hirota-method},
one could then seek a solution of (\ref{eq1opermkdv}) and (\ref{eq2opermkdv}) using
\begin{eqnarray}
\label{hirotaexpansionfgforf}
f &=& f^{(0)} + \epsilon f^{(1)} + \epsilon^2 f^{(2)} + \ldots,
\\
\label{hirotaexpansionfgforg}
g &=& g^{(0)} + \epsilon g^{(1)} + \epsilon^2 g^{(2)} + \ldots \, .
\end{eqnarray}
Based on the interchangeability of $f$ and $g$, one can either
take $f^{(0)} = g^{(1)} = 0$ and $g^{(0)} = 1$, or equivalently,
$g^{(0)} = f^{(1)} = 0$ and $f^{(0)} = 1$.
In either case,
${\cal L} {\mathrm e}^{\theta_i}
= {\cal L} {\mathrm e}^{k_i x - \omega_i t + \delta_i}
= 0$
determines the dispersion relation $\omega_i = k_i^3$.
Proceeding with the former case but skipping the details of
the computations we summarize the results.
\vskip 3pt
\noindent
\leftline{\bf One-soliton solution of the mKdV equation}
\vskip 2pt
\noindent
With
\begin{equation}
\label{mkdv1solfg}
f = \e^{\theta} = \e^{k x - k^3 t + \delta} \;\; {\mathrm{and}} \;\; g = 1,
\end{equation}
one gets
\begin{eqnarray}
u & = & \frac{f_x}{1+f^2}
  = k \frac{\e^{\theta}}{1 + \e^{2\theta}}
  = \tfrac{1}{2} k \, \sech\, \theta
\nonumber \\
  & = & \tfrac{1}{2} k \, \sech\, ({k x - k^3 t + \delta})
    = K \sech \left( 2 K x - 8 K^3 t + \delta \right)
\end{eqnarray}
with $K = \frac{k}{2}$.
\vskip 2pt
\noindent
\leftline{\bf Two-soliton solution of the mKdV equation}
\vskip 2pt
\noindent
Now
\begin{eqnarray}
\label{mkdv2solfg}
f &=& \e^{\theta_1} + \e^{\theta_2},
\nonumber \\
g &=& 1 - a_{12} \e^{\theta_1 + \theta_2},
\end{eqnarray}
with $\theta_i = k_i x - k_i^3 t + \delta_i$ and
$a_{12} = \left( \frac{k_1-k_2}{k_1+k_2} \right)^2$.
Then,
\begin{equation}
u = \frac{k_1 \e^{\theta_1} + k_2 \e^{\theta_2}
  + a_{12} \,
    (k_1 \e^{\theta_2} + k_2 \e^{\theta_1})
        \e^{\theta_1 + \theta_2}}{
    1 + \e^{2\theta_1} + \e^{2\theta_2}
  + \displaystyle{\frac{8 k_1 k_2}{(k_1 + k_2)^2}}\, \e^{\theta_1+\theta_2}
  + a_{12}^2 \,
    \e^{2\theta_1+2\theta_2}}.
\end{equation}
\vskip 2pt
\noindent
\leftline{\bf Three-soliton solution of the mKdV equation}
\vskip 2pt
\noindent
After some computations one finds that
\begin{eqnarray}
\label{mkdv3solfg}
f &=& \e^{\theta_1} + \e^{\theta_2} + \e^{\theta_3}
- b_{123} \e^{\theta_1 + \theta_2 + \theta_3},
\nonumber \\
g &=& 1 - a_{12} \, \e^{\theta_1 + \theta_2}
- a_{13} \, \e^{\theta_1 + \theta_3} - a_{23}\,\e^{\theta_2 + \theta_3},
\end{eqnarray}
with $\theta_i = k_i x - k_i^3 t + \delta_i,\,$
$a_{ij} = \left( \frac{k_i-k_j}{k_i+k_j} \right)^2$,
and $b_{123} = a_{12}a_{13}a_{23}$.
\vskip 3pt
\noindent
\leftline{\bf {\it N}-soliton solution of the mKdV equation}
\vskip 3pt
\noindent
A concise
formula \cite{hietarinta-jmp-1987b,ito-jpsjpn-1980,matsuno-jpsjpn-1980}
for the function $\tilde{F} = g + i\,f$
leading to the $N$-soliton solution
$u = \tfrac{1}{2} i \,
\left(\ln \left(\frac{\tilde{F}}{{\tilde{F}}^{\star}} \right) \right)_x$
of the mKdV equation is
given\footnote{Recall that the roles of $f$ and $g$ can be
interchanged because $-u$ solves (\ref{orgmkdv})
whenever $u$ does.
$\tilde{F}$ is $F$ with the roles of $f$ and $g$ reversed.}
by
\begin{equation}
\label{nsolitonmkdv}
\tilde{F} = \sum_{\mu = 0,1}
\e^{ \left[ \sum_{i < j}^{(N)} \mu_{i} \mu_{j} A_{ij}
+ \sum_{j=1}^{N} \mu_{j} (\theta_{j} + i \frac{\pi}{2}) \right]},
\end{equation}
where the summations have the same meaning as in (\ref{nsolitonkdv})
and again $a_{ij} = \e^{A_{ij}}$.
The extra $i \frac{\pi}{2}$ takes care of the complex coefficients
and sign reversals.

The $N$-soliton solution can be written
\cite{hirota-jpsjpn-1972,wadati-jpsjpn-1973,wadati-sawada-jpsjpn-1980b} as
\begin{equation}
\label{mkdvNsoliton}
u(x,t) =
\frac{1}{2i} \left(\ln \frac{{\mathrm{det}}(I + i M)}
{{\mathrm{det}}(I - i M)} \right)_{x},
\end{equation}
where $I$ is the $N \times N$ identity matrix and
\begin{equation}
\label{Mmatrixmkdv}
M_{\ell m} = \frac{\e^{\Theta_\ell + \Theta_m}}{K_\ell + K_m}
\;\;\; {\mathrm{with}} \; \Theta_{\ell} = K_{\ell} x - 4 K_{\ell}^3 t +\Delta_{\ell}.
\end{equation}
Note that $\frac{{\mathrm{det}}(I + i M)}{{\mathrm{det}}(I - i M)}$
matches
$\frac{\tilde{F}}{{\tilde{F}}^{\star}} = \frac{g + i \, f}{g - i\, f}$
with $f$ and $g$ in (\ref{mkdv1solfg}), (\ref{mkdv2solfg}), and (\ref{mkdv3solfg})
when $k_i = 2 K_i$ and $\delta_i = 2 \Delta_i - \ln (2 K_i)$
with $K_i > 0$.
%
\section{Application to Non-solitonic PDEs}
\label{non-solitonic-equations}

\subsection{The Fisher equation with convection}
\label{fisher-equation-convection}
One of the examples discussed in \cite{hayek-amc-2011} is
the Fisher equation with convection term
\cite{mimura-ohara-hmj-1985,murray-book-1977},
\begin{equation}
\label{orgfisherconv}
u_t + \alpha u u_x - u_{xx} - u (1-u) = 0, \quad \alpha \ne 0,
\end{equation}
where $\alpha$ is the convection coefficient.
This equation can also be viewed as a Burgers equation with quadratic source term.
Motivated by a truncated Laurent series, use
\begin{equation}
\label{tffisherconv}
u = -\tfrac{2}{\alpha} (\ln f)_{x}
       = - \tfrac{2}{\alpha} \left(\frac{f_x}{f}\right)
\end{equation}
to replace (\ref{orgfisherconv}) with a homogeneous equation
of second degree
\begin{equation}
\label{fisherconvinf}
f ( f_{3x} + f_x - f_{xt} ) + f_x (f_t - f_{xx} + \tfrac{2}{\alpha} f_{x})
   \equiv f \, {\cal L} f + {\cal N} (f,f) = 0,
\end{equation}
where
\begin{eqnarray}
\label{loperfisherconv}
{\cal L} f &=& f_{3x} + f_x - f_{xt},
\\
\label{noperfisherconv}
{\cal N} (f,g) &=& f_x (g_t - g_{xx} + \tfrac{2}{\alpha} g_{x} ).
\end{eqnarray}
Seeking a solution of type (\ref{hirotaexpansionf}),
${\cal L} f^{(1)} = {\cal L}(\sum_{i=1}^N \e^{\theta_i})$ yields
$\omega_i = - (1 + k_i^2 ).$
The second equation in (\ref{newbilinearscheme}) then becomes
\begin{equation}
\label{fisherconv2ndeq}
{\cal L} f^{(2)} =
  -\sum_{i=1}^{N} k_i \left( 1 + \tfrac{2}{\alpha} k_i \right) \, \e^{2 \theta_i}
  -\sum_{1 \le i< j \le N}
    \left( k_i + k_j + \tfrac{4}{\alpha} k_i k_j \right) \e^{\theta_i + \theta_j}.
\end{equation}
If we were to include the terms $\e^{2 \theta_i}$ in $f^{(2)}$
the perturbation scheme would not terminate.
Hence, we are forced to set all wave numbers equal to
$k_i = -\tfrac{\alpha}{2} \; (i = 1, 2, \ldots, N)$.
Thus, $N=1$ and only a solitary solution can be obtained.
Note that both sums in (\ref{fisherconv2ndeq}) vanish when
$k_i = -\tfrac{\alpha}{2}$.
Hence, $f^{(2)} = 0$ and
\begin{equation}
\label{fisherconvf1sol}
f(x,t)
 = 1 + \e^{\theta}
 = 1 + \e^{-\tfrac{\alpha}{2} x + \tfrac{1}{4} (4+\alpha^2) t + \delta}.
\end{equation}
Finally, from (\ref{tffisherconv})
\begin{equation}
\label{fisherconvu1sol}
u(x,t)
= \frac{\e^{\theta}}{1 + \e^{\theta}}
= \tfrac{1}{2} \left( 1 - \tanh \left[ \tfrac{1}{2}
  \left(\tfrac{\alpha}{2} x -\tfrac{1}{4}(4 + \alpha^2) t
  - \delta \right) \right] \right),
\end{equation}
since $k = -\tfrac{\alpha}{2}$.
The graphs of the kink solution (\ref{fisherconvu1sol})
in 2D and 3D are similar to those in Fig.~\ref{kinksolutionburgerseq}.

\subsection{The Fisher equation}
\label{fisher-equation}
A transformation to homogenize the Fisher equation
\cite{fisher-ae-1937,murray-book-1989} without convection,
\begin{equation}
\label{orgfisher}
u_t - u_{xx} - u (1 - u) = 0,
\end{equation}
is remarkably different from (\ref{tffisherconv}).
Indeed, a truncated Laurent series suggests 
\begin{equation}
\label{tffisher}
u = - 6 (\ln f)_{xx} + \tfrac{6}{5} (\ln f)_t,
\end{equation}
which yields
\begin{eqnarray}
\label{fisherinf}
&& f (f_{4x} + f_{xx} - \tfrac{6}{5} f_{xxt} + \tfrac{1}{5} f_{tt} - \tfrac{1}{5} f_t)
   - 4 f_x f_{3x} + 3 f_{xx}^2 - f_x^2
\nonumber \\
&& - \tfrac{6}{5} f_t f_{xx} + \tfrac{12}{5} f_{x} f_{xt} + \tfrac{1}{25} f_t^2
\equiv f \, {\cal L} f + {\cal N} (f,f) = 0.
\end{eqnarray}
Here,
\begin{eqnarray}
\label{loperfisher}
{\cal L} f \!&=&\!
f_{4x} + f_{xx} - \tfrac{6}{5} f_{xxt} + \tfrac{1}{5} f_{tt}
- \tfrac{1}{5} f_t,
\\
\label{noperfisher}
{\cal N} (f,g) \!&=&\! - 4 f_x g_{3x} + 3 f_{xx} g_{xx} - f_x g_x
- \tfrac{6}{5} f_t g_{xx} + \tfrac{12}{5} f_{xt} g_{x}
+ \tfrac{1}{25} f_t g_t.
\end{eqnarray}
Solving (\ref{newbilinearscheme}) with
$f^{(1)} = \sum_{i=1}^{N} \e^{\theta_i}
         = \sum_{i=1}^{N}  \e^{k_i \, x - \omega_i \, t + \delta_i}$
as a starting point, one gets
$\omega_i = -5 k_i^2$ or $\omega_i = -(1 + k_i^2).$
\vskip 3pt
\noindent
{\bf Case 1:}
For $\omega_i = -5 k_i^2$ the second equation in (\ref{newbilinearscheme}) reads
\begin{equation}
\label{fisher2ndeqchoice1}
{\cal L} f^{(2)}
= \sum_{i=1}^{N} k_i^2 (1 - 6 k_i^2 ) \, \e^{2 \theta_i}
  + 2 \!\!\!\sum_{1 \le i< j \le N} c_{ij} \, \e^{\theta_i + \theta_j},
\end{equation}
where $c_{ij} = k_i k_j \left[ 1 + 2 k_i k_j - 4(k_i^2 + k_j^2) \right]$.
If we put terms $\e^{2 \theta_i}$ in $f^{(2)}$ 
the perturbation scheme does not terminate.
Hence, $k_i = \pm \frac{1}{\sqrt{6}} \; (i = 1, 2, \ldots, N)$ 
which also makes $c_{ij} = 0$.
This leads us to conclude that a multi-soliton solution does not
exist and $N=1$.
With $k = \pm \tfrac{1}{\sqrt{6}}$ we have $\omega = -\tfrac{5}{6}$.
Using (\ref{tffisher}) with $f = 1 + \e^{\theta}$ gives
\begin{equation}
u(x,t) = \frac{\e^{2\theta}}{(1 + \e^{\theta})^2}
       = \frac{1}{(1 + e^{-\theta})^2}
       = \tfrac{1}{4} \left( 1 + \tanh \tfrac{\theta}{2} \right)^2,
\end{equation}
where each of these forms of the solution appears in the literature
(see, e.g., \cite{ablowitz-zeppetella-bmb-1979}).
Explicitly, for $k = -\tfrac{1}{\sqrt{6}}$,
\begin{equation}
\label{fishersol1}
u(x,t) =
\tfrac{1}{4} \left( 1 - \tanh \left[ \tfrac{1}{2}
\left( \tfrac{1}{\sqrt{6}} x - \tfrac{5}{6}t - \delta \right) \right] \right)^2,
\end{equation}
which is a wave traveling to the right.
The graph of this kink solution is the same as in Fig.~\ref{kinksolutionburgerseq}
but with a steeper slope due to the square in (\ref{fishersol1}).
For $k = \tfrac{1}{\sqrt{6}}$,
\begin{equation}
\label{fishersol2}
u(x,t) =
\tfrac{1}{4} \left( 1 + \tanh \left[ \tfrac{1}{2}
\left( \tfrac{1}{\sqrt{6}} x + \tfrac{5}{6}t + \delta \right) \right] \right)^2,
\end{equation}
which is a left-traveling wave, a bit steeper than the one shown in
Fig.~\ref{kinksolutionburgerseq} after a vertical flip.
Note that (\ref{fishersol2}) does not follow from (\ref{fisherconvu1sol})
in the limit for $\alpha \rightarrow 0.$
\vskip 3pt
\noindent
{\bf Case 2:}
For $\omega_i = -(1 + k_i^2)$ the second equation in (\ref{newbilinearscheme}) becomes
\begin{equation}
\label{fisher2ndeqchoice2}
{\cal L} f^{(2)}  =
  -\tfrac{1}{25} \left(
  \sum_{i=1}^{N} (1 + k_i^2) ( 1 + 6 k_i^2 )\, \e^{2 \theta_i}
  + 2 \!\!\! \sum_{1 \le i< j \le N} c_{ij} \, \e^{\theta_i + \theta_j} \right),
\end{equation}
where
$c_{ij} = \left[ 1 + 35 k_i k_j + 46 k_i^2 k_j^2
          - 2(k_i^2 + k_j^2)(7 + 10 k_i k_j) \right].$
So, for real wave numbers $k_i$ the terms $\e^{2 \theta_i}$ do not vanish.
No solitary wave solutions or solitons can be obtained in this case.

\subsection{The FitzHugh-Nagumo equation with convection}
\label{fitzhugh-nagumo-equation-convection}
The FHN equation with convection term \cite{kawahara-tanaka-pla-1983},
\begin{equation}
\label{orgfhnconv}
u_t + \alpha u u_x - u_{xx} + u (1-u) (a-u) = 0,
\end{equation}
where $\alpha$ denotes the convection coefficient and $a$ is an
arbitrary constant, is also called the Burgers-Huxley equation \cite{ozis-aslan-znf-2009}.

A truncated Laurent series suggests two possible transformations,
namely,
\begin{equation}
\label{tffhnconv1}
u = \sqrt{m} \, (\ln f)_{x}
  =  \sqrt{m} \left(\frac{f_x}{f}\right), \quad m > 0.
\end{equation}
and
\begin{equation}
\label{tffhnconv2}
u = -\frac{2}{\sqrt{m}} \, (\ln f)_{x}
  = -\frac{2}{\sqrt{m}} \left(\frac{f_x}{f}\right), \quad m > 0,
\end{equation}
where we have replaced $\alpha$ by $\tfrac{m - 2}{\sqrt{m}}$ in
(\ref{orgfhnconv}) to simplify their forms and the computations below.
Using (\ref{tffhnconv1}), (\ref{orgfhnconv}) transforms into
\begin{eqnarray}
\label{orgfhnconvinf}
& & f (f_{3x} - a f_x - f_{xt}) + f_x \left(f_t - (m+1) f_{xx} +
\sqrt{m} (1+a) f_{x} \right)
\nonumber \\
& & \equiv f\, {\cal L} f + {\cal N} (f,f) = 0,
\end{eqnarray}
where
\begin{eqnarray}
\label{loperfhnconv}
{\cal L} f &=& f_{3x} - a f_x - f_{xt},
\\
\label{noperfhnconv}
{\cal N} (f,g) &=& f_x \left( g_t - (m+1) g_{xx} + \sqrt{m} (1+a) g_{x} \right).
\end{eqnarray}
To compute a single solitary wave solution we take $f = 1 + \e^{\theta}$.
Then, ${\cal L} \e^{\theta} = 0$ yields $\omega = a - k^2$.
Next, ${\cal N} (\e^{\theta}, \e^{\theta}) = 0$ determines $k = \tfrac{1}{\sqrt{m}}$
or $k = \tfrac{a}{\sqrt{m}}.$
Thus, $\omega = \tfrac{a m - 1}{m}$ or $\omega = \tfrac{a (m - a)}{m},$ respectively.
Returning to $u$, we obtain the solitary wave solutions
\begin{equation}
\label{singlesolitarysol1FHN}
u(x,t) =
\tfrac{1}{2} \left(1 + \tanh \left[ \tfrac{1}{2} \left(\frac{1}{\sqrt{m}} x
- \frac{(a m - 1)}{m} t + \delta \right) \right] \right)
\end{equation}
and
\begin{equation}
\label{singlesolitarysol2FHN}
u(x,t) =
\tfrac{1}{2} a \left(1 + \tanh \left[ \tfrac{1}{2} \left(\frac{a}{\sqrt{m}} x
- \frac{a (m - a)}{m} t + \delta \right) \right] \right).
\end{equation}
Although it is impossible to find a two-soliton solution, a so-called
{\it bi-soliton} solution can be computed which describes coalescent wave fronts.
Indeed, taking $ f = 1 + \e^{\theta_1} + \e^{\theta_2},$ with
$\omega_i = a - k_i^2 \; (i = 1, 2)$, after some computations one gets
\begin{equation}
\label{bisolitonFHNeqconv}
u(x,t) = \frac{\e^{\theta_1}
         + a \, \e^{\theta_2}}{1 + \e^{\theta_1} + \e^{\theta_2}},
\end{equation}
where
\begin{equation}
\label{fhnconvthetaset1}
\theta_1 =
   \tfrac{1}{\sqrt{m}} x - \left(\tfrac{a m - 1}{m} \right) t + \delta_1, \;\;
\theta_2 =
   \tfrac{a}{\sqrt{m}} x - \left( \tfrac{a (m - a)}{m} \right) t + \delta_2.
\end{equation}
Since $\alpha = \tfrac{m-2}{\sqrt{m}}$,
possible\footnote{For any positive value of $m$, the pair $(\alpha,m)$
must still satisfy $\alpha = \tfrac{m-2}{\sqrt{m}}$.}
values for $m$ are
\begin{equation}
\label{mvaluesFHNeqconv}
m =
\tfrac{1}{2} \left(4 + \alpha^2 \pm \alpha \sqrt{8 + \alpha^2} \right),
\quad m > 0.
\end{equation}
This solution can be found in \cite{hereman-nuseir-mcs-1997}
and \cite{kawahara-tanaka-pla-1983}
where it was obtained with a different method.

\begin{figure}[htb]
\begin{center}
\begin{tabular}{ccc}
\includegraphics[width=2.1975in,height=1.9715in]{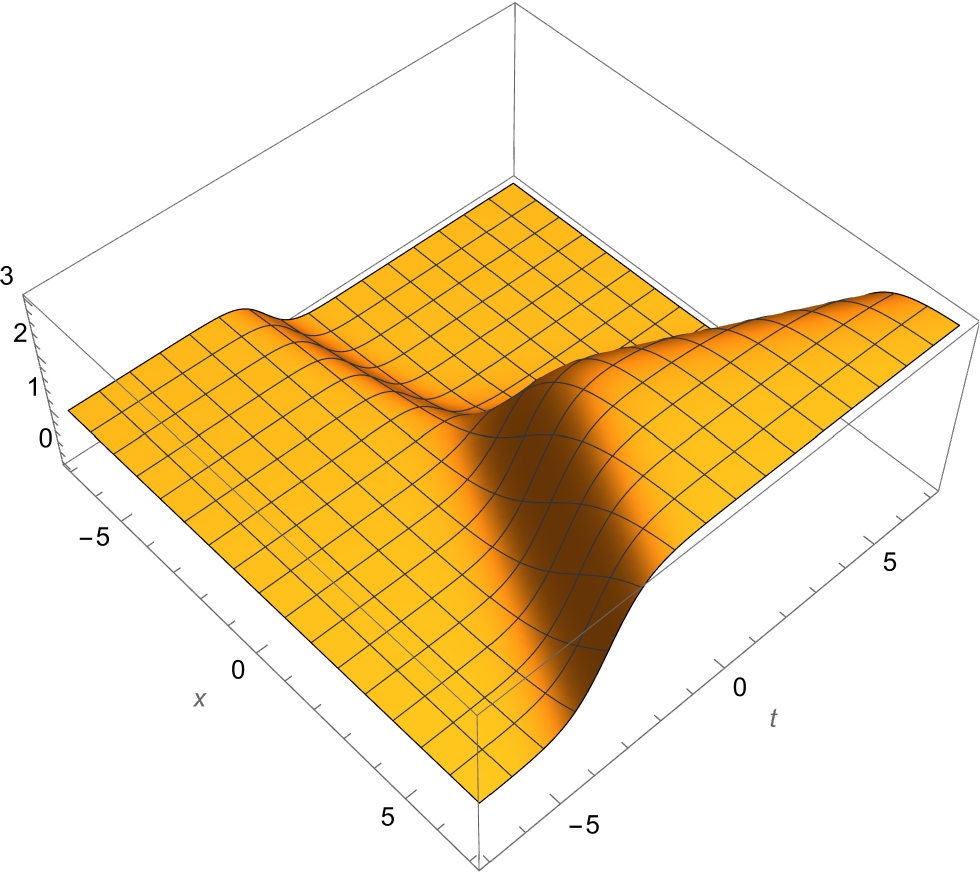}
& \phantom{x} & 
\includegraphics[width=2.1875in,height=1.9625in]{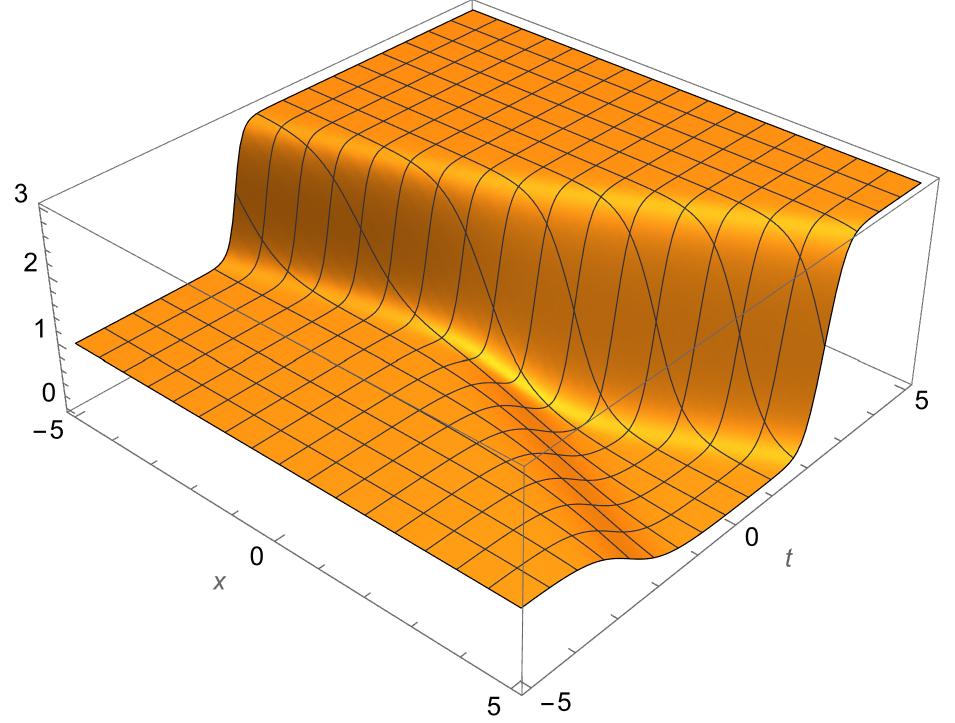}
\end{tabular}
\end{center}
\caption{3D graphs of solution (\ref{bisolitonFHNeqconv}) with
(\ref{fhnconvthetaset1}) (left) and (\ref{fhnconvthetaset2}) (right)
both for $a = 3, \alpha = 1$ (i.e., $m = 4$), and $ \delta_1 = \delta_2 = 0 $.}
\label{Graph-CoalescentWaveFrontsFHNeq}
\end{figure}

Skipping the details, with (\ref{tffhnconv2}) one obtains the following solutions
\begin{equation}
\label{singlesolitarysol1FHNtf2}
u(x,t) =
\tfrac{1}{2} \left(1 - \tanh \left[ \tfrac{1}{2} \left(\frac{\sqrt{m}}{2} x
 + \frac{(4 a - m)}{4} t - \delta \right) \right] \right)
\end{equation}
and
\begin{equation}
\label{singlesolitarysol2FHNtf2}
u(x,t) =
\tfrac{1}{2} a \left(1 - \tanh \left[ \tfrac{1}{2} \left(\frac{a \sqrt{m}}{2} x
 + \frac{a (4 - a m)}{4} t - \delta \right) \right] \right),
\end{equation}
with $m$ given in (\ref{mvaluesFHNeqconv}).
The bi-soliton solution corresponding to (\ref{tffhnconv2}) is
(\ref{bisolitonFHNeqconv}) with
\begin{equation}
\label{fhnconvthetaset2}
\theta_1 =
  -\tfrac{\sqrt{m}}{2} x - \left(\tfrac{4 a - m}{4} \right) t + \delta_1, \;\;
\theta_2 =
  -\tfrac{a\sqrt{m}}{2} x - \left( \tfrac{a (4 - a m)}{4} \right) t + \delta_2.
\end{equation}
Solution (\ref{bisolitonFHNeqconv}) with either (\ref{fhnconvthetaset1}) or
(\ref{fhnconvthetaset2}) describes the coalescence of two wave fronts
pictured in Fig.~\ref{Graph-CoalescentWaveFrontsFHNeq}.
\vskip 0.00001pt
\noindent
\vskip 0.0001pt
\noindent
Finally, for $m=2$ (i.e., $\alpha = 0$), one gets a solitary wave
solution of the FHN equation without convection \cite{aronsonetal-lnm446-1975}.

\subsection{A Burgers equation with a cubic source term}
\label{burgers-equation-cubic-nonlinearity}
\begin{figure}[htb]
\begin{center}
\begin{tabular}{ccc}
\includegraphics[width=2.1875in, height=1.9375in]{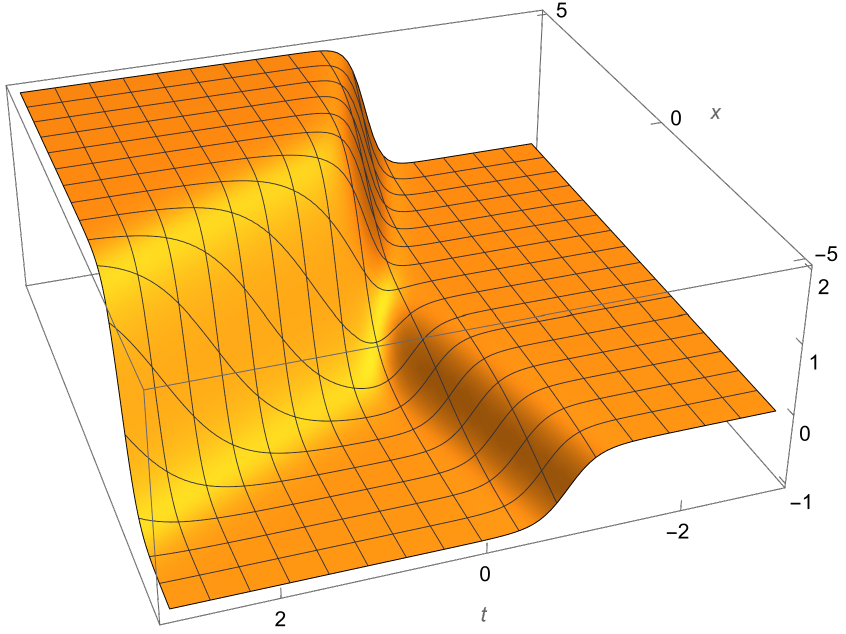}
& \phantom{x} &
\includegraphics[width=2.1875in, height=1.9375in]{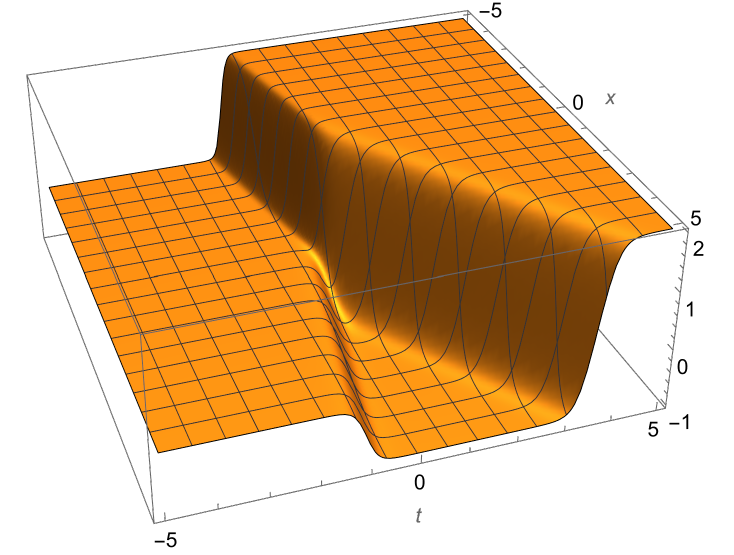}
\end{tabular}
\end{center}
\caption{3D graphs of (\ref{bisolitonBurgerscubicspecialtf1}) (left)
and (\ref{bisolitonBurgerscubicspecialtf2}) (right) for $\delta_1 = - \delta_2 =1$.}
\label{3DgraphsCoalescentWaveFrontsBurgersCubic}
\end{figure}

Consider the Burgers equation with a polynomial source term of third degree,
\begin{equation}
\label{burgerscubic}
u_t + \alpha u u_x - u_{xx} =  3 u (2 - u) (u + 1),
\end{equation}
which is of the kind treated in \cite[Eq.\ (26)]{vladimirov-maczka-repmathphys-2007}.
Eq.\ (\ref{burgerscubic}) can also be viewed as an equation of FitzHugh-Naguma-type 
with convection term\footnote{Except that $u-1$ is now replaced by $u+1$.}.
Such equations are known to have coalescent wave fronts
\cite{hereman-kluwer-1990,kawahara-tanaka-pla-1983}.
Based on a truncated Laurent series,
there are potentially two homogenizing transformations:
\begin{equation}
\label{tfburgerscubic1}
u = \sqrt{m} \,(\ln f)_{x}
  =  \sqrt{m} \left(\frac{f_x}{f}\right), \quad m > 0.
\end{equation}
and
\begin{equation}
\label{tfburgerscubic2}
u = -\frac{2}{3\sqrt{m}} \, (\ln f)_{x}
  = -\frac{2}{3\sqrt{m}} \left(\frac{f_x}{f}\right), \quad m > 0,
\end{equation}
where we used $\alpha = \tfrac{3 m - 2}{\sqrt{m}}$ in (\ref{burgerscubic})
to simplify their forms.
Starting with (\ref{tfburgerscubic1}), substitution into (\ref{burgerscubic})
yields
\begin{eqnarray}
\label{burgerscubicforf}
& & f \left(6 f_x - f_{xt} + f_{3x} \right)
 + f_{x} \left( f_t +3 \sqrt{m} f_x -  \left(1 + 3 m\right) f_{xx} \right)
\\
& & \equiv f \, {\cal L} f + {\cal N} (f,f) = 0.
\end{eqnarray}
Here, ${\cal L}$ and ${\cal N}$ are defined by
\begin{eqnarray}
\label{loperburgerscubic}
{\cal L} f &=& 6 f_x - f_{xt} + f_{3x},
\\
\label{noperburgerscubic}
{\cal N} (f,g) &=&
   f_{x} \left( g_t +3 \sqrt{m} g_x - \left(1 + 3 m \right) g_{xx} \right).
\end{eqnarray}
For the single solitary wave solution, ${\cal L} \e^{\theta} = 0$ yields
$\omega = -(k^2 + 6)$.
Next, ${\cal N} (\e^{\theta}, \e^{\theta}) = 0$ determines $k = -\tfrac{1}{\sqrt{m}}$
or $k = \tfrac{2}{\sqrt{m}}.$
Thus, $\omega = -\tfrac{6 m + 1}{m}$ or $\omega = -\tfrac{2 (3 m + 2)}{m},$ respectively.
So, with $f = 1 + \e^{\theta}$ we obtain the solitary wave solutions
\begin{equation}
\label{solitarysol1burgerscubictf1}
u(x,t) =
 - \tfrac{1}{2} \left( 1 - \tanh \left[ \tfrac{1}{2 \sqrt{m}} x
 - \frac{(6 m + 1)}{2 m} t  - \frac{\delta}{2} \right] \right)
\end{equation}
and
\begin{equation}
\label{solitarysol2burgerscubictf1}
u(x,t) =
1 + \tanh \left[ \frac{1}{\sqrt{m}} x
+ \frac{(3 m + 2)}{m} t + \frac{\delta}{2} \right],
\end{equation}
where, with regard to $\alpha = \tfrac{3m-2}{\sqrt{m}}$, possible
values\footnote{For any positive value of $m$, the pair $(\alpha,m)$
must still satisfy $\alpha = \tfrac{3m-2}{\sqrt{m}}$.}
for $m$ are
\begin{equation}
\label{mvaluesBurgerscubic}
m = \tfrac{1}{18}
\left(12 + \alpha^2 \pm \alpha \sqrt{24 + \alpha^2} \right),
\quad m > 0.
\end{equation}
As with the FHN equation with convection term, no two-soliton solution
exists but a bi-soliton solution can be found which describes
coalescent wave fronts.
Indeed, taking $ f = 1 + \e^{\theta_1} + \e^{\theta_2}$
where $\omega_i = -(k_i^2 + 6) \; (i = 1, 2)$ one gets
\begin{eqnarray}
\label{bisolitonBurgerscubictf1}
u(x,t) &=& \frac{\sqrt{m} (k_1 \e^{\theta_1} + k_2  \e^{\theta_2})}
{1 + \e^{\theta_1} + \e^{\theta_2}}
\nonumber \\
&=& \frac{2 \, \e^{ \frac{2}{\sqrt{m}} x + \frac{2 (3 m + 2)}{m} t + \delta_1}
- \e^{-\frac{1}{\sqrt{m}} x + \frac{(6 m + 1)}{m} t + \delta_2}}
  {1 + \e^{\frac{2}{\sqrt{m}} x + \frac{2 (3 m + 2)}{m} t + \delta_1}
+ \e^{-\frac{1}{\sqrt{m}} x + \frac{(6 m + 1)}{m} t + \delta_2}},
\end{eqnarray}
because $k_1 =\tfrac{2}{\sqrt{m}}$ and $k_2 = -\tfrac{1}{\sqrt{m}}$
with $m$ in (\ref{mvaluesBurgerscubic}).
For $m=1$, a solution of (\ref{burgerscubic}) with $\alpha = 1$ then reads
\begin{equation}
\label{bisolitonBurgerscubicspecialtf1}
u(x,t) = \frac{2 \e^{2 x + 10 t + \delta_1}
- \e^{-x + 7 t + \delta_2}}{1 + \e^{2 x + 10 t + \delta_1} + \e^{-x + 7 t + \delta_2}}.
\end{equation}
The solution procedure using (\ref{tfburgerscubic2}) is similar
and leads to
\begin{equation}
\label{solitarysol1burgerscubictf2}
u(x,t) = - \tfrac{1}{2} \left( 1 + \tanh \left[ \tfrac{3\sqrt{m}}{4} x
 + \frac{3 (3 m + 8)}{8} t  + \frac{\delta}{2} \right] \right)
\end{equation}
and
\begin{equation}
\label{solitarysol2burgerscubictf2}
u(x,t) = 1 - \tanh \left[ \frac{3\sqrt{m}}{2} x
 - \frac{3 (3 m + 2)}{2} t - \frac{\delta}{2} \right],
\end{equation}
with $m$ given in (\ref{mvaluesBurgerscubic}).
The bi-soliton solution corresponding to (\ref{tfburgerscubic2}) reads
\begin{eqnarray}
\label{bisolitonBurgerscubictf2}
u(x,t) &=& -\frac{2}{3\sqrt{m}} \frac{(k_1 \e^{\theta_1} + k_2  \e^{\theta_2})}
{(1 + \e^{\theta_1} + \e^{\theta_2})}
\nonumber \\
&=& - \frac{
  \e^{\frac{3\sqrt{m}}{2} x + \frac{3 (3 m + 8)}{4} t + \delta_1}
  - 2 \, \e^{-3 \sqrt{m} x + 3 (3 m + 2) t + \delta_2}}
  {1 + \e^{\frac{3\sqrt{m}}{2} x + \frac{3 (3 m + 8)}{4} t + \delta_1}
+ \e^{-3 \sqrt{m} x + 3 (3 m + 2) t + \delta_2}},
\end{eqnarray}
because $k_1 = \frac{3\sqrt{m}}{2}$ and $k_2 = -3 \sqrt{m}$
with $m$ in (\ref{mvaluesBurgerscubic}).
For $m=1$, a bi-soliton solution of (\ref{burgerscubic}) with $\alpha = 1$
then becomes
\begin{equation}
\label{bisolitonBurgerscubicspecialtf2}
u(x,t) = - \frac{\e^{\frac{3}{2} x + \frac{33}{4} t + \delta_1}
- 2 \e^{-3 x + 15 t + \delta_2}}
{1 + \e^{\frac{3}{2} x + \frac{33}{4} t + \delta_1} + \e^{-3 x + 15 t + \delta_2}}.
\end{equation}
Solutions (\ref{bisolitonBurgerscubicspecialtf1}) and (\ref{bisolitonBurgerscubicspecialtf2}),
describing two coalescent wave fronts, are shown in
Fig.~\ref{3DgraphsCoalescentWaveFrontsBurgersCubic}.

Returning to $\alpha$ via (\ref{mvaluesBurgerscubic}) also allows one to consider
the case $\alpha = 0$ (i.e., $m = \tfrac{2}{3}$),
leading to solutions of (\ref{burgerscubic}) with a cubic source but without convection.
\vskip 0.00001pt
\noindent

\subsection{A wave equation with cubic source term}
\label{generalized-burgers-equation}
Consider the wave equation,
\begin{equation}
\label{generalizedburgers}
 \tfrac{1}{8} u_{tt} + u_t + u u_x - \tfrac{1}{8} u_{xx} =  u (u - 1) (u + 2),
\end{equation}
which is a special case of an equation investigated in
\cite[Eq.\ (2)]{vladimirov-maczka-repmathphys-2007}.
The Laurent series solution suggests the transformation
\begin{equation}
\label{bothtfgeneralizedburgers}
u = \tfrac{1}{2} \kappa \left[ (\ln f)_{t} - \kappa (\ln f)_{x} \right]
  =  \tfrac{1}{2} \kappa \left( \frac{f_t - \kappa f_x}{f} \right),
\end{equation}
with $\kappa = \pm 1.$
We first consider the case where $\kappa = 1$.
Using
\begin{equation}
\label{tfgeneralizedburgers}
u = \tfrac{1}{2} \left( \frac{f_t - f_x}{f} \right),
\end{equation}
allows one to replace (\ref{generalizedburgers}) by
\begin{eqnarray}
\label{generalizedburgersinf}
& & f \left( 16 f_t + 8 f_{tt} + f_{3t}  - 16 f_x -8 f_{xt}
    - f_{xtt} - f_{xxt} + f_{3x} \right)
\nonumber \\
& & -\left(3 f_t - f_x \right) \left(4 f_t +f_{tt} - 4 f_x
-2 f_{xt} + f_{xx}\right)
\nonumber \\
& & \equiv f \, {\cal L} f + {\cal N} (f,f) = 0,
\end{eqnarray}
with
\begin{eqnarray}
\label{lopergeneralizedburgers}
{\cal L} f &=& 16 f_t + 8 f_{tt} + f_{3t} - 16 f_x
       - 8 f_{xt} -f_{xtt} -f_{xxt} + f_{3x},
\\
\label{nopergeneralizedburgers}
{\cal N} (f,g) &=&
-\left(3 f_t - f_x \right) \left(4 g_t + g_{tt}
- 4 g_x -2 g_{xt} + g_{xx} \right).
\end{eqnarray}
Then, ${\cal L} \e^{\theta} = 0$ yields $\omega = -k,\,
\omega = 4-k, $ or $\omega = 4+k.$
With $f = 1 + \e^{\theta},$ one readily obtains
\begin{equation}
\label{solitarywavegenBurgers}
u(x,t) =
 - \tfrac{1}{2} (\omega+k) \left(\frac{\e^{\theta}}{1 + \e^{\theta}}\right)
 = - \tfrac{1}{4} (\omega+k) \left(1+\tanh \tfrac{\theta}{2} \right).
\end{equation}
Obviously, the choice $\omega = -k$ must be rejected.
For $\omega = 4-k,$ one finds that
${\cal N} (\e^{\theta}, \e^{\theta}) \equiv 0$
and
\begin{equation}
\label{solitarywavegenBurgerscase1}
u(x,t) = - \left( 1 + \tanh \left[ \tfrac{1}{2}
\left( k x - (4-k) t + \delta \right) \right] \right)
\end{equation}
with arbitrary $k$ and $\delta$.
For $\omega = 4+k$, ${\cal N} (\e^{\theta}, \e^{\theta}) = 0$
determines $k = -2$ or $k =  -3$ resulting in $\omega = 2$ or $\omega = 1,$ 
respectively.
The case $k = -2$ (i.e., $\omega = 2$) is rejected for it leads to
$u(x, t) \equiv 0.$
For $k = -3$ (i.e., $\omega = 1$) one gets
\begin{equation}
\label{solitarywavegenBurgerscase2}
u(x,t) =
\tfrac{1}{2} \Big( 1 - \tanh \left[ \tfrac{1}{2} \left(3 x + t - \delta \right) \right] \Big),
\end{equation}
which is different from (\ref{solitarywavegenBurgerscase1})
when $k = -3$.

Attempting to find a solution of type (\ref{hirotaexpansionf}),
${\cal L} f^{(1)} = {\cal L}(\sum_{i=1}^N \e^{\theta_i})$ determines
$\omega_i  = -k_i, 4 - k_i,$ or $4 + k_i.$
As with the solitary wave solution, to avoid trivial solutions we continue
with $\omega_i = 4 - k_i$ and $\omega_i  = 4 + k_i$.

\begin{figure}[htb]
\begin{center}
\includegraphics[width=2.45in, height=2.198in]
{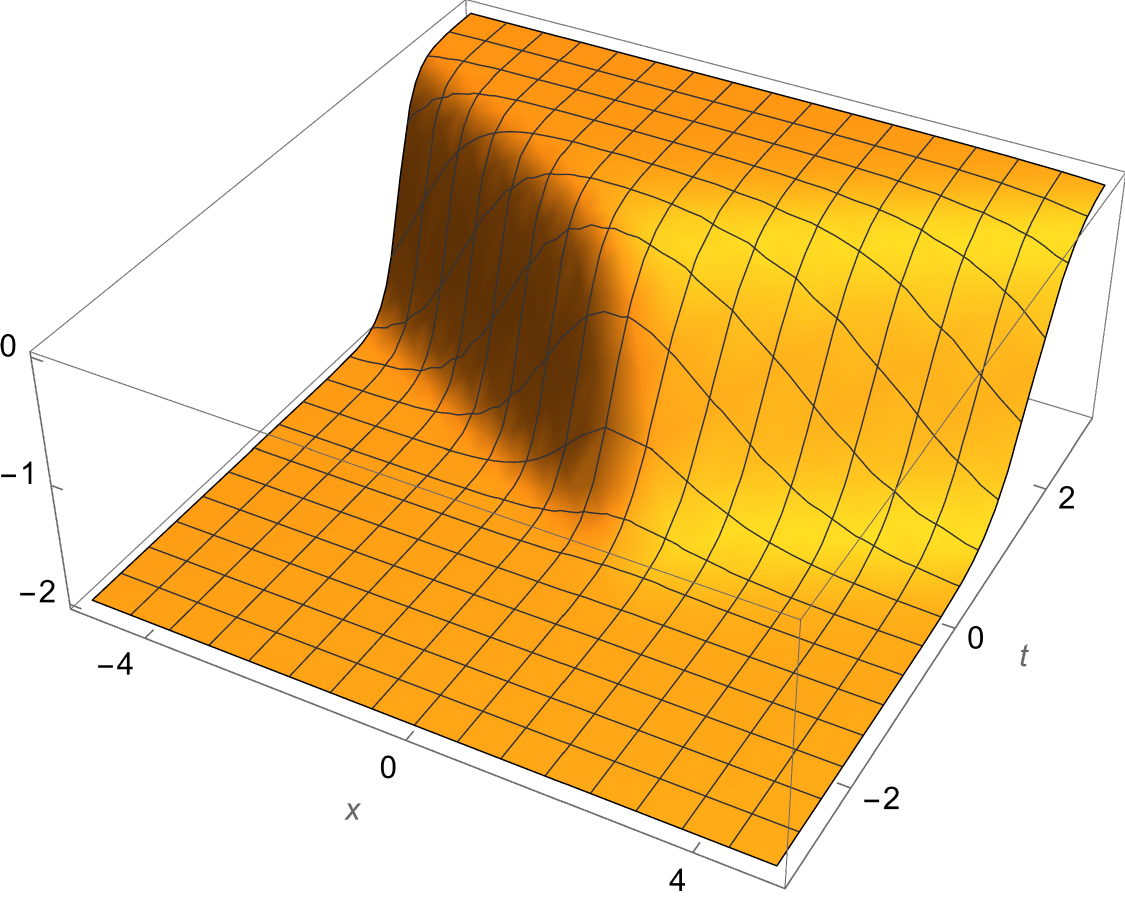}
\end{center}
\caption{3D graph of (\ref{bisolitongeneralizedBurgers})
with $k_1 = 1, k_2 = -2$, and $\delta_1 = \delta_2 = 0$.}
\label{Graph-CoalescentWaveFrontsGeneralizedBurgers}
\end{figure}

Here again, it is impossible to find a two-soliton solution
but a bi-soliton solution can be computed.
Indeed, taking
$f = 1 + \e^{\theta_1} + \e^{\theta_2} + a_{12} \e^{\theta_1 + \theta_2},$
leads to $a_{12} = 0.$
Then, for $\omega_i = 4 - k_i \; (i = 1, 2)$, after some computation one gets
\begin{equation}
\label{bisolitongeneralizedBurgers}
u(x,t) =
-2 \left(\frac{\e^{\theta_1} + \e^{\theta_2}}{1 + \e^{\theta_1} + \e^{\theta_2}}\right),
\end{equation}
where
$\theta_1 = k_1 x - (4-k_1) t + \delta_1$
and
$\theta_2 = k_2 x - (4-k_2) t + \delta_2$.
Solution (\ref{bisolitongeneralizedBurgers}) agrees with the result in
\cite[Eq.\ (37)]{vladimirov-maczka-repmathphys-2007}.
As shown in Fig.~\ref{Graph-CoalescentWaveFrontsGeneralizedBurgers},
(\ref{bisolitongeneralizedBurgers}) describes two coalescent wave fronts.
For  $\omega_i  = 4 + k_i,$ after some computations one gets
$k_1 = -2, k_2 = -3$,
and $a_{12} = 1$, resulting in (\ref{solitarywavegenBurgerscase2})
with $\delta$ replaced by $\delta_2$.

The computations for $\kappa = -1$ in (\ref{bothtfgeneralizedburgers})
are similar but only lead to
\begin{equation}
\label{solitarywavegenBurgerscasebranch2}
u(x,t) =
\tfrac{1}{2} \Big( 1 + \tanh \left[ \tfrac{1}{2} \left(x - 3 t + \delta \right) \right] \Big)
\end{equation}
which does not follow from (\ref{solitarywavegenBurgerscase1})
when $k = 1$.
\vskip 0.0001pt
\noindent

\subsection{A combined KdV-Burgers equation}
\label{kdv-burgers-equation}
\begin{figure}[htb]
\begin{center}
\begin{tabular}{ccc}
\includegraphics[width=2.31in,height=1.425in]
{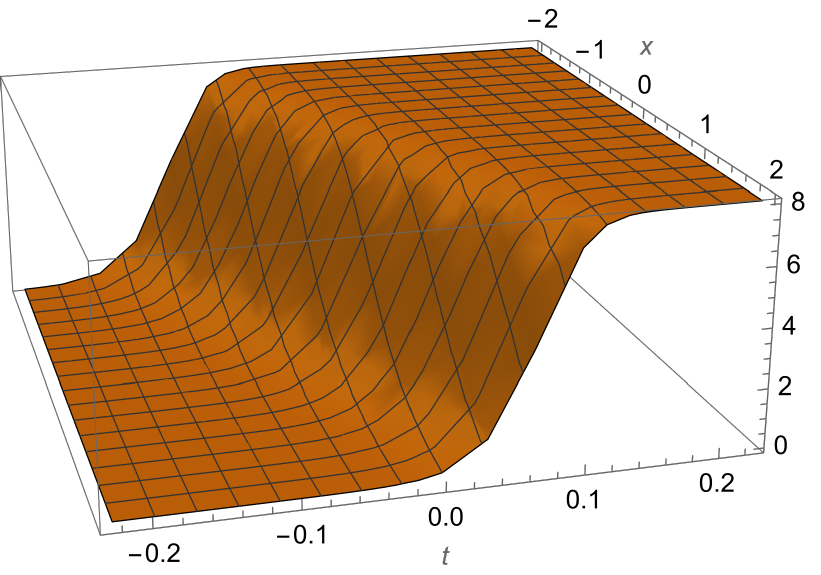}
& \phantom{x} &
\includegraphics[width=2.31in,height=1.425in]
{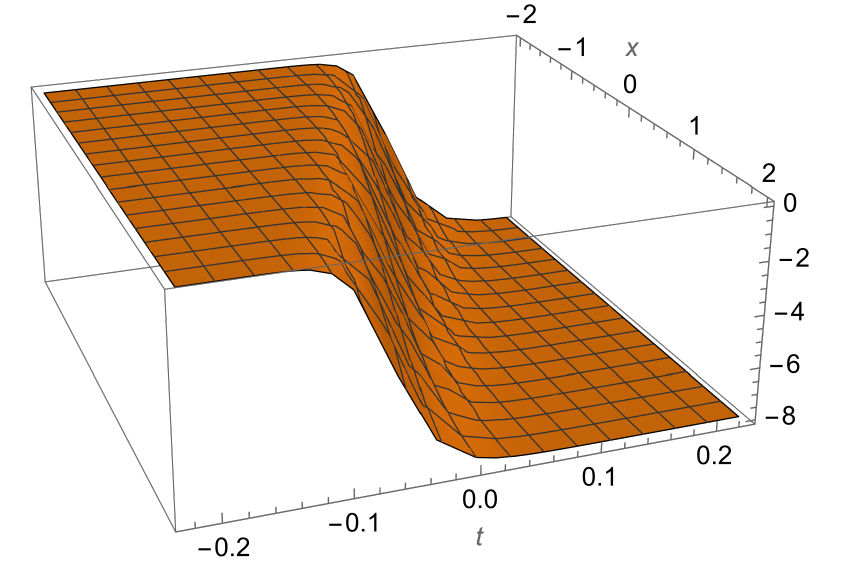}
\end{tabular}
\end{center}
\caption{3D graphs of (\ref{kinkkdvburgerscase1}) (left) and 
(\ref{kinkkdvburgerscase2}) (right) for $\beta=2$ and $\delta=0$.}
\label{kinksolutionkdvburgerseqbothcases}
\end{figure}

A combined KdV-Burgers equation \cite{su-gardner-jmp-1969},
\begin{equation}
\label{orgkdvburgerseq}
u_t + 6 u u_x + u_{3x} - 5 \beta u_{xx} = 0,
\end{equation}
where $\beta > 0$, is used in models where both dispersive and dissipative
effects are relevant.
A Laurent series of the solution of (\ref{orgkdvburgerseq})
suggests the transformation
\begin{equation}
\label{tfkdvburgerseq}
u = 2 (\ln f)_{xx} - 2 \beta (\ln f)_x,
\end{equation}
which we substitute into the integrated KdV-Burgers equation,
\begin{equation}
\label{intkdvburgerseq}
\partial_t \left( \int^x u \, dx \right) + 3 u^2 + u_{xx} - 5 \beta u_x = 0
\end{equation}
to get
\begin{eqnarray}
\label{intkdvburgerseqinf}
&& f ( f_{xt} - \beta f_t + 5 \beta^2 f_{xx} - 6 \beta f_{3x} + f_{4x} )
\nonumber \\
&& - f_x f_t + \beta^2 f_{x}^2 + 6 \beta f_x f_{xx}
+ 3 f_{xx}^2 - 4 f_x f_{3x} = 0.
\end{eqnarray}
This homogeneous equation is of the form $f \, {\cal L} f + {\cal N} (f,f) = 0.$
Therefore, we can proceed as in the KdV case and solve (\ref{newbilinearscheme})
step-by-step with
\begin{eqnarray}
\label{loperkdvburgerseq}
{\cal L} f &=& f_{xt} - \beta f_t + 5 \beta^2 f_{xx} - 6 \beta f_{3x}
+ f_{4x},
\\
\label{noperkdvburgerseq}
{\cal N} (f,g) &=& - f_x g_t + \beta^2 f_x g_x + 6 \beta f_x g_{xx}
 + 3 f_{xx} g_{xx} - 4 f_x g_{3x}.
\end{eqnarray}
We summarize the results.
First, ${\cal L} \e^\theta = {\cal L} \e^{k x - \omega t + \delta} = 0$
yields $(\beta - k)(\omega - k^3 + 5 \beta k^2) =0 $.
Thus, two cases have to be considered.
\vskip 3pt
\noindent
{\bf Case 1}: When $\omega = k^2 (k - 5 \beta)$ and $k \ne \beta$,
${\cal N} (\e^{\theta}, \e^{\theta}) = 0$ determines $k = -\beta$. So, $\omega = -6 \beta^3$.
Inserting $f = 1 + \e^{\theta}$ into (\ref{tfkdvburgerseq}) yields
\begin{equation}
\label{kinkkdvburgerscase1}
u(x,t)
  = 2 \beta^2 \left( \frac{\e^{\theta} (2 + \e^{\theta})}{ (1 + \e^{\theta})^2 } \right)
  = \tfrac{1}{2} \beta^2 \left( 3 -\tanh \tfrac{\theta}{2} \right)
    \left(  1 + \tanh \tfrac{\theta}{2} \right),
\end{equation}
with $\theta = - \beta x + 6 \beta^3 t + \delta$.
\vskip 3pt
\noindent
{\bf Case 2}: When $k =\beta$,
${\cal N} (\e^{\theta}, \e^{\theta}) = 0$ determines $\omega = -6 \beta^3$,
yielding
\begin{equation}
\label{kinkkdvburgerscase2}
u(x,t)
 = - 2 \beta^2 \left( \frac{\e^{2 \theta}}{(1 + \e^{\theta})^2} \right)
 = - \tfrac{1}{2} \beta^2 \left( 1 + \tanh \tfrac{\theta}{2} \right)^2,
\end{equation}
with $\theta = \beta x + 6 \beta^3 t + \delta$.
Solutions (\ref{kinkkdvburgerscase1}) and (\ref{kinkkdvburgerscase2}) are shown in
Fig.~\ref{kinksolutionkdvburgerseqbothcases} for $\beta = 2$ and $\delta = 0$.
\vskip 0.00001pt
\noindent
\vskip 0.00001pt
\noindent
An attempt to find a multi-soliton or bi-soliton solutions based
on (\ref{hirotaexpansionf}) failed.
Assuming $k_i \ne \beta$ (discussed in Case 2) and working with (\ref{hirotaexpansionf}),
${\cal L} f^{(1)} = {\cal L}(\sum_{i=1}^N \e^{\theta_i})$ determines
$\omega_i = k_i^2 (k_i - 5 \beta)$.
The second equation in (\ref{newbilinearscheme}) then becomes
\begin{equation}
\label{kdvburgers2ndeq}
{\cal L} f^{(2)}
= - \beta \sum_{i=1}^{N} k_i^2 (k_i + \beta) \, \e^{2 \theta_i}
  - \!\!\!\sum_{1 \le i< j \le N} c_{ij} \, \e^{\theta_i + \theta_j},
\end{equation}
where $c_{ij} =
k_i k_j \left[ 2 \beta^2 + \beta (k_i + k_j) + 6 k_i k_j -3(k_i^2 + k_j^2)\right].$
Putting terms $\e^{2 \theta_i}$ in $f^{(2)}$ prevents the 
perturbation scheme from terminating.
Hence, $k_i = - \beta \; (i = 1, 2, \ldots, N)$ which also makes $c_{ij} = 0$.
But if the wave numbers have to be equal then $N=1$ and that brings
us back to Case 1 and (\ref{kinkkdvburgerscase1}).

\subsection{An equation due to Calogero}
\label{calogero-equation}
\begin{figure}[htb]
\begin{center}
\begin{tabular}{ccc}
\includegraphics[width=2.31in,height=1.43in]{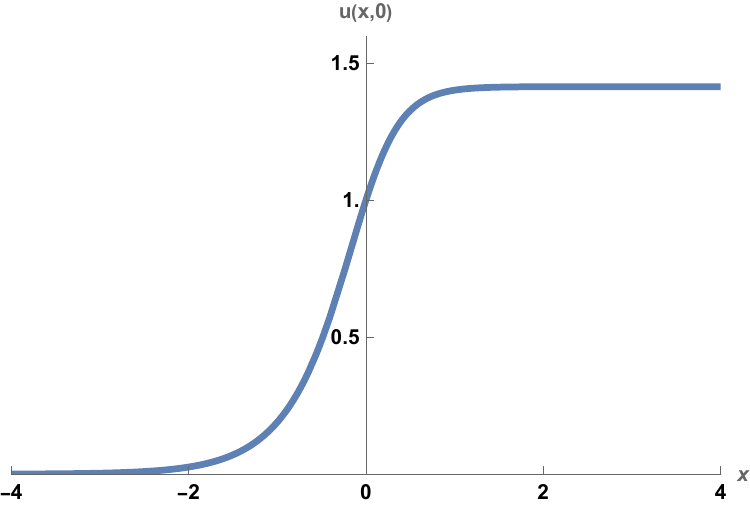}
& \phantom{x} &
\includegraphics[width=2.31in,height=1.54in]
{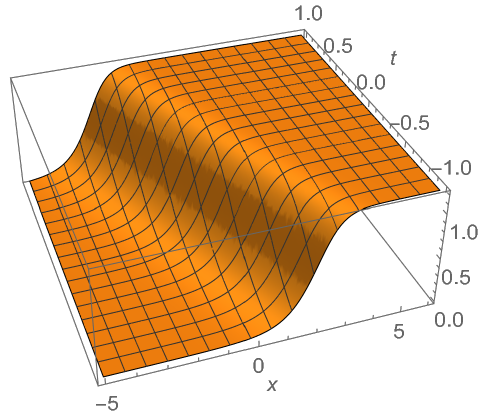}
\end{tabular}
\end{center}
\caption{2D and 3D graphs of (\ref{solitarywavecalogeroeq}) for $k=4$ and $\delta=0$.}
\label{graphstanhsolcalogeroeq}
\end{figure}

For the equation
\begin{equation}
\label{orgcalogeroeq}
u_t  - 3 (3 u u_x^2 + u^4 u_x + u^2 u_{xx}) - u_{3x} = 0
\end{equation}
due to Calogero \cite{calogero-jmp-1987},
the Laurent series (\ref{laurent}) has $\alpha = -\tfrac{1}{2}$.
Therefore, to apply the simplified Hirota method we first change
the dependent variable.
Setting $u = \sqrt{v}$ with $v > 0$ gives
\begin{equation}
\label{changedcalogeroeq}
4 v^2 v_t - 3 v_x^3 - 12 v^2 v_x^2 - 12 v^4 v_x + 6 v v_x v_{xx}
  - 12 v^3 v_{xx}  - 4 v^2 v_{3x} = 0,
\end{equation}
which looks more complicated than (\ref{orgcalogeroeq}) but has a
truncated Laurent series with $\alpha = -1$.
Then, with the transformation
\begin{equation}
\label{tfchangedcalogeroeq}
v = \tfrac{1}{2} (\ln f)_x
  = \tfrac{1}{2} \left( \frac{f_x}{f} \right)
\end{equation}
(\ref{changedcalogeroeq}) can be replaced by an equation of fourth degree,
\begin{eqnarray}
\label{calogeroeqinf}
&& f ( 4 f_x^2 f_{xt} - 3 f_{xx}^3 + 6 f_x f_{xx} f_{3x} - 4 f_x^2 f_{4x} )
- f_x^2 ( 4 f_x f_t + 3 f_{xx}^2 - 4 f_x f_{3x} )
\nonumber \\
&& \equiv f \, {\cal N}_1 (f,f,f) + {\cal N}_2 (f,f,f,f) = 0,
\end{eqnarray}
with
\begin{eqnarray}
\label{nonloper1calogeroeq}
{\cal N}_1 (f,g,h) &=& 4 f_x g_x h_{xt} - 3 f_{xx} g_{xx} h_{xx}
      + 6 f_x g_{xx} h_{3x} - 4 f_x g_x h_{4x},
\\
\label{nonloper2calogeroeq}
{\cal N}_2 (f,g,h,j) &=&
     - f_x g_x ( 4 h_x j_t + 3 h_{xx} j_{xx} - 4 h_x j_{3x} ).
\end{eqnarray}
If one seeks a solution to (\ref{calogeroeqinf}) of type (\ref{hirotaexpansionf}),
then ${\cal N}_1 ( \e^\theta, \e^\theta, \e^\theta)$ with
$ \theta = k x - \omega t + \delta$ yields $\omega = -\tfrac{1}{4} k^3$.
Fortuitously, if the dispersion law holds then
${\cal N}_2 ( \e^\theta, \e^\theta, \e^\theta, \e^\theta) = 0$
and, therefore, $f = 1 + \e^{\theta}$ solves (\ref{calogeroeqinf}).
Using (\ref{tfchangedcalogeroeq}) and $u = \sqrt{v}$,
after some algebra one gets
\begin{equation}
\label{solitarywavecalogeroeq}
u =  \tfrac{1}{2} \sqrt{k} \,
 \sqrt{1+\tanh\left[ \tfrac{1}{8}
 \left(4 k x + k^3 t + 4 \delta \right) \right]},
\end{equation}
where $k > 0$.
This solution was computed in \cite{hereman-nuseir-mcs-1997}
with a different method.
It is graphed in Fig.~\ref{graphstanhsolcalogeroeq} for $k = 4$ and $\delta = 0$.

If one tries to find a multi-soliton solution with
$f^{(1)} = \sum_{i=1}^N \e^{\theta_i} $ with
$\theta_i = k_i x + \tfrac{1}{4} k_i^3 t + \delta_i,$ then
${\cal N}_1 (f^{(1)},f^{(1)},f^{(1)})$ only vanishes if the wave numbers are equal.
\vskip 0.00001pt
\noindent
%
\section{An Equation with Two but not Three Solitons}
\label{two-but-not-three-solitons}
Equations that have two-soliton but not three-soliton solutions have been discovered.
The best known example is the sine-Gordon equation in two space variables
which already appears in early work by Hirota \cite{hirota-jpsjpn-1973}
and was later studied in greater generality in \cite{kobayashi-izutsu-jpsjpn-1976}.
Another example is a $(3+1)$-dimensional eight-order equation due to
Kac-Wakimoto \cite{pekcan-arxiv-2016,Wang-etal-physcr-2020}.

With respect to equations in $(1+1)$ dimensions,
Hietarinta (see \cite{hietarinta-manchester-1989,hietarinta-kluwer-1990}
and references therein) did an extensive search of bilinear forms for which
the necessary condition to have a three-soliton solution is violated.
Although the appropriate bilinear forms are given explicitly,
the equations in the original field variable $u$ are not always
available in his papers.
Reversing the process, i.e., finding the nonlinear PDE (or a system thereof)
that leads to a (known) bilinear form is not straightforward.
Consult, e.g., \cite{hietarinta-manchester-1989,hietarinta-kluwer-1990,hietarinta-lnp767-2009,ma-pdeam-2022} for strategies and explicit examples.

Taking a different example, we study the soliton solutions of
a polynomial equation in $(1+1)$ dimensions,
\begin{equation}
\label{7thordereqIlinetal}
u_t + \tfrac{15}{784} u^3 u_x + \tfrac{15}{28} u u_x u_{xx}
+ \tfrac{15}{56} u^2 u_{3x}
+ \tfrac{5}{2} u_{xx} u_{3x} + u_x u_{4x} + u u_{5x} + u_{7x} = 0,
\end{equation}
which appears in \cite[Eq.\ (19) for $K=56$]{ilinetal-csf-2015}.
The authors claim that this equation has at most a two-soliton solution.
However, they do not give the constraint on the wave numbers $k_i$
that prevents the existence of, e.g., a three-soliton solution.
We therefore investigate (\ref{7thordereqIlinetal}) in more detail.

Based on a truncated Laurent series, we substitute
\begin{equation}
\label{tf7thordereqIlinetal}
u = 56 \,(\ln f)_{xx}
  = 56 \left(\frac{f f_{xx} - {f_x}^2}{f^2}\right)
\end{equation}
into the integrated form of (\ref{7thordereqIlinetal}),
\begin{equation}
\label{int7thordereqIlinetal}
\partial_t \left( \int^x u \, dx \right)
 + \tfrac{15}{3136} u^4 + \tfrac{15}{56} u^2 u_{xx}
 + \tfrac{5}{4} u_{xx}^2 + u u_{4x} + u_{6x} = 0,
\end{equation}
yielding
\begin{equation}
\label{7thordereqinf}
f \left(f_{xt} + f_{8x} \right)
  -f_x f_t +35 {f_{4x}}^2 -56 f_{3x} f_{5x} +28 f_{xx} f_{6x}
  -8 f_x f_{7x} = 0,
\end{equation}
which is of the form $f {\cal L} f + {\cal N} (f, f) = 0$ with
\begin{eqnarray}
\label{loper7thordereq}
{\cal L} f &=& f_{xt} + f_{8x},
\\
\label{noper7thordereq}
{\cal N} (f,g) &=&
 -f_x g_t +35 f_{4x} g_{4x} -56 f_{3x} g_{5x} +28 f_{xx} g_{6x} -8 f_x g_{7x}.
\end{eqnarray}
As usual, ${\cal L} \e^{\theta} = 0$ yields the dispersion relation
$\omega = k^7.$
So, with $f = 1 + \e^{\theta}$ we obtain the solitary wave solution
\begin{equation}
\label{solitarysol1for7thordereqIlinetal}
u(x,t)
   =  14 k^2 \sech^2 \left[ \tfrac{1}{2} \left(k x - k^7 t + \delta \right) \right].
\end{equation}
Seeking a solution of the form (\ref{hirotaexpansionf}),
as before
${\cal L} f^{(1)} = {\cal L}(\sum_{i=1}^N \e^{\theta_i}) = 0 $ with
$\theta_i = k_i x - \omega_i t + \delta_i$ yields $\omega_i  = {k_i}^7$.

For the two-soliton solution, taking
$ f = 1 + \e^{\theta_1} + \e^{\theta_2} + a_{12} \e^{\theta_1 + \theta_2}$,
after some computations one gets
\begin{equation}
\label{a12for7thordereqIlinetal}
a_{12} =  \left( \frac{(k_1-k_2)
\left({k_1}^2-k_1 k_2 + {k_2}^2\right)}
{(k_1+k_2) \left({k_1}^2+k_1 k_2 + {k_2}^2\right)} \right)^2
\end{equation}
and, then from (\ref{tf7thordereqIlinetal}),
\begin{eqnarray}
\label{soliton2for7thordereqIlinetal}
u(x, t) &=&  56 \left(
\frac{{k_1}^2 \e^{\theta_1} + {k_2}^2 \e^{\theta_2} + a_{12}
\e^{\theta_1 + \theta_2} (k_1 + k_2)^2}
{1+ \e^{\theta_1} + \e^{\theta_2} + a_{12} \e^{\theta_1 + \theta_2}} \right.
\nonumber \\
& & \left. - \frac{\left(k_1 \e^{\theta_1} + k_2 \e^{\theta_2}
+ a_{12}  \e^{\theta_1 + \theta_2} (k_1 + k_2) \right)^2}
{(1+ \e^{\theta_1} + \e^{\theta_2} + a_{12} \e^{\theta_1 + \theta_2})^2} \right)
\end{eqnarray}
with $\theta_i = k_i x - {k_i}^7 t + \delta_i$.

\begin{figure}[htb]
\begin{center}
\begin{tabular}{ccc}
\includegraphics[width=1.52in, height=1.76in]{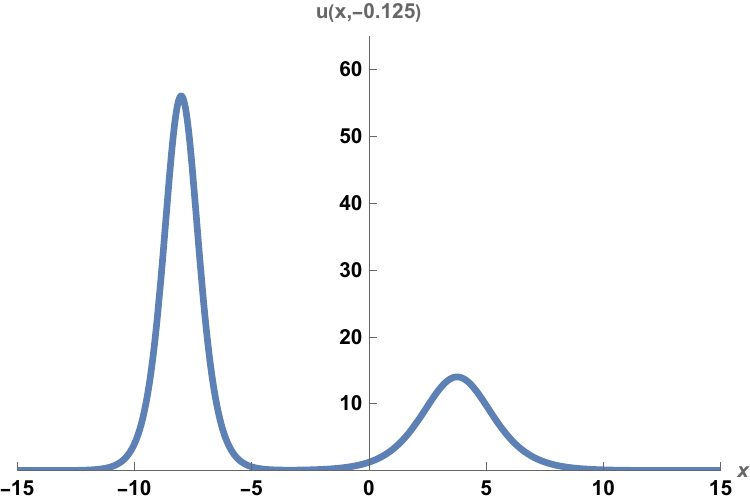} &
\includegraphics[width=1.52in, height=1.76in]{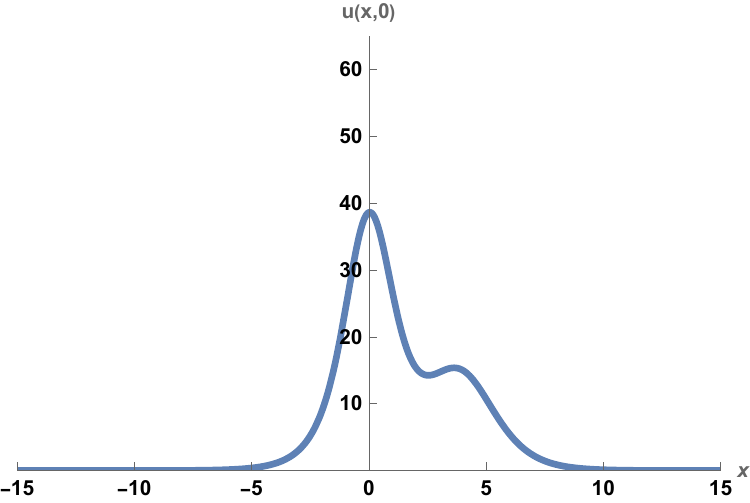} &
\includegraphics[width=1.52in, height=1.76in]{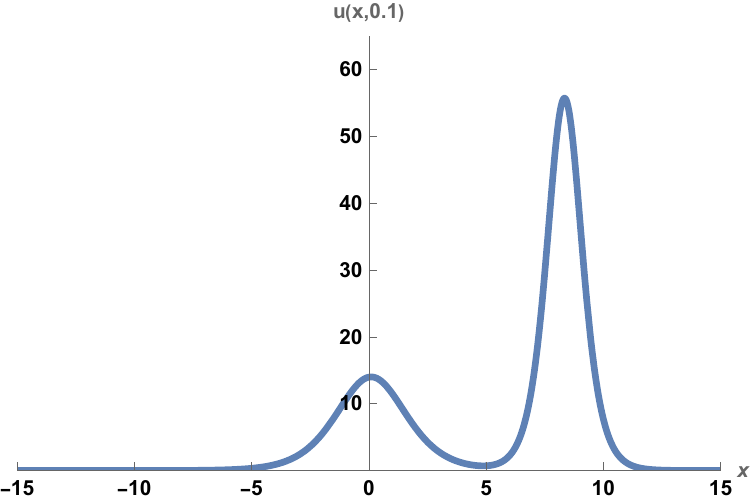}
\end{tabular}
\end{center}
\caption{
Graph of the two-soliton solution (\ref{soliton2for7thordereqIlinetal}) of 
(\ref{7thordereqIlinetal}) at three different moments in time.}
\label{7thorderIlin-two-solitons-in-time}
\end{figure}

\begin{figure}[htb!]
\begin{center}
\includegraphics[width=3.20in, height=2.29in]{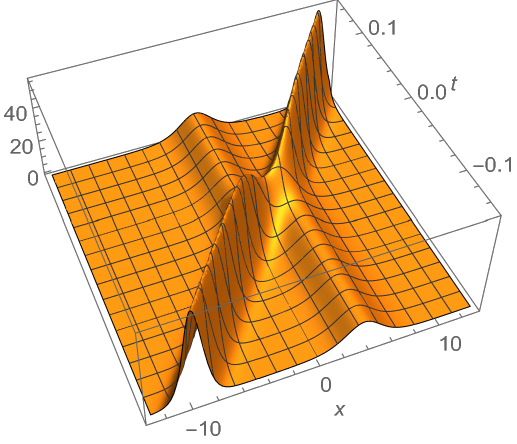}
\end{center}
\caption{Bird's eye view of the collision of two solitons for equation (\ref{7thordereqIlinetal}).
Notice the phase shift after the collision.}
\label{7thorderIlin-two-solitons-in-3D}
\end{figure}

The collision of two solitons for equation (\ref{7thordereqIlinetal})
is shown in Figs.~\ref{7thorderIlin-two-solitons-in-time}
and~\ref{7thorderIlin-two-solitons-in-3D} for $k_1 = 1,\, k_2 = 2,$ and 
$\delta_1 = \delta_2 = 0$.
\vskip 0.000001pt
\noindent
\noindent
 
The existence of a two-soliton solution comes as no surprise because (\ref{7thordereqinf})
can be written in bilinear form as
\begin{equation}
\label{bilinearfor7thordereqIlinetal}
\left( D_x D_t + D_x^8 \right) (f \mathbf{\cdot} f) = 0,
\end{equation}
which satisfies the
conditions\footnote{For a derivation of such conditions see,
e.g., \cite{matsuno-book-1984}.}
for the existence of a two-soliton solution
(see, for example, \cite[Eq.\ (22)]{hietarinta-jmp-1987a} and
\cite[Eq.\ (5.47)]{hirota-solitons-1980}).

In an attempt to find a three-soliton solution, one would take
\begin{equation}
\label{soliton3for7thordereqIlinetal}
 f =
 1 + \e^{\theta_1} + \e^{\theta_2} + \e^{\theta_3} + a_{12} \e^{\theta_1 + \theta_2} + a_{13} \e^{\theta_1 + \theta_3}
 + a_{23} \e^{\theta_2 + \theta_3} + b_{123} \e^{\theta_1 + \theta_2 + \theta_3}
\end{equation}
with
\begin{equation}
a_{ij} =  \left( \frac{(k_i-k_j) \left({k_i}^2-k_i k_j + {k_j}^2\right)}
{(k_i+k_j) \left({k_i}^2+k_i k_j + {k_j}^2\right)} \right)^2
\end{equation}
and $b_{123} = a_{12} a_{13} a_{23}$ and substitute it into (\ref{7thordereqinf}).
A lengthy computation shows that the equation is only satisfied if
the wave numbers are equal or zero.
Actually, this agrees with Hietarinta's earlier studies of equations
that have a bilinear form.
Indeed, for an $N$-soliton solution to exist, the condition \cite{hirota-solitons-1980,matsuno-book-1984}
\begin{equation}
\label{nsolitoncondition}
S[P,n] \!=\!
 \sum_{\sigma=\pm 1} \! P \left(\sum_{i=1}^{n}\sigma_{i}k_i,
 - \sum_{i=1}^{n}\sigma_{i}\omega_i \!\right)
   \!\prod_{i < j}^{(n)} P(\sigma_{i}k_i-\sigma_{j}k_j,
 - \sigma_{i}\omega_i+\sigma_{j}\omega_j)\sigma_{i}\sigma_{j} = 0
\end{equation}
must hold for $n = 2, 3, \ldots, N$.
In (\ref{nsolitoncondition}), 
$P$ is the polynomial corresponding to the bilinear operator $B$, 
$\sum_{\sigma = \pm 1}$ indicates the
summation over all possible combinations of
$\sigma_1 = \pm 1, \; \sigma_2 = \pm 1,\; \ldots, \; \sigma_n = \pm 1$
and $\prod_{i<j}^{(n)}$ means the product of all possible combinations of
the $n$ elements with $i < j,$
and all $k_i, \omega_i$ subject to the dispersion law $\omega_i(k_i).$
Note that (\ref{nsolitoncondition}) is a condition for $P$ and not for the $k_i$.
Also, all $\omega_i$ are replaced in terms of the $k_i$ because
(\ref{nsolitoncondition}) is evaluated on the dispersion law.

For (\ref{bilinearfor7thordereqIlinetal}), 
$P(D_x, D_t) = B = D_x D_t + D_x^8 $
and the three-soliton condition $S[P,3] = 0$
(see, \cite[Eq.\ (28)]{hietarinta-jmp-1987a})
gives\footnote{Based on symmetry considerations,
simplified expressions of (\ref{nsolitoncondition})
are given in \cite[Eq.\ (2.9)]{ma-ijnsns-2021}
and \cite[Eqs.\ (4.3) and (4.4)]{newell-yunbo-jmp-1986}.
A computer implementation can be found in
\cite[pp.\ 27-29 and p.\ 82]{zhuang-ms-thesis-1991}.}
\begin{eqnarray}
\label{obstruction3SolitonConditionfor7thordereqIlinetal}
&&
(k_1 k_2 k_3)^4 \left[(k_1^2 - k_2^2)(k_1^2 - k_3^2)(k_2^2 - k_3^2)\right]^2
\nonumber \\
&& (k_1^2 k_2^2 + k_1^2 k_3^2 + k_2^2 k_3^2)(k_1^4 + k_2^4 + k_3^4 + k_1^2 k_2^2 + k_1^2 k_3^2
+ k_2^2 k_3^2) = 0.
\end{eqnarray}
Thus, the wave numbers must be either equal, each other's opposites, or zero.
In conclusion, the non-existence of a three-soliton solution agrees
with the claim in \cite{ilinetal-csf-2015}.
%
\section{Soliton Solutions in Multiple Space Dimensions}
\label{multiple-space-dimensions}
\subsection{The Kadomtsev-Petviashvili equation}
\label{KP-equation}
Arguably, the KP equation
\cite{ablowitz-clarkson-book-1991,drazin-johnson-book-1989,kp-spd-1970},
\begin{equation}
\label{KPequation}
(u_t + 6 u u_x + u_{3x})_x + 3 \sigma u_{yy} = 0
\end{equation}
for $u(x,y,t)$ and $\sigma = \pm 1$,
is one of the most studied soliton equations
involving more than one space variable.
We only consider the so-called KPII equation \cite{biondini-pelinovsky-scholarpedia-2008,kodama-book-2018}
where $\sigma = 1$.
A Laurent series of its solution suggests the transformation
$u = 2 \, (\ln f)_{xx}$.
We therefore integrate (\ref{KPequation}) twice,
\begin{equation}
\label{intkp}
\partial_t \left( \int^x u \, dx \right) + 3 u^2 + u_{xx}
+ 3 \partial_y^2 \left( \int^x \left(\int^x u \, dx \right) \, dx \right)
= 0,
\end{equation}
before replacing $u$ in terms of $f$.
The resulting equation,
\begin{equation}
\label{kpquadratic}
f (f_{xt} + f_{4x} + 3 f_{yy})
   - f_x f_t + 3 f_{xx}^2 - 4 f_x f_{3x} - 3 f_y^2 = 0,
\end{equation}
can be written in bilinear form
\begin{equation}
\label{hirotakp}
\left( D_x D_t + D_x^4 + 3 D_y^2 \right) (f \mathbf{\cdot} f) = 0,
\end{equation}
where $D_y$ is the Hirota operator defined in a similar way as
$D_x$ and $D_t$ in (\ref{Dxoper}) and (\ref{Dtoper}), respectively.

Continuing with (\ref{kpquadratic}), the computation of
soliton solutions is similar to the KdV case in 
Section~\ref{korteweg-devries-equation}.
Indeed, the forms of $f(x,y,t)$ for multi-soliton
solutions remain the same except that
$\theta_i = k_i x + l_i y -\omega_i t + \delta_i$
with $\omega_i = \frac{k_i^4 + 3 l_i^2}{k_i}$.
Setting $l_i = k_i P_i$ simplifies matters.
Then
$\theta_i
 = k_i \left( x + P_i y - (k_i^2 + 3 P_i^2) t \right) + \delta_i$
and the phase factors are
\begin{equation}
\label{kpaij}
a_{ij} =
\frac{(k_i - k_j)^2 - (P_i - P_j)^2}{(k_i + k_j)^2 - (P_i - P_j)^2}
\end{equation}
and $b_{123} = a_{12} \, a_{13} \, a_{23}$.

Setting $k = 2 K$ and $\delta = 2 \Delta$,
we obtain the solitary wave solution
\begin{equation}
\label{singlesolitarysol1KP}
u(x,y,t)
= 2 K^2 \sech^2 \left[ K\left(x + P y - (4 K^2 + 3 P^2) t \right)
  + \Delta \right]
\end{equation}
which is essentially one-dimensional.

\begin{figure}[h]
\begin{center}
\includegraphics[width=3.12in, height=2.04in]{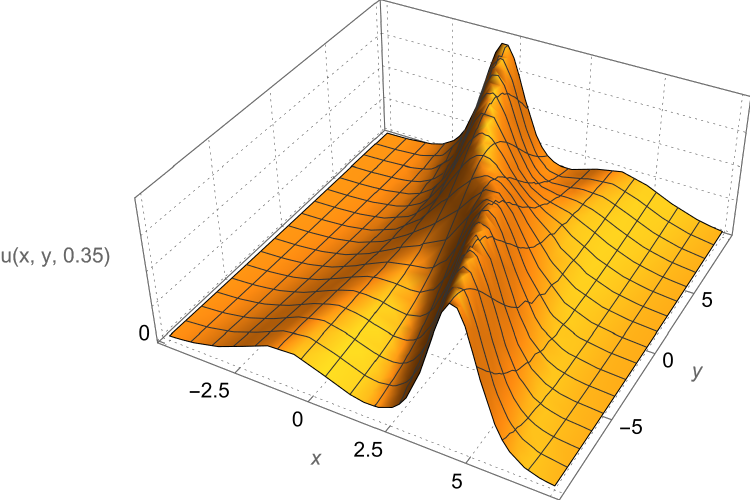}
\end{center}.
\caption{Snapshot of a two-soliton solution for the KP equation.}
\label{KP-two-solitons-in-3D}
\end{figure}

The lengthy expressions for the two- and three-soliton solutions
are not shown for brevity.
A graph of the two-soliton solution of (\ref{KPequation}) at $t = 0.35$
for $K_1 = \frac{1}{2}, K_2 = 1, P_1 = -\frac{1}{8}, P_2 = \frac{3}{16}$
and $\Delta_1 = \Delta_2 = 0$ is shown in Fig.~\ref{KP-two-solitons-in-3D}.
Various types of soliton interactions have been reported in the
literature and observed at flat beaches
\cite{ablowitz-baldwin-pre-2012}.
%

\subsection{A $(3+1)$-dimensional evolution equation}
\label{Geng-Ma-Equation}
Consider the $(3+1)$-dimensional evolution equation \cite{geng-ma-physlettA-2007},
\begin{equation}
\label{gengmainw}
3 W_{xz} -2  \left(2 W_t + W_{3x} -2 W W_x \right)_y
+ 2 \left(W_x \partial_x^{-1} W_y \right)_x = 0,
\end{equation}
which can be written as
\begin{equation}
\label{gengmainu}
3 u_{xxz} - \left(2 u_{xt} + u_{4x} - 2 u_x u_{xx} \right)_y
+ 2 ( u_{xx} u_y )_x = 0
\end{equation}
after substituting $W = u_x$. 
Integrating (\ref{gengmainu}) twice with respect to $x$, yields
\begin{equation}
\label{integratedgengmainu}
3 u_z - 2 \partial_t \left( \int^x u_y \, dx \right)
- u_{xxy} + 2 u_x u_y = 0.
\end{equation}
A Laurent series solution of (\ref{gengmainu}) suggests
the transformation
$u = -3\,(\ln f)_{x} $
which indeed allows one to replace (\ref{integratedgengmainu})
by a homogeneous equation, 
\begin{equation}
\label{op1integratedgengmainu}
f (-2 f_{yt} + 3 f_{xz} - f_{xxxy})
+ 2 f_y f_t
- 3 (f_x f_z + f_{xx} f_{xy} - f_x f_{xxy}) + f_{3x} f_y
= 0,
\end{equation}
of the form $ f {\cal L} f + {\cal N} (f,f) = 0$ with
\begin{eqnarray}
\label{op2integratedgengmainu}
{\cal L} f &=& -2 f_{yt} + 3 f_{xz} - f_{xxxy},
\\
\label{op3integratedgengmainu}
{\cal N} (f,g) &=&
2 f_y g_t - 3 (f_x g_z + f_{xx} g_{xy} - f_x g_{xxy}) + f_{3x} g_y.
\end{eqnarray}
To compute a single solitary wave solution we set
$f = 1 + \e^{\theta}$, where
${\theta} = k x + l y + m z -\omega t + \delta$.
Then, ${\cal L} \e^{\theta} = 0$ yields
$\omega = \frac{k (k^2 l-3m)}{2 l}$.
Since ${\cal N} (\e^{\theta}, \e^{\theta}) = 0$
we readily obtain a solitary wave solution
\begin{equation}
\label{singlesolitarysol1gengma}
u(x,t) =
-\frac{3}{2} k \left(1 + \tanh \left[ \frac{1}{2}
\left(k  x + l y + m z
- \frac{k (k^2 l -3 m) t}{2 l} + \delta \right) \right] \right) .
\end{equation}
For a two-soliton solution we take
$ f = 1 + \e^{\theta_1} + \e^{\theta_2}
+ a_{12} \e^{\theta_1+\theta_2},$
with $\theta_i = k_i x + l_i y + m_i z -\omega_i t + \delta_i$ and
$\omega_i = \frac{k_i (k_i^2 l_i - 3 m_i)}{2 l_i}. $
After some computations
\begin{equation}
\label{rulesa12gengma}
a_{12}
= \frac{k_1 k_2 l_1 l_2 (k_1 - k_2) (l_1 - l_2)
 - (k_1 l_2 - k_2 l_1)(l_1 m_2 - l_2 m_1)}
 {k_1 k_2 l_1 l_2 (k_1 + k_2) (l_1 + l_2)
 - (k_1 l_2 - k_2 l_1)(l_1 m_2 - l_2 m_1)}.
\end{equation}
Thus, a two-soliton solution exists without having to
impose any restrictions on the components $(k_i, l_i, m_i)$
of the wave vector.
In \cite{geng-ma-physlettA-2007}, the authors took
$l_i = k_i$ and $m_i = k_i^3$ from the outset and therefore
only computed a special two-soliton solution for which
$\omega_i = - k_i^3$.
Assuming a traveling frame
$\theta_i = k_i (x+y) + k_i^3 (t+z)$ from the start is
too restrictive.
Indeed, by a change of variables $(x,y,z,t) \rightarrow (\xi,\eta)$
with $\xi = x + y$ and $\eta = t + z$,
one can readily show that after two integrations with respect
to $\xi$ equation (\ref{gengmainw}) becomes 
the integrated KdV equation, that is, (\ref{intkdv}) with $t$ replaced by 
$\eta$, $x$ by $\xi$, and $u(x,t)$ by $u(\xi,\eta)$.

Moving on with our computations, a three-soliton solution does
not exist for arbitrary wave numbers.
Indeed,
\begin{equation}
\label{gengf3sol}
 f = 1 + \e^{\theta_1} + \e^{\theta_2}+ \e^{\theta_3} + a_{12} \e^{\theta_1+\theta_2}
+ a_{13} \e^{\theta_1+\theta_3} + a_{23} \e^{\theta_2+\theta_3}
+ b_{123} \e^{\theta_1 + \theta_2 + \theta_3}
\end{equation}
only yields a three-soliton solution if $l_i = k_i$ with $m_i$
still arbitrary (and a lengthy computation shows that the 
same holds for a four-soliton solution).
The dispersion law and coefficients then simplify into 
$\omega_i = \frac{1}{2}(k_i^3 - 3 m_i)$,  
$ a_{ij} = ((k_i - k_j)/(k_i + k_j))^2 $,
and $b_{123} = a_{12} a_{13} a_{23}$, 
which are the same as for the KdV equation.

\begin{figure}[h]
\begin{center}
\includegraphics[width=3.20in, height=2.88in]{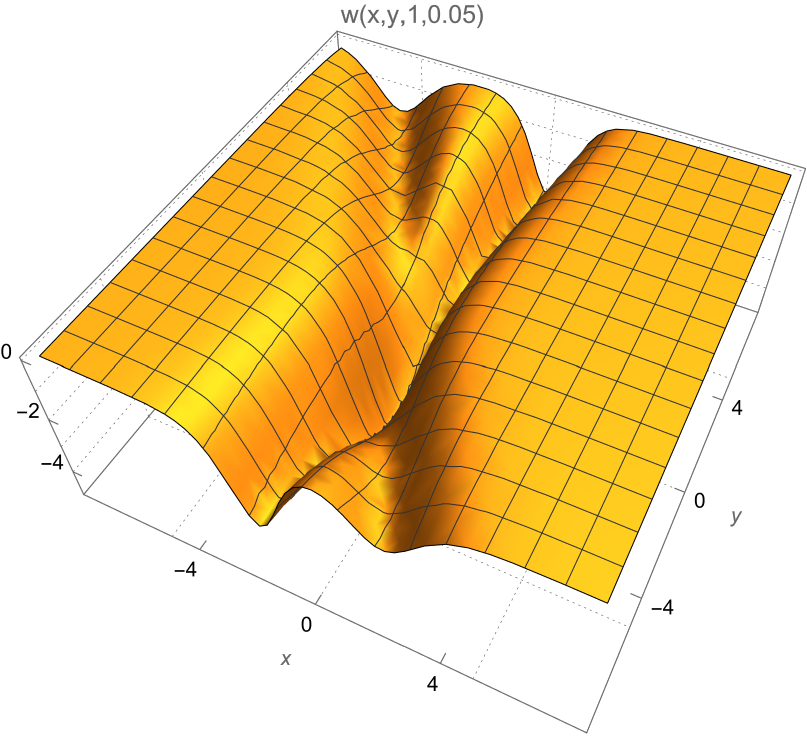}
\end{center}.
\caption{Plot of a two-soliton solution for (\ref{gengmainw})
at $t=0.05$ and  $z=1$ with $k_1 = 2, k_2 = \tfrac{3}{2},
l_1 = -\tfrac{1}{4}, l_2 = \tfrac{3}{4}, m_1 = 4, m_2 = \tfrac{9}{4},$ 
and $\delta_1 = \delta_2 = 0$.}
\label{GengMa-two-solitons-in-3D}
\end{figure}

A graph of a two-soliton solution of (\ref{gengmainw})
at $t = 0.05$ and $z = 1$ with $k_1 = 2, k_2 = \tfrac{3}{2},
l_1 = -\tfrac{1}{4}, l_2 = \tfrac{3}{4}, m_1 = 4,
m_2 = \tfrac{9}{4}$, and $\delta_1 = \delta_2 = 0$
is shown in Fig.~\ref{GengMa-two-solitons-in-3D}.
%
%
\section{Symbolic Software}
\label{software}
Symbolic software for Hirota's method comes in two flavors:
(i) code that aims at finding the bilinear form of a nonlinear PDE and
(ii) code to compute soliton solutions with and without the use of
the bilinear form.

\subsection{Early developments of soliton software}
\label{early-developments}
As part of the design of symbolic software for soliton theory,
in the early 1990s Hereman and Zhuang \cite{hereman-acm-1992,hereman-zhuang-bookames-1992,hereman-zhuang-report-1994,hereman-zhuang-aam-1995} implemented the Hirota method in {\it Macsyma},
a commercial computer algebra system now superseded by
{\it Maxima}\footnote{{\it Maxima} is freely available
from SourceForge at \url{https://maxima.sourceforge.io/}.},
a descendant of the original {\it DOE Macsyma} system.
The code \verb|HIROTA_SINGLE.MAX| is able to automatically compute
up to three-soliton solutions of well-known nonlinear PDEs that
can be transformed into a single bilinear equation of KdV-type \cite{hereman-zhuang-bookames-1992,hietarinta-jmp-1987a},
including the KdV, Boussinesq, KP, SK, and shallow water wave equations.
To compute soliton solutions of these mostly $(1+1)$-dimensional PDEs,
the bilinear form must be given explicitly.
The code can also verify condition (\ref{nsolitoncondition})
for the existence of three- or four-soliton solutions ($n=3$ or $4$).
To cover bilinear equations of mKdV-type \cite{hietarinta-jmp-1987b},
Hereman and Zhuang made \verb|HIROTA_SYSTEM.MAX| \cite{hereman-acm-1992,zhuang-ms-thesis-1991}
which was
applied to various extensions of the mKdV equation taken from
\cite{ito-jpsjpn-1980}.
Codes for the sine-Gordon equation, NLS equations, and various other
types of soliton equations which have quite complicated bilinear forms
\cite{hietarinta-kluwer-1990} were not developed.
The code \verb|HIROTA_SINGLE.MAX| was converted into {\em Mathematica}
syntax and released under the name \verb|hirota.m|.
Further details about these open source
codes\footnote{The codes are still available at \url{https://inside.mines.edu/~whereman}.}
can be found in \cite{hereman-zhuang-report-1994,zhuang-ms-thesis-1991}.

Although the simplified Hirota method (which does not use the
bilinear form) was already published in \cite{hereman-nuseir-mcs-1997,nuseir-phd-thesis-1995},
its implementation did not start until 2012 and is still ongoing.
Cook {\em et al.} \cite{cook-etal-code-2012} developed the
{\em Mathematica} code
\verb|Homogenize-And-Solve.m| to automate the computation of
the soliton solutions discussed in Section~\ref{fifth-order-equation}
and other soliton equations in $(1+1)$-dimensions.
That code is now superseded by \verb|PDESolitonSolutions.m|
\cite{goktas-hereman-code-2023}.
\subsection{Implementation and Limitations of PDESolitonSolutions.m}
\label{implementation-limitations}
The current version  of \verb|PDESolitonSolutions.m| \cite{goktas-hereman-code-2023}
computes up to three-soliton solutions for a given
single PDE in one dependent variable (called $u$ below)
which is function of up to three space variables $(x, y, z)$
and time $(t)$.
The PDE must have polynomial terms with constant coefficients.
Presently, the code can not handle systems of PDEs.
The algorithm largely follows the steps of Section~\ref{simplified-Hirota-method}:
\begin{itemize}

\item [(i)]
The PDE is integrated with respect to $x$ as many times as possible.

\item[(ii)]
The code first attempts to find a transformation to homogenize
the given PDE based on the (truncated) Laurent series expansion of
its solution.
If unsuccessful, the code tries a transformation of type
$u = c \, (\ln f)_{nx},$ with integer
$1 \le n \le n_{\rm max}$ (with default value $n_{\rm max} = 4$)
and constant $c$.
Starting with $n=1$, the code seeks the lowest value of $n$
and matching $c$ so that the PDE is transformed into an
equation that is homogeneous in $f$.

\item[(iii)]
A solution of type
$f(x, y, z, t) =
1 + \sum_{n=1}^{p} \epsilon^n \, f^{(n)}(x, y, z, t)$
is sought where $1 \le p \le p_{\rm max}$
(with default value $p_{\rm max} = 8$).
The bookkeeping parameter $\epsilon$ helps with splitting expressions
into single exponentials, products of two exponentials, etc.
Substituting the above sum for $f$ into a homogeneous equation
for $f$ (of degree $d$) yields an expression of
degree $\ell_{\rm max} = d \, p_{\rm max}$ in $\epsilon$.

\item[(iv)]
Starting with $f^{(1)} = \sum_{i=1}^{N} \phi_i(x,y,z,t)$, 
where the natural number $N$ refers to the $N$-soliton solution one 
aims to compute and 
$\phi_i(x,y,z,t) \equiv \e^{\theta_i}$
$ = {\mathrm e}^{k_i x + l_i y + m_i z - \omega_i t + \delta_i}$,  
at order $\epsilon$ the code balances the linear terms in $\phi_i$
to determine the dispersion relation $\omega_i(k_i,l_i,m_i)$.

\item[(v)]
Next, based on the monomials in the functions $\phi_i$
that occur at order $\epsilon^2$, the code builds
$f^{(2)} = \sum_{i,j} a_{ij} \phi_i \phi_j $
and computes the coefficients $a_{ij}$
(and possible constraints for $k_i, l_i,$ and $m_i$)
by balancing like products of two exponentials.
Note that $i = j$ is allowed to account for terms
in $\phi_i^2$.

\item[(vi)]
At the next orders in $\epsilon$, expressions for
$f^{(3)}, f^{(4)}$, etc., are computed the same way.
If at some order $n < p_{\rm max}$ in $\epsilon$
the function $f^{(n)}$ becomes identically zero,
the code verifies that $f^{(n+1)}, \ldots, f^{(p_{\rm max})}$
can be set to zero.
It also verifies whether or not the coefficients of
$\epsilon^{n+1}, \ldots , \epsilon^{\ell_{\rm max}}$
in the expression mentioned in (iii) all vanish.
For non-solitonic equations this may lead to (additional)
constraints on the wave numbers.
If both verifications are successful, the code returns the
solutions after explicitly verifying that the final $f$
indeed satisfies the homogenized PDE.
If none of the $f^{(n)}$ become zero, the code reports that
a $N$-soliton solution could not be computed.
The code will return a solitary wave solution for $N=1$
and a bi-soliton solution for $N=2$, provided such solutions
exist.
\end{itemize}
Some remarks are warranted:
\begin{itemize}
\item[(i)]
The current code only considers integration with respect to $x$
ignoring the possibility to integrate the given PDE
with respect to $y$ or $z$.

\item[(ii)]
In addition to transformations based on a truncated Laurent series,
currently only single-term logarithmic derivative transformations
with respect to $x$ up to fourth-order are used.
At present only transformations involving one new dependent variable
$(f)$ are considered.
Therefore, the current code can not find solutions of,
for example, the mKdV equation.

\item[(iii)]
With regard to the growing complexity of $f^{(n)}$ as $n$ increases,
$p_{\rm max} = 8$ has been set as default value.

\item[(iv)]
The current code is limited to three space variables and time.
To prevent long expressions and avoid {\em Mathematica}'s conversion
of products of exponentials into a single exponential,
the explicit form of $\phi_i(x,y,z,t)$ is never used.
Instead, the code uses rules for derivatives of $\phi_i(x,y,z,t)$,
such as
$\phi_i(x,y,z,t)_{nx} = k_i^n \phi_i(x,y,z,t)$ and
$\phi_i(x,y,z,t)_{mt} = (-\omega_i)^m \phi_i(x,y,z,t)$.

\item[(v)]
For example for the two-soliton case,
$f^{(2)} =
a_{11} \phi_1^2 + a_{12} \phi_1 \phi_2 + a_{22} \phi_2^2$
where some of these terms might not be included.
Indeed, after substitution of
$f = 1 + f^{(1)} = 1 + \phi_1 + \phi_2$ into the homogeneous
equation, the code generates the list of monomials of type
$\phi_i \phi_j $ (including $\phi_i^2$)
that occur at order ${\epsilon}^2$
and makes a linear combination of those monomials with
undetermined coefficients $a_{ij}$ to create $f^{(2)}$
with the minimal number of terms.
The coefficients $a_{ij}$ are then computed by requiring
that like terms in $\phi_i \phi_j$ vanish.

The same procedure is used to compute $N$-soliton solutions.
Starting from
$f = 1 + f^{(1)} = 1 + \phi_1 + \phi_2 + \ldots + \phi_N$,
the code constructs the minimal expressions for all
subsequent $f^{(n)}$ in which each term is a product of $n$
(not necessarily distinct) functions taken from
$\{\phi_1, \phi_2, \ldots, \phi_N \}$.
The code determines which of these products are
actually needed and combines them with undetermined coefficients.

\item[(vi)]
For homogeneous equations of high degree, some symbolic
verifications can be quite slow.
To speed things up, the code does no longer {\em symbolically}
verify that coefficients of higher orders in $\epsilon$ in the
perturbation scheme
vanish as soon as two consecutive coefficients of lower orders terms
already vanished identically.
Once two consecutive expressions are determined to be zero,
the code {\em numerically} tests if the expressions at higher order
are also zero.
This applies to both the computation of the 
$f^{(n)}$ as well as the coefficients of $\epsilon$ in the perturbation scheme.

Furthermore, verifying that the (lengthy) expressions of $f$
indeed solve the homogeneous equation can be time consuming,
especially for cubic and quartic equations.
Indeed, checking that (\ref{kkfsol3fin}) satisfies
(\ref{kkeqinf}) is computationally very expensive.
Therefore, after the solution is substituted into the homogeneous
equation, all independent variables, wave numbers $(k_i, l_i, m_i)$,
phase constants $\delta_i$, and parameters in the PDE (if present)
are repeatedly replaced by random real numbers in $[-2,2]$.
In each case it is checked if the resulting expression is zero
within machine precision.
Likewise, the solitary wave and one-soliton solutions for
$u(x,y,z,t)$ are tested symbolically but the numerical approach
is used to verify that the 
often long expressions of two- and three-soliton solutions $u(x,y,z,t)$ 
indeed solve the original PDE.
\end{itemize}
\subsection{Other software packages for Hirota's method}
\label{other-software}
As early as 1988, Ito \cite{ito-cpc-1988} designed code in REDUCE to
interactively investigate nonlinear PDEs with Hirota's bilinear and
Wronskian operators.

In \cite{zhouetal-amc-2006}, Zhou {\it et al.} introduced the
{\em Maple} package \verb|Bilinearization|
to convert (mainly) nonlinear evolution equations of KdV-type into their 
bilinear form using logarithmic-derivative transformations.
To cover the mKdV and nonlinear Schr\"{o}dinger equations,
they later extended the algorithm to work for $\arctan$
and rational transformations.
They also added the code \verb|Multisoliton|
to compute up to three-soliton solutions for single bilinear equations
and simple systems of bilinear equations.
Ye {\em et al.} \cite{yeetal-amc-2011,ye-zhou-jmp-2013} presented
a more efficient method to do the same but only with
logarithmic-derivative transformations.
Their method is also implemented in {\it Maple}.
When successful, these {\it Maple} codes return the bilinear form
explicitly.

Yang and Ruan \cite{yang-ruan-ctp-2009,yang-ruan-ctp-2011,yang-ruan-amc-2013}
have produced the {\it Maple} packages \verb|HBFTrans|, \verb|HBFTrans2|,
and \verb|HBFGen|
to transform nonlinear PDEs into their bilinear forms, again based on 
logarithmic derivative transformations.
In their newest algorithms, they take advantage of the properties
of the Hirota operators and the scaling
invariance\footnote{Dilation or scaling symmetry is a special
Lie-point symmetry shared by many integrable PDEs 
\cite{heremanetal-bookchapter-2009}.}
of the original equation.
Doing so, makes their codes more efficient and faster.

Based on the Bell polynomial approach \cite{ma-rmp-2013,ma-fmc-2013},
Miao {\em et al.} \cite{miaoetal-cpc-2014} developed the
{\em Maple} package \verb|PDEBellII|
to compute bilinear forms,
bilinear B\"acklund transformations, Lax pairs, and conservation
laws of KdV-type equations.
In contrast to \verb|PDEBell|,
developed earlier by Yang and Chen,
\verb|PDEBellII| does no longer use scaling invariance
to make it applicable to a broader class of nonlinear PDEs.

For completeness, we mention the new computational method of
Kumar {\em et al.} \cite{kumar-mohan-pdfam-2022} for
the construction of bilinear forms which, as far as we know,
has not been implemented yet.
%
\section{Conclusions and Future Work}
\label{conclusions}
Hirota's bilinear method is an effective method to construct soliton
solutions of completely integrable nonlinear PDEs.
In this paper we discussed a simplified version of Hirota's method
(which does not use Hirota's bilinear operators)
and used it to construct solitary and soliton solutions of various
soliton equations as well as 
some nonlinear polynomial equations that do not have solitons.

We showed that the Hirota transformation is crucial to obtain a PDE
that is homogeneous of degree (in the new dependent variable).
We focused on logarithmic derivative transformations but,
as we saw with the mKdV equation, rational and $\arctan$
transformations might be required, or combinations thereof.
To homogenize, e.g., the Davey-Stewartson system, one needs
a mixture of rational and logarithmic derivative transformations.
There is no systematic way for finding these transformations
but the first few terms of a Laurent series solution and
scaling invariance of the PDE can help determine a suitable
candidate thereby reducing the guess work.

The actual recasting of the transformed PDE into bilinear form in
terms of Hirota's operators, which assumes a quadratic equation
or a tricky decoupling into quadratic equations,
is not required to compute solitary wave solutions or solitons.
Indeed, without bilinear forms, 
exact solutions of the transformed equation can still be constructed
straightforwardly by solving a perturbation-like scheme on the computer
using a symbolic manipulation package.

The simplified version of Hirota's method is largely algorithmic
and now available as the
{\em Mathematica} program \verb|PDESolitonSolutions.m|.
In future releases a broader class of transformations
(likely involving two functions $f$ and $g$) will be considered to
make the code applicable to a large set of PDEs   
including mKdV-type equations.
A future version of the code might follow the algorithm presented
in this paper even closer.
It will use the perturbation schemes involving the linear and
nonlinear operators which will automatically be generated by
splitting the homogeneous equations into linear and nonlinear pieces.
This ``divide-and-conquer" strategy 
is expected to make the computations faster.
An extension of the algorithm and code to systems of PDEs is being investigated. 
%
\section*{Acknowledgements}
One of the authors (WH) thanks Solomon Manukure and Wen-Xiu Ma for the
invitation to give a plenary talk on this subject at the 
6th International Workshop on Nonlinear and Modern Mathematical Physics (NMMP2022).
Both authors are grateful for the opportunity to write this chapter
for the Proceedings of NMMP2022.
We thank Jarmo Hietarinta for providing a photograph of Prof.\ Hirota.

We gratefully acknowledge Wuning Zhuang (Master student) who helped with early
implementations of the Hirota method in {\it Macsyma} and 
Ameina Nuseir (Ph.D.\ student)
who first computed the soliton solutions of the Kaup-Kuperschmidt equation.
We also thank Andrew Cook (REU student) who helped write {\em Mathematica}
code for the simplified Hirota method.
%
\section*{Appendix}
In the derivations below we use that ${\cal L}(f)$ is linear in $f$,
${\cal N}_1(f,g) $ is bilinear (i.e., linear in both $f$ and $g$),
${\cal N}_2$ is trilinear, and ${\cal N}_3$ is quadrilinear.
\vskip 3pt
\noindent
\leftline{\bf Bilinear scheme}
\vskip 2pt
\noindent
For the derivation of the perturbation scheme for an equation of type (\ref{kdvoperform})
we need Cauchy's product formula (to regroup terms in powers of $\epsilon$),
\begin{equation}
\label{cauchyquadratic}
\left( \sum_{p=1}^{\infty} \epsilon^p \, a_p \right)
\left( \sum_{q=1}^{\infty} \epsilon^q \, b_q \right)
= \sum_{n=2}^{\infty} \epsilon^n \sum_{j=1}^{n-1} a_{j} \, b_{n-j}.
\end{equation}
Then,
\begin{eqnarray}
\label{fLfgen}
f \, {\cal L} f
&=& \left( 1 + \sum_{r=p}^{\infty} \epsilon^p \, f^{(p)} \right)
  \, {\cal L} \left( \sum_{q=1}^{\infty} \epsilon^q \, f^{(q)} \right)
\nonumber \\
&=& \left( 1 + \sum_{p=1}^{\infty} \epsilon^p \, f^{(p)} \right)
  \, \sum_{q=1}^{\infty} \epsilon^q \, {\cal L} f^{(q)}
\nonumber \\
&=& \sum_{q=1}^{\infty} \epsilon^q \, {\cal L} f^{(q)}
  + \left( \sum_{p=1}^{\infty} \epsilon^p \, f^{(p)} \right)
  \left( \sum_{q=1}^{\infty} \epsilon^q \, {\cal L} f^{(q)} \right)
\nonumber \\
&=& \sum_{n=1}^{\infty} \epsilon^n \, {\cal L} f^{(n)}
    + \sum_{n=2}^{\infty} \epsilon^n \sum_{j=1}^{n-1} f^{(j)} {\cal L} f^{(n-j)},
\end{eqnarray}
where we have applied (\ref{cauchyquadratic}) with $a_p = f^{(p)}$
and $b_q = {\cal L} f^{(q)}.$
Similarly, we compute
\begin{eqnarray}
\label{Nffgen}
{\cal N}(f,f)
&=& {\cal N}\left( 1 + \sum_{p=1}^{\infty} \epsilon^p \, f^{(p)}, \,
              1 + \sum_{q=1}^{\infty} \epsilon^q \, f^{(q)} \right)
\nonumber \\
&=& {\cal N}\left( \sum_{p=1}^{\infty} \epsilon^p \, f^{(p)}, \,
              \sum_{q=1}^{\infty} \epsilon^q \, f^{(q)} \right)
\nonumber \\
&=& \sum_{n=2}^{\infty} \epsilon^n \sum_{j=1}^{n-1} {\cal N}(f^{(j)}, f^{(n-j)}),
\end{eqnarray}
where again we applied (\ref{cauchyquadratic}) and used the 
bilinearity of ${\cal N}(f,g).$
Adding (\ref{fLfgen}) and (\ref{Nffgen}), the coefficient of $\epsilon^n$ is
\begin{equation}
\label{ordernquadratic}
{\cal L} f^{(n)} + \sum_{j=1}^{n-1}
   \left( f^{(j)} {\cal L}f^{(n-j)}
   + {\cal N}(f^{(j)}, f^{(n-j)}) \right) = 0,
   \;\;\; n \ge 2.
\end{equation}
\vskip 3pt
\noindent
\leftline{\bf Trilinear scheme}
\vskip 2pt
\noindent
For the derivation of the perturbation scheme for equations of 
type (\ref{laxoperform}) we need Cauchy's product formula for three sums:
\begin{equation}
\label{cauchycubic}
\left( \sum_{p=1}^{\infty} \epsilon^p \, a_p \right)
\left( \sum_{q=1}^{\infty} \epsilon^q \, b_q \right)
\left( \sum_{r=1}^{\infty} \epsilon^r \, c_r \right)
= \sum_{n=3}^{\infty} \epsilon^n \sum_{j=2}^{n-1}
  \sum_{\ell=1}^{j-1} a_{\ell} \, b_{n-j} \, c_{j-\ell}.
\end{equation}
Substituting (\ref{hirotaexpansionf}) into (\ref{laxoperform}) and applying
(\ref{cauchyquadratic}) and (\ref{cauchycubic}) yields the following
term in $\epsilon^n$:
\begin{eqnarray}
\label{orderncubic}
&& {\cal L} f^{(n)} + \sum_{j=1}^{n-1}
   \left(
   2 f^{(j)} {\cal L}f^{(n-j)} + {\cal N}_1(f^{(j)}, f^{(n-j)})
   \right)
   + \sum_{j=2}^{n-1} \sum_{\ell=1}^{j-1}
   \left(
   f^{(\ell)} f^{(n-j)} {\cal L}f^{(j-\ell)} \right.
\nonumber \\
&& + \left. f^{(\ell)} {\cal N}_1(f^{(n-j)}, f^{(j-\ell)})
 + {\cal N}_2(f^{(\ell)}, f^{(n-j)}, f^{(j-\ell)})
 \right)
 = 0, \;\;\; n \ge 3.
\end{eqnarray}
\vskip 3pt
\noindent
\leftline{\bf Quadrilinear scheme}
\vskip 2pt
\noindent
Setting up the perturbation scheme for equations of type (\ref{kkoperform})
requires the formula
\begin{eqnarray}
\label{cauchyquartic}
&&
\left( \sum_{p=1}^{\infty} \epsilon^p \, a_p \right)
\left( \sum_{q=1}^{\infty} \epsilon^q \, b_q \right)
\left( \sum_{r=1}^{\infty} \epsilon^r \, c_r \right)
\left( \sum_{s=1}^{\infty} \epsilon^s \, d_s \right)
\nonumber \\
&& = \sum_{n=4}^{\infty} \epsilon^n \sum_{j=3}^{n-1}
\sum_{\ell=2}^{j-1} \sum_{m=1}^{\ell-1} a_{m} \, b_{n-j} \, c_{j-\ell} \, d_{\ell-m}.
\end{eqnarray}
Substituting (\ref{hirotaexpansionf}) into (\ref{kkoperform}) and applying
(\ref{cauchyquadratic}), (\ref{cauchycubic}), and (\ref{cauchyquartic}) yields
the following at $O(\epsilon^n)$:
\begin{eqnarray}
\label{ordernquartic}
&& \!\!{\cal L}f^{(n)} + \sum_{j=1}^{n-1}
   \left(
   3 f^{(j)} {\cal L}f^{(n-j)} + {\cal N}_1(f^{(j)}, f^{(n-j)})
   \right)
\nonumber \\
&& \!\!\!+ \sum_{j=2}^{n-1} \sum_{\ell=1}^{j-1}
   \left(
   \!3 f^{(\ell)} f^{(n-j)} {\cal L}f^{(j-\ell)}
   +\!2 f^{(\ell)} {\cal N}_1(f^{(n-j)}, f^{(j-\ell)})
   +\!{\cal N}_2(f^{(\ell)}, f^{(n-j)}, f^{(j-\ell)})
 \!\right)
\nonumber \\
&& \!\!+ \sum_{j=3}^{n-1} \sum_{\ell=2}^{j-1} \sum_{m=1}^{\ell-1}
  \left(
 f^{(m)} f^{(n-j)} f^{(j-\ell)} {\cal L}f^{(\ell-m)}
 + f^{(m)} f^{(n-j)} {\cal N}_1(f^{(j-\ell)}, f^{(\ell-m)})
 \right.
\nonumber \\
&& \left.
\!\!+ f^{(m)} {\cal N}_2(f^{(n-j)}, f^{(j-\ell)}, f^{(\ell-m)})
 + {\cal N}_3(f^{(m)},f^{(n-j)}, f^{(j-\ell)}, f^{(\ell-m)}) \right) =0,
\nonumber \\
&& \quad\quad\quad\quad\quad\quad\quad\quad\quad\quad\quad\quad\quad\quad
\quad\quad\quad\quad\quad\quad\quad\quad\quad\quad\quad\quad\quad\quad\quad\;
n \ge 4.
\end{eqnarray}
%

\end{document}